\documentclass[a4paper,11pt]{amsart}

\pdfoutput=1
\usepackage{booktabs}
\usepackage{pst-all}
\usepackage{graphicx}
\usepackage{natbib}   

\usepackage{geometry}                		
\bibpunct{(}{)}{,}{a}{}{,}     
\newcommand{\bs}[1]{\boldsymbol{#1}}

\graphicspath{{figs/}}

\title{MCMC algorithms for Bayesian variable selection in the
logistic regression model for large-scale genomic applications}

\author{Manuela Zucknick$^1$, Sylvia Richardson$^2$}
\date{January 2014, \emph{Addresses:} $^1$Division of Biostatistics, German Cancer Research Center (DKFZ), Im Neuenheimer Feld 280, 69120 Heidelberg, Germany, $^2$ MRC Biostatistics Unit, Institute of Public Health, University Forvie Site, Robinson Way, Cambridge CB2 0SR, United Kingdom, \emph{Note:} This manuscript is an update of an older version from March 2009.}

\begin{document}
\maketitle

\begin{abstract}
In large-scale genomic applications vast numbers of molecular features are
scanned in order to find a small number of candidates which are linked to a
particular disease or phenotype. This is a variable selection problem in the
``large $p$, small $n$'' paradigm where many more variables than samples are
available. Additionally, a complex dependence structure is often observed among
the markers/genes due to their joint involvement in biological processes and
pathways.

Bayesian variable selection methods that introduce sparseness through
additional priors on the model size are well suited to the problem. However,
the model space is very large and standard Markov chain Monte Carlo (MCMC) algorithms such as a Gibbs
sampler sweeping over all $p$ variables in each iteration are often
computationally infeasible. We propose to employ the dependence structure in
the data to decide which variables should always be updated together and which
are nearly conditionally independent and hence do not need to be considered
together.

Here, we focus on binary classification applications. We follow the
implementation of the Bayesian probit regression model by \citet{albert93} and the Bayesian logistic regression model by \citet{holmes06}
which both lead to marginal Gaussian distributions. We investigate several MCMC
samplers using the dependence structure in different ways. The mixing and
convergence performances of the resulting Markov chains are evaluated and
compared to standard samplers in two simulation studies and in an application
to a real gene expression data set.
\end{abstract}

\section{Introduction}
Advances in high-throughput technologies in the medical and
biosciences since the mid-1990s have resulted in a shift towards datasets with a vast
number of variables $p$ and a comparably small sample size $n$. In this setup we typically have many more variables than samples, i.e. $p>>n$. It is often reasonable to assume that only a very small subset of all measured variables is sufficient to predict the biological condition or phenotype of interest. This leads to the introduction of sparse regression models where the estimated
regression coefficients $\beta_i$ ($i = 1,...,p$) are assumed to be zero for most input variables. In a Bayesian framework this can be achieved easily by introducing a binary indicator vector $\bs{\gamma} = (\gamma_1,...,\gamma_p)$, which indicates whether a variable $\bs{x}_i$ is considered to be included in the model ($\gamma_i = 1$) or not ($\gamma_i = 0$). Sparsity is induced by setting the prior probability of including a variable to a small value, reflecting the expected model size.

Here, we are particularly interested in applying this Bayesian variable selection (BVS) framework in the binary regression context for modelling the effect of a high-dimensional gene expression data matrix $\bs{x} \in \mathbb{R}^{n\times p}$ (with $p>>n$) on a dichotomous outcome vector $\bs{y} \in \{0,1\}^n$ such as treatment response (response versus non-response) or the categorisation of samples into tumour samples versus healthy tissue. Throughout this manuscript we sometimes refer to covariates $\bs{x}_i$ ($i=1,...,p$) as \emph{genes} or \emph{probe sets}. Other authors who have used BVS in this context
have focussed on the probit regression model for which an auxiliary
variable implementation is available that leads to conjugate
Gaussian priors \citep{albert93}. 
In recent years there have been many attempts to develop similar data augmentation methods for the logistic regression model, for example \citet{holmes06, fruehwirth10, gramacy12, polson13} (see \citet{polson13} for a recent overview). In this manuscript we apply the approach by \citet{holmes06}, which leads to marginal Gaussian distributions after the introduction of an additional layer of parameters in the Bayesian hierarchical model.

Because of the vastness of the model space, posterior inference by
Markov chain Monte Carlo (MCMC) using standard samplers such as full Gibbs
sampling is computationally very demanding, which is a big hurdle
for practical applications. A Bayesian variable
selection model with Gibbs sampling has therefore not been used often in an
application setting with several thousand variables, one example being an application based on microarray gene
expression data  for binary classification by probit regression by \citet{lee03}. With a view to these practical limitations of the Gibbs sampler
\citet{hans07} proposed the shotgun stochastic search (SSS)
algorithm as an alternative method. SSS is
related to MCMC but does not sample from the full posterior
distribution. Instead it performs a stochastic search in the model
space to quickly hone into the regions of high posterior
probability. 

Other approaches have remained in the MCMC setting and attempted to
replace the Gibbs sampler by faster MCMC algorithms. In particular,
an add/delete(/swap) Metropolis-Hastings algorithm has been
proposed, where in each iteration the state of one variable is
proposed to be swapped from $\gamma_i = 0$ to $\gamma_i = 1$ (add
move) or from $\gamma_i = 1$ to $\gamma_i = 0$ (delete move)
\citep{brown98_2,sha04}. Sometimes a swap move is included as well
where the states of two variables $\gamma_i$ and $\gamma_k$ are
proposed to be exchanged. The add/delete(/swap) algorithm is
computationally very fast. However, it has been noted that such
proposals experience problems if $p >> n$ in that the acceptance
probability for deleting variables tends to zero \citep{hans07}.
Also, mixing is a problem, since only one or two randomly selected variables are proposed to be updated in each iteration. This is
especially problematic with $p >> n$, where one usually assumes
sparseness, i.e. only a small number of covariates are
related to the response variable, while most do not carry information
regarding the response. In this situation, the randomly selected
covariates are very unlikely to be related to the response and will
thus not be updated in most MCMC iterations. In addition, the
sampler does not make use of the correlation structure among the
covariates, which increases the likelihood of the sampler getting
stuck: Imagine a situation where two covariates $\bs{x}_i$ and $\bs{x}_k$ are
moderately correlated with each other, and $\bs{x}_i$ has a strong effect
on the response while $\bs{x}_k$ only has a comparably small effect on
$\bs{y}$. If $\bs{x}_k$ is included first by the Metropolis-Hastings sampler then it might prevent the inclusion $\bs{x}_i$ as long as it remains in the model, although inclusion of $\bs{x}_i$ might result in a better model fit.

A complex dependence structure is often observed among genes in high-throughput biological data due to their joint
involvement in biological processes and pathways. Luckily it
is often reasonable to assume that the conditional dependence
structure for such data is sparse, that is each of the variables is
only correlated with a small number of covariates when conditioning
on all other variables in the data set \citep[e.g.][]{west03}. Thus
it might not be necessary to do full Gibbs sampling updating all
covariates in each iteration in order to avoid mixing problems as
the one described above. Rather, one could enjoy the same fast
mixing by exploiting the dependence structure in the data to decide
which variables should always be updated together and which are
nearly conditionally independent and hence do not need to be
considered together. Such a sampler is not as computationally
demanding as full Gibbs sampling and should result in better mixing
relative to the required computation time per iteration.

All these MCMC samplers can be implemented within a parallel tempering framework
\citep{geyer91}. Parallel tempering methods are designed to help overcome local optima by
running different Markov chains at higher temperatures in parallel with the
original Markov chain and proposing to switch the states of the chains in an
additional Metropolis-Hastings step. The aim is to trickle the faster mixing effects in
the higher-temperature chains down to the original sampler, while
still maintaining the proper target invariant distribution for the original
chain. 

In the following section the Bayesian variable selection model for
logistic regression is presented. Then, the MCMC algorithm for
sampling from the logistic variable selection model is described
including the add/delete and Gibbs samplers. We develop our
alternative samplers (which we call neighbourhood samplers) using the dependence structure between
covariates and we outline how we estimate the dependence structure.
The mixing and convergence performances of the MCMC samplers are
evaluated and compared in two simulation studies in Section \ref{sim}. The add/delete sampler and one representative neighbourhood Gibbs sampler are applied to a real gene expression data set in Section \ref{realdata}, where we combine these samplers with parallel tempering. The paper concludes with a discussion of our findings.

The software for sampling from the logistic as well as the probit BVS model is available as \texttt{MATLAB} \citep{matlab06} toolbox \texttt{BVS} (\texttt{http://www.bgx.org.uk/software.html}). Most options of the logistic BVS model are also implemented in the \texttt{R} statistical computing environment \citep{R13} with computationally intensive parts of the MCMC algorithm outsourced to \texttt{C}, in the \texttt{bvsflex} package available on R-forge\\(\texttt{http://bvsflex.r-forge.r-project.org}) \citep{bvsflex13}.

\section{Bayesian variable selection for logistic regression}\label{BVS}
A conjugate formulation for a Bayesian binary regression model with
response $\bs{y} \in \{0,1\}^n$ was first developed for the probit model
by \citet{albert93} by introducing a latent variable $z_j$ for all $j=1,...,n$, which has a
normal prior distribution and hence a conjugate normal posterior
distribution. The binary response $y_j$ for subject $j=1,...,n$ is modelled by the
probit link in a deterministic manner:
\begin{eqnarray}\label{albert1993}
y_j &=& \left\{\begin{array}{ll}1 & \mbox{if }z_j>0\\
0 & \mbox{otherwise}\end{array}\right.\\\nonumber z_j &=& \bs{x}_j \bs{\beta}
+ \epsilon_j\\\nonumber \epsilon_j &\sim& N(0,1)\\\nonumber \bs{\beta}
&\sim& N(\bs{b}, \bs{v}),
\end{eqnarray}
where the regression coefficient vector $\bs{\beta} = (\beta_i)_{i=1}^p$ has a normal prior distribution with mean vector $\bs{b}$ and covariance matrix $\bs{v}$.

Often, the logistic regression model is preferred over the probit
model in statistical applications, as it provides regression
coefficients that are more interpretable due to their connection to
odds ratios. \Citet{holmes06} have developed an auxiliary variable
formulation of the logistic model in the Bayesian context, which uses a latent variable $\bs{z}$ with conjugate
normal priors:
\begin{eqnarray}
y_{j} & = & \left\{\begin{array}{ll}1 & \mbox{if }z_{j}>0\\0 &
\mbox{otherwise}\end{array}\right.\\ \nonumber
z_{j} & = & \bs{x}_{j}\bs{\beta} + \epsilon_{j}\\
\nonumber
\epsilon_{j} & \sim & N(0,\lambda_{jj})\\
\nonumber \lambda_{jj} & = & (2\phi_{j})^2\\ \nonumber \phi_{j} &
\sim & \mbox{Kolmogorov-Smirnov (i.i.d.)}\\ \nonumber \bs{\beta} & \sim &
N(\bs{b}, \bs{v}).
\end{eqnarray}
The auxiliary variables $\phi_{j}$, $j=1,...,n$, are independent
random variables following the Kolmogorov-Smirnov distribution. This
leads to a normal scale mixture distribution for $\epsilon_{j}$
resulting in a marginal logistic distribution, so that this model is
equivalent to a Bayesian logistic regression model
\citep{andrews74}. Since the prior distribution of $\bs{\beta}$ is
normal, the posterior distribution of $\bs{\beta}$ is still normal with
mean $\bs{B}$ and covariance matrix $\bs{V}$, according to standard Bayesian
modelling theory \citep[e.g.][]{lindley72}:
\begin{eqnarray}\label{betapost}
\bs{\beta}|\bs{z},\bs{\lambda} &\sim& N(\bs{B},\bs{V})\\\nonumber 
\bs{B} &=& \bs{V}(\bs{v}^{-1}\bs{b}+\bs{x}' \bs{\lambda}^{-1}\bs{z})\\\nonumber 
\bs{V} &=& (\bs{v}^{-1}+\bs{x}' \bs{\lambda}^{-1}\bs{x})^{-1}\\\nonumber 
\bs{\lambda}^{-1} &=& \mbox{diag}(\lambda_{11},...,\lambda_{nn})^{-1}.
\end{eqnarray}

\Citet{holmes06} extend their Bayesian logistic regression model to
incorporate variable selection by including a covariate indicator
variable $\bs{\gamma} \in \{0,1\}^p$. We denote the size of the active
covariate set by $p_\gamma = \sum_{i=1}^p{I(\gamma_i=1)}$, where $I$
is an indicator function. Then, the Bayesian logistic model for
variable selection is given by
\begin{eqnarray}\label{BVSmodel}
y_{j} & = & \left\{\begin{array}{ll}1 & \mbox{if }z_{\gamma j}>0\\0 & \mbox{otherwise}\end{array}\right.\\\nonumber 
z_{\gamma j} & = & \bs{x}_{\gamma j}\bs{\beta}_\gamma + \epsilon_{j}\\\nonumber 
\epsilon_{j} & \sim & N(0,\lambda_{jj})\\\nonumber 
\lambda_{jj} & = & (2\phi_{j})^2\\\nonumber 
\phi_{j} & \sim & \mbox{Kolmogorov-Smirnov (i.i.d.)}\\\nonumber 
\bs{\beta}_\gamma & \sim & p(\bs{\beta}_\gamma)\\\nonumber 
\bs{\gamma} & \sim & p(\bs{\gamma}) = \prod_{i=1}^p{\pi_i^{\gamma_i}(1-\pi_i)^{1-\gamma_i}}
\end{eqnarray}

The $\bs{\gamma}$ subscripts indicate that the model is only defined for
those components $i$ for which $\gamma_i = 1$. The prior on the
model space is specified in terms of the prior distribution
$p(\bs{\gamma})$, which in this case is a binomial prior with individual
prior probabilities $\pi_i$ for each indicator variable $\gamma_i$.
Throughout this paper we assume constant prior probabilities $\pi_i =
p^*/p$ for all $\gamma_i$ so that the expected number of covariates
\emph{a priori} is $p^*$. The prior distribution on the regression coefficient
vector $\bs{\beta}_\gamma$ is $N(\bs{b}_\gamma, \bs{v}_\gamma)$, where  $\bs{b}_\gamma = \bs{0}_{p_\gamma}$ is typically chosen ($\bs{0}_{p_\gamma}$ denotes the zero vector of length $p_\gamma$). Throughout this manuscript we use the independence prior, which is defined by $\bs{v}_\gamma = c^2 \bs{I}_{p_\gamma}$, where $\bs{I}_{p_\gamma}$ is the identity matrix of dimension $p_\gamma \times p_\gamma$. An alternative would be Zellner's g-prior, i.e. if $\bs{v}_\gamma = c^2(\bs{x}_\gamma' \bs{x}_\gamma)^{-1}$ is chosen, where $\bs{x}_\gamma$ is a sub-matrix of $\bs{x}$ where only those columns $i \in \{1,...,p\}$ are kept for which $\gamma_i = 1$.

The hierarchical logistic regression model (\ref{BVSmodel}) leads to
the following joint posterior distribution for $\{\bs{\beta}_\gamma,
\bs{\gamma}, \bs{z}, \bs{\lambda}\}$ \citep{holmes06}:
\begin{eqnarray}\label{jointpost}
p(\bs{\beta}_\gamma, \bs{\gamma}, \bs{z}, \bs{\lambda} | \bs{x}, \bs{y}) & \propto &
p(\bs{\beta}_\gamma, \bs{\gamma}, \bs{z}, \bs{\lambda}, \bs{y} | \bs{x}) \\\nonumber & = &
p(\bs{y}|\bs{z}) p(\bs{z}|\bs{\lambda}, \bs{\beta}, \bs{\gamma}, \bs{x}) p(\bs{\beta}_\gamma | \bs{\gamma})
p(\bs{\gamma}) p(\bs{\lambda})
\end{eqnarray}
where $$p(\lambda_{jj}) \sim \frac{1}{4\sqrt{\lambda_{jj}}}KS(0.5
\sqrt{\lambda_{jj}})$$ and $$p(\bs{z}|\bs{\lambda}, \bs{\beta}, \bs{\gamma}, \bs{x}) =
N(\bs{x}_\gamma \bs{\beta}_\gamma, \bs{\lambda}).$$

\section{MCMC algorithms}\label{mcmcalgo}

Variables are quite highly correlated due to the way they are constructed, especially
$\bs{\beta}_\gamma$ with $\bs{\gamma}$ and $\bs{z}$ with $\bs{\lambda}$. We can implement the MCMC sampler efficiently by using a blocked Gibbs sampler where $\{\bs{z},\bs{\lambda}\}$ and $\{\bs{\gamma}, \bs{\beta}_{\gamma}\}$ are updated jointly, respectively. Such a sampler has the additional advantage that it allows for efficient updating within the neighbourhoods, because sampling from all distributions involved can be done in a fast manner, see Table \ref{gibbs} below. 

\begin{table}[!ht]
\caption{\label{gibbs}Outline of MCMC sampling algorithm: Gibbs sampling from full conditional distributions $p(\bs{z},\bs{\lambda}|\bs{\beta},\bs{\gamma},\bs{x},\bs{y})$ and $p(\bs{\beta}_\gamma,\bs{\gamma}|\bs{z},\bs{\lambda},\bs{x})$. Note that by $\mbox{Logistic}(\mu,\sigma)$ we denote a logistic distribution with location parameter $\mu$ and scale parameter $\sigma > 0$.}
\centering
\begin{tabular}{lll}
\toprule
&&\\
(1)& Sample from
$p(\bs{z},\bs{\lambda}|\bs{\beta},\bs{\gamma},\bs{x},\bs{y}) =
	p(\bs{\lambda}|\bs{z},\bs{\beta},\bs{\gamma},\bs{x})p(\bs{z}|\bs{\beta},\bs{\gamma},\bs{x},\bs{y})$&\\
\cmidrule(l){2-3}
&&\\
(a)& $p(\lambda_{jj}|z_j,\bs{\beta},\bs{\gamma},\bs{x}_j) \propto p(z_j|\lambda_{jj},\bs{\beta},\bs{\gamma},\bs{x}_j)p(\lambda_{jj})$ $\forall j$, where &Rejection sampling\\
&\hspace*{0em}$p(z_j|\lambda_{jj},\bs{\beta},\bs{\gamma},\bs{x}_j) = N(\bs{x}_{\gamma,j}' \bs{\beta}_\gamma, \lambda_{jj})$ and $p(\lambda_{jj}) = \frac{1}{4\sqrt{\lambda_{jj}}}\mbox{KS}(0.5 \sqrt{\lambda_{jj}})$&\\
(b)& $p(z_j|\bs{\beta},\bs{\gamma},\bs{x},\bs{y}) =
\left\{\begin{array}{ll}\mbox{Logistic}(\bs{x}_{\gamma,j}'\bs{\beta}_\gamma, 1) I(z_j > 0),    &y_j=1\\
                        \mbox{Logistic}(\bs{x}_{\gamma,j}'\bs{\beta}_\gamma, 1) I(z_j \leq 0), &y_j=0
\end{array}\right.$ $\forall j$&Inversion method\\
&&\\
\midrule
&&\\
(2)& Sample from $p(\bs{\beta}_\gamma,\bs{\gamma}|\bs{z},\bs{\lambda},\bs{x}) = p(\bs{\gamma}|\bs{z},\bs{\lambda},\bs{x})p(\bs{\beta}_\gamma|\bs{\gamma},\bs{z},\bs{\lambda},\bs{x})$&\\
\cmidrule(l){2-3}
&&\\
(a)& $p(\bs{\gamma} | \bs{z},\bs{\lambda},\bs{x})  \propto p(\bs{z}|\bs{\lambda},\bs{x},\bs{\gamma})p(\bs{\gamma})$ & Various samplers\\
&& (see Section \ref{blocks}) \\
&&\\
(b)& $p(\bs{\beta}_\gamma|\bs{\gamma},\bs{z},\bs{\lambda},\bs{x}) = N(\bs{B}_\gamma, \bs{V}_\gamma)$, where & Direct sampling\\
&\hspace*{0em}$\bs{B}_\gamma = \bs{V}_\gamma \bs{x}_{\gamma}' \bs{\lambda}^{-1}\bs{z}$ and
$\bs{V}_\gamma =(\bs{x}_{\gamma}'\bs{\lambda}^{-1}\bs{x}_{\gamma} + \bs{v}_{\gamma}^{-1})^{-1}$ &\\
&&\\
\bottomrule
\end{tabular}
\end{table}

\Citet{holmes06} propose to use an add/delete proposal distribution
$q(\bs{\gamma})$ in the Metropolis-Hastings step for updating
$p(\bs{\beta}_\gamma,\bs{\gamma}|\bs{z},\lambda,\bs{x})$, which is similar to the
add/delete/swap algorithm proposed by for example \citet{brown98_2}.
That is, the proposal distribution for a randomly selected $\gamma_i$ is defined as
\begin{equation}
q(\gamma^*_i) = \left\{\begin{array}{ll}1 & \mbox{if }\gamma_i = 0\\
0 & \mbox{if }\gamma_i = 1\end{array}\right..
\end{equation}
This results in the following Metropolis-Hastings acceptance probability for $\gamma_i$ \citep{holmes06}:
\begin{equation}\label{alpha}
\alpha = \min\left\{1, \frac{|\bs{V}_{\bs{\gamma}^*}|^{1/2}
|\bs{v}_{\gamma}|^{1/2}}{|\bs{V}_{\gamma}|^{1/2} |\bs{v}_{\gamma^*}|^{1/2}}
\frac{\exp(0.5\bs{B}'_{\gamma^*}\bs{V}^{-1}_{\gamma^*}\bs{B}_{\gamma^*})}{\exp(0.5\bs{B}'_{\gamma}\bs{V}^{-1}_{\gamma}\bs{B}_{\gamma})}
\frac{1 - \pi_i}{\pi_i}\right\}
\end{equation}
Note that $\bs{\beta}_\gamma$ and $\bs{\beta}_{\gamma^*}$ do not occur in the
acceptance probability and hence only need to be sampled if the move
is accepted. The add/delete sampler is fast and efficient, but
because only one randomly selected covariate is proposed to be
updated per iteration it results in very slow mixing of the Markov
chain if the number of covariates $p$ is large.

The other extreme is the use of an ``inner'' Gibbs sampler for
$\bs{\gamma}$, which updates all $\gamma_i$ ($i = 1,...,p$) in each
iteration of the ``outer'' Gibbs sampler by sampling from the full
conditional distributions
\begin{equation}
p(\gamma_i|\bs{\gamma}_{-i},\bs{z},\bs{x},\bs{\lambda}) \propto
p(\bs{z}|\bs{\lambda},\bs{x},\bs{\gamma})p(\gamma_i) = N(\bs{0}_{n}, \bs{\lambda} + \bs{x}_\gamma
\bs{v}_\gamma \bs{x}_\gamma')\pi_i^{\gamma_i}(1-\pi_i)^{1-\gamma_i}
\end{equation}
It fits the algorithm outlined in Table \ref{gibbs} by setting $\bs{I} = \{1,...,p\}$. Note that here also
$\bs{\beta}_\gamma$ only needs to be updated once after $\bs{\gamma}$ has been
updated, which saves computation time. This sampler has also been
applied to large-scale gene expression data \citep{lee03}. It is
much more computationally intensive per iteration than the
add/delete sampler, but it also results in better mixing of the
Markov chains.

We are interested in comparing the chain mixing relative to required
CPU time for these two MCMC sampler in sparse $p >> n$ situations,
and consequently whether an improved sampler can be constructed
which is more efficient in mixing relative to CPU time than both of
these vanilla samplers. For improving the sampler we assume that $\bs{\gamma}$ has a sparse dependence structure and that
hence there is no need to update all variables together in each
iteration, but only covariates which are related. 
In the following section we outline how we estimate the dependence structure between covariates
in this study and in Section \ref{blocks} we describe the alternatives which are compared with respect to their relative improvements in Markov chain mixing relative to computation time.

Throughout this manuscript, $N$ denotes the overall number of iterations for
which an MCMC sampler was evaluated; $B$ is the length of the
burn-in period, i.e. the number of MCMC iterations in the
initial period where the sampler has not yet converged to the
target distribution. For posterior inference, only the $M = N -
B$ iterations after burn-in are used, where the MCMC samples
are considered to be from the target distribution. In this
context we denote by $(\theta_{i,m})_{m=1}^M$ the vector of
MCMC samples (after burn-in) of any variable $\theta_i$. In
order to simplify the notation, the vector
$(\theta_{i,m})_{m=1}^M$ is also sometimes written as
$\theta_i$. The meaning should always be clear from the
context.


\section{Estimating the dependence structure}\label{dependence}
%

One can assess the dependence of the variables in terms of
their covariance matrix $\bs{S} = (s_{ik})_{i,k=1,...,p}$ and
corresponding correlation matrix $\bs{R} = (r_{ik})_{i,k=1,...,p}$.
Recall that under the assumption that the matrix of covariates $\bs{x}$ follows a multivariate
normal distribution, a correlation of zero between two
covariates $\bs{x}_i$ and $\bs{x}_k$ implies that they are marginally independent.
However, we are rather interested in the conditional
independence of variables, and so the above relationship cannot
be used directly. The matrix of partial correlations
$(\rho_{ik})_{i,k=1,...,p}$, on the other hand, can be used to
infer conditional independences, as under the assumption of
normal distributions of the covariates a partial correlation
$\rho_{ik}$ of zero implies that variables $i$ and $k$ are
conditionally independent given all the other variables $j\neq
i,k$. Note that the partial correlation matrix is related to
the inverse of the standard covariance matrix $\bs{S}$ in the
following way \citep{whittaker90}:
\begin{equation}\label{parcor}
\rho_{ik} = -\frac{s^{-1}_{ik}}{\sqrt{s^{-1}_{ii}s^{-1}_{kk}}},
\end{equation}
where $\bs{S}^{-1} = (s^{-1}_{ik})_{i,k=1,...,p}$ is the inverse of
the covariance matrix $\bs{S} = (s_{ik})_{i,k=1,...,p}$. 

In the $p>>n$ paradigm, the classic maximum-likelihood and
related empirical covariance matrix estimators $\bs{\hat{S}}_{ML}$
and $\bs{\hat{S}}_E = \frac{n}{n-1}\bs{\hat{S}}_{ML}$ can be greatly
improved upon by using biased shrinkage estimators, where a
small introduced bias, e.g. towards a target matrix $\bs{T}$ with
imposed restrictions, can result in a much reduced mean squared
error \citep[e.g.][]{stein56,efron75,schaefer05_2}. The
restricted target matrix has assumptions imposed, which result
in a smaller number of parameters to be estimated and thus in a
reduced dimensionality. 
Here, we follow the approach proposed by \citet{schaefer05_2}
who use the linear shrinkage equation
\begin{equation}\label{shrink}
\bs{\hat{S}} = (1-\Lambda)\bs{\hat{S}}_{E} + \Lambda \bs{T},
\end{equation}
where the estimate is a linear combination of the unbiased
empirical covariance estimate and the target matrix $\bs{T}$. Note
that if $\bs{T}$ is the identity matrix, this is very close to the
ridge estimator with penalty parameter $\Lambda$, which results
from equation (\ref{shrink}) when the unbiased estimator
$\bs{\hat{S}}_E$ is replaced by the maximum-likelihood estimator.
Following \citet{schaefer05_2} we use a slightly more general
target matrix, which is also diagonal but allows for unequal
variance entries on the diagonal. This implies that only the
off-diagonal elements of $\bs{\hat{S}}$ are shrunken. Because of
this, it is more convenient to parametrise the covariance
matrix $\bs{S}$ in terms of variances $s_{ii}$ ($i = 1,...,p$) and
correlations with
\begin{equation}\label{rik}
s_{ik} = r_{ik}\sqrt{s_{ii}s_{kk}}.
\end{equation}
\Citet{schaefer05_2} propose to determine the shrinkage parameter
$\Lambda$ analytically using the lemma of \citet{ledoit03}, minimising the risk function $R(\Lambda)$ associated with the mean squared error loss
\begin{equation}
    \bs{R}(\Lambda) = E(\sum_{i=1}^p(\bs{\hat{s}}_{i} -\bs{ s_{i}})^2),
\end{equation}
where $\bs{\hat{s}}_{i}$ and $\bs{s}_{i}$ are the column vectors of the shrinkage estimator $\bs{\hat{S}}$ in equation (\ref{shrink}) and the covariance matrix $\bs{S}$, respectively. In the case of our target matrix $\bs{T}$ this results in the following optimal value for $\Lambda$:
\begin{equation}
\Lambda^* = \frac{\sum_{i\ne k}{\mbox{Var}(\hat{r}_{ik})}}{
\sum_{i\ne k}{\hat{r}^2_{ik}}},
\end{equation}
where $\hat{r}_{ik}$ is estimated from the empirical covariance
matrix $\bs{\hat{S}}_E = (\hat{s}_{Eik})$ plugged into equation
(\ref{rik}). In practice, $\mbox{Var}(\hat{r}_{ik})$ is being
substituted by an unbiased estimate
$\widehat{\mbox{Var}}(\hat{r}_{ik})$ \citep{schaefer05_2}.
The \texttt{R} package \texttt{corpcor} \citep{corpcor07} was used for the estimation of correlation and partial correlation matrices. 

The estimated correlation and partial correlation
matrices are used to determine which
variables should be updated together in the MCMC algorithm. This could be done by testing $r_{ik} = 0$ (or $\rho_{ik} = 0$)
for all pairs of variables $\bs{x}_i \neq \bs{x}_k$. All coefficient
entries in the correlation matrix $\bs{R} = (r_{ik})_{i,k=1,...,p}$,
which are not considered to be significantly different from
zero, can be interpreted as implying marginal independence
between the corresponding variables $\bs{x}_i$ and $\bs{x}_k$ when we
assume that all variables follow a normal distribution. Under
the same assumption, all partial correlation coefficients
$\rho_{ik}$, which are not significantly different from zero,
can be seen as conditionally independent. 

Instead of estimating the covariance or correlation matrix and then inferring partial correlation estimates
by using equation (\ref{parcor}) and then inducing sparsity by setting non-significant partial correlations to zero, one can also use the fact that partial correlations can also be estimated directly by linearly
regressing each variable on all others. This results in a very
large set of regression equations, effectively one for each
partial correlation coefficient. Sparseness can be introduced
by combining the regression analysis with variable selection.
\Citet{dobra04} have implemented a Bayesian variable selection
approach, while recently \citet{meinshausen06} have used lasso
\citep{tibshirani96} to reduce the number of non-zero
coefficient estimates.


Since we want to use the correlations and partial correlations only to guide the determination of neighbourhoods to decide for which covariate indices the corresponding entries of $\gamma$ should be updated together in a Gibbs sampler, we are not interested in the statistical significance of (partial) correlations in itself. Rather, we use the statistical test results for each pair of covariates to sort them (e.g. by raw p-values or p-values adjusted for multiple testing), and then we apply several
threshold values covering a range of average neighbourhood sizes in our
simulation studies in the following section to allow for comparisons of performances of MCMC samplers with varying neighbourhood
sizes. This gives us an insight into how the
average neighbourhood size relates to the mixing performance of the
Markov chain relative to CPU time per iteration. We
characterise the threshold values $C$ in terms of percentiles
of the distributions of (partial) correlation
coefficients. All pairs of variables, for which the absolute
value of the pairwise partial correlation coefficient value
$|\rho_{ik}|$ is below the threshold, are treated as if they
were conditionally independent, and the corresponding $\bs{\gamma}$
values are not updated together in the MCMC algorithm. In
addition, for comparison within the simulation studies in Section \ref{sim}, the pairwise absolute correlation
values $|r_{ik}|$ are also used to construct the neighbourhoods,
although they only relate to marginal rather than conditional
independence. By replacing all those correlation or partial
correlation matrix entries, for which the absolute values are
below the threshold $C$, by zero, a sparse matrix is created.
Finally, in one of the simulation scenarios in Section \ref{sim}, we will also
construct neighbourhoods simply by randomly drawing variables, matching the neighbourhood sizes with the mean neighbourhood sizes
observed for the partial-correlation-based and
correlation-based neighbourhood structures for comparison, in order to
see whether the structure of the neighbourhoods influences mixing,
rather than neighbourhood size alone.

\begin{figure}[!ht]
\begin{center}
\resizebox{5in}{5in}{\includegraphics{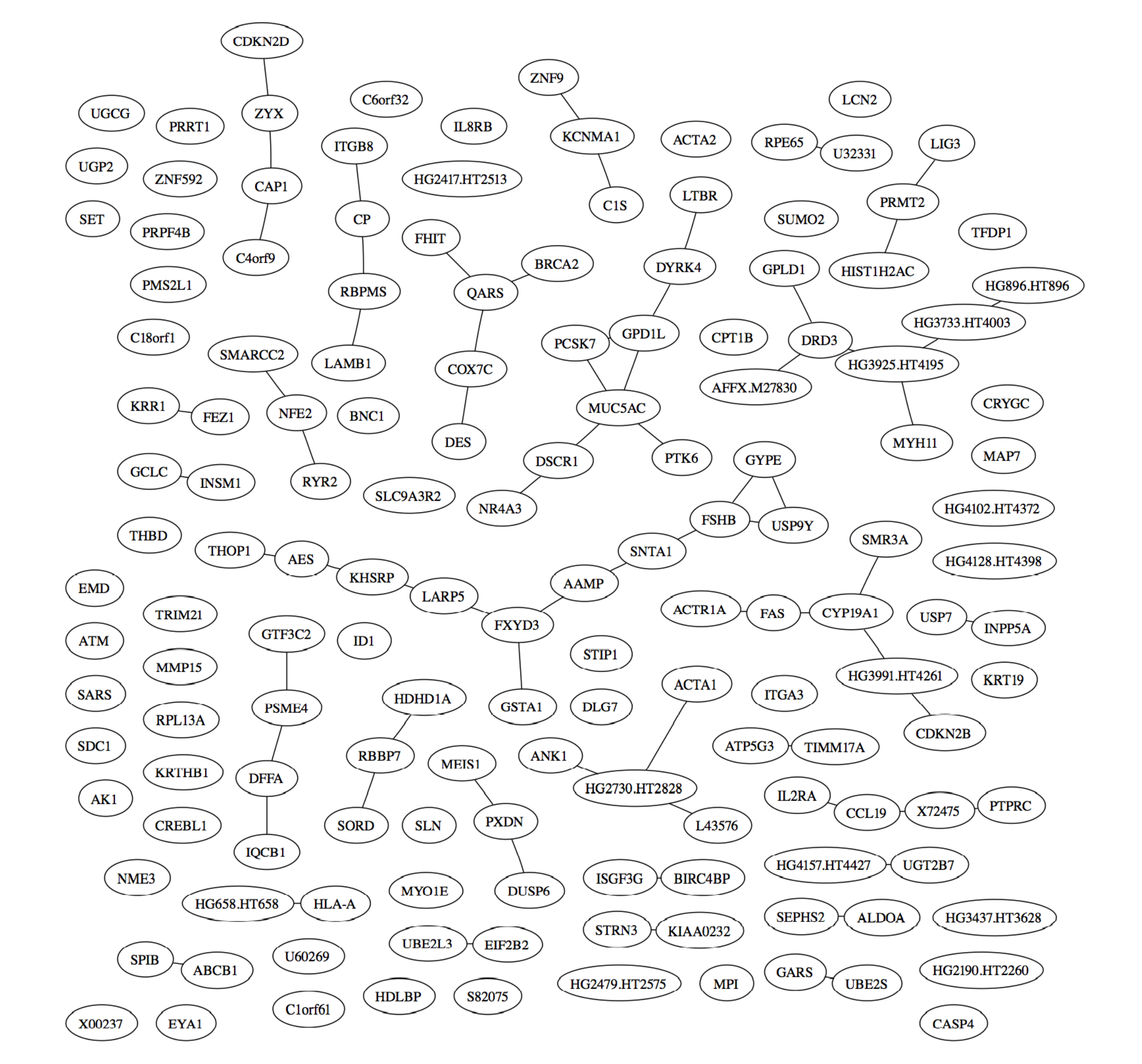}}
\caption[Conditional independence graph for the \citet{schwartz02} gene expression data set
(random subset of 150 probe sets)]{\small \label{graph}Conditional independence graph for the \citet{schwartz02} gene expression data set
(random subset of 150 probe sets), based on a sparse shrinkage estimate of the partial correlation matrix.
Only the partial correlations with absolute values larger than the $C=50\%$
percentile are considered significantly different from zero and shown as edges.
The nodes represent probe sets, which are labelled by the corresponding gene symbols if known,
otherwise they are identified by their Affymetrix probe set ID's.}
\end{center}
\end{figure}

A sparse matrix can be illustrated by a graph where all
non-zero entries represent edges between the nodes which
represent the variables (see for example Figure \ref{graph}). A graph corresponding to a sparse
covariance or correlation matrix is commonly referred to as a
relevance network and a graph representing a partial
correlation matrix is known as a conditional independence graph
\citep[e.g.][]{whittaker90}. 

\section{\label{blocks}MCMC samplers for the covariate indicator $\bs{\gamma}$}
Based on the dependence structure estimated in the way
described above, the covariate indicator vector $\bs{\gamma}$ in the
Bayesian variable selection model could be updated in each MCMC
iteration by first selecting a variable at random and then
updating this variable and in addition all those in the same
neighbourhood, that is the variables which are considered to be related
based on the estimated dependence matrix. In a straight-forward
implementation of the graph structure described above, one
could use all separate sub-graphs as neighbourhoods, which would
produce a natural neighbourhood structure. This is especially the case
when constructing the graph based on the partial correlations
$\rho_{ik}$, since then all nodes (i.e. gene variables), which
are not connected through edges, can be considered
conditionally independent. However, these conditional
independence graphs constructed from gene expression data tend
to consist of a few large sub-graphs (neighbourhoods) and many very
small neighbourhoods, most of them singletons (see Figure \ref{graph} for a small-scale example). This would mean that
whenever a gene in one of the largest sub-graphs is selected
for sampling, this iteration would take quite long and genes in
these sub-graphs would be covered by the MCMC algorithm much
more often than genes, which are in small sub-graphs. Based on
the results of preliminary test runs where we assessed Markov
chain mixing relative to required CPU time, an alternative
approach for neighbourhood-building is preferred here: only the direct
neighbours of a variable, defined as all nodes to which it is
directly connected via an edge in the graph, are considered to
be in a neighbourhood with this variable. Note that
this implies that there is no fixed structure of
non-overlapping neighbourhoods. We have implemented and tried other variations of
this neighbourhood approach, in particular the
possibility to use not only the first-order neighbours but also
a random selection of up to $k$ second-order neighbours. Since
preliminary test runs did not yield promising results, this was
not pursued further, but the implementation is available in the \texttt{MATLAB} toolbox \texttt{BVS}.

For each MCMC iteration, the elements of $\bs{\gamma}$ within the
selected neighbourhood of variables $i \in \bs{I}$ are proposed to be
updated. This can be done by any MCMC sampler. Here we propose
the univariate Gibbs sampler, updating each $\gamma_i$ by
sampling from its full conditional distribution
$p(\gamma_i|\bs{\gamma}_{-i}, \bs{z}, \bs{x},\bs{\lambda})$. In addition, one can
argue that a joint update for all $\gamma_i$ ($i \in \bs{I}$),
sampling from the joint conditional distribution
$p(\bs{\gamma}_I|\bs{\gamma}_{-I}, \bs{z}, \bs{x},\bs{\lambda})$, might be advantageous,
especially here, where the variables within a neighbourhood are
selected because they are considered to be related. Hence, in
the simulation studies in Section \ref{sim}, the following MCMC
algorithms are assessed and compared with respect to mixing and
convergence performances relative to CPU time per iteration:
\begin{enumerate}
\item \textbf{Neighbourhood samplers}: select $\gamma_k$ randomly,
    find the set of neighbours $nb(k)$ and within neighbourhood
    $\bs{I}_k = \{k\} \cup nb(k)$ update using:
\begin{enumerate}
\item \textbf{Univariate Gibbs} ($Gibbs$): for
    each $\gamma_i \in \bs{I}_k$ sample from its full
    conditional distribution $p(\gamma_i|\bs{\gamma}_{-i},
    \bs{z}, \bs{x},\bs{\lambda})$.

\item \textbf{Restricted joint Gibbs}
    ($Joint<d>$): for vector $\bs{\gamma}_{I_{kd}}$ ($\bs{I}_{kd}
    \subseteq \bs{I}_k$) sample from joint full conditional
    distribution $p(\bs{\gamma}_{I_{kd}}|\bs{\gamma}_{-I_{kd}},
    \bs{z}, \bs{x},\bs{\lambda})$ . The size of $\bs{I}_{kd}$ is restricted
    to $d$ for computational reasons, by randomly
    sampling $\min(d, \#\bs{I}_k)$ variables from the set
    $\bs{I}_k$, where $\#\bs{I}_k$ denotes the size of $\bs{I}_k$.

\item \textbf{Restricted univariate Gibbs}
    ($RGibbs<d>$): like univariate Gibbs, but
    only considering $\gamma_i$ with $i \in \bs{I}_{kd}$ in
    order to allow direct comparison with $Joint<d>$.
\end{enumerate}

\item Vanilla samplers for comparison:
\begin{enumerate}
\item \textbf{Add/delete} ($AD$): select one
    $\gamma_i$ at random and propose to change state with a Metropolis-Hastings step

\item \textbf{Full Gibbs} ($Full$): update the
    entire vector $(\gamma_i)_{i=1}^p$ in each MCMC
    iteration by sampling from the respective full conditional distributions $p(\gamma_i|\bs{\gamma}_{-i},
    \bs{z}, \bs{x},\bs{\lambda})$ for all $i=1,...,p$.
\end{enumerate}
\end{enumerate}



\subsection{Evaluation of the performance of MCMC algorithms}
Our main aim is to improve the mixing performance of MCMC
samplers with respect to $\bs{\gamma}$. The mixing of candidate MCMC
samplers is assessed visually by plotting the traces of the
model deviance (i.e. $-2\times$ log-likelihood), the current
size of the model $p_\gamma$, and most importantly the
$\bs{\gamma}$ vector itself. Also, mixing is measured by the
effective sample sizes $\mbox{ESS}(\gamma_i)$
\citep{neal93,kass98} of the indicator variables $\gamma_i$. The effective sample size is based on the autocorrelations
between MCMC steps and intends to assess to what sample size
the observed MCMC sample size would correspond to, in terms of
information contained in the sample, if the samples were
independent observations from the target distribution rather
than highly dependent MCMC samples. For each $\gamma_i$ it is
defined as
\begin{equation}
\mbox{ESS}(\gamma_i)=\frac{M}{\tau(\gamma_i)},
\end{equation}
where
\begin{equation}\label{tau}
\tau(\gamma_i) = 1 + 2\sum_{\kappa=1}^\infty{\varrho_{\kappa}(\gamma_i)}
\end{equation}
is the integrated auto-correlation for estimating $\gamma_i$
using the Markov chain, with $\varrho_ {\kappa}(\gamma_i)$
denoting the auto-correlation at lag $\kappa$. This definition
is motivated by the fact, that $\tau(\gamma_i)$ is equal to one
iff all auto-correlations $\varrho_{\kappa}(\gamma_i)$ are
equal to zero, that is if the samples were independent.
Usually, an MCMC sampler will provide strongly positively
correlated samples, resulting in a reduction of
$\mbox{ESS}(\gamma_i)$ compared to the sample size $M$.

The effective sample sizes are estimated using the R package
\texttt{coda} \citep{coda06}. In \texttt{coda}, in order to
provide robust estimators of the integrated auto-correlation,
the Markov chain is viewed as a time series and an
autoregressive model $AR(k)$ of order $k$ is fitted, assuming
the following relationship between the MCMC sample of
$\gamma_i$ in iteration $m$ and its $k$ previous MCMC samples:
\begin{equation}
\gamma_{i,m} = \alpha_{i1} \gamma_{i,m-1} + ... + \alpha_{ik} \gamma_{i,m-k} + \epsilon_{i,m}.
\end{equation}
The auto-correlations are then estimated from the fitted
$AR(k)$ model and plugged into (\ref{tau}) in order to estimate
$\tau(\gamma_i)$ with $\hat{\tau}(\gamma_i) = 1 +
2\sum_{\kappa=1}^k{\hat{\varrho}_{\kappa}(\gamma_i)}$. The
order $k$ of the autoregressive model is determined via the
Akaike Information Criterion. However, the
maximum possible order that can be fitted is restricted to
$10\log_{10}(M)$, as is suggested by \citet{coda06}, to reduce
the computational burden as well as reduce the variance of the
estimator by removing the small and highly unstable
auto-correlation estimates of high lag $\kappa$. Note that the
stochastic process $(\epsilon_{i,m})_{m=1}^M$ is assumed to be a
white-noise process and autoregressive processes are commonly
used to model continuous normally distributed data. Hence, the
$AR(k)$ process is not completely appropriate for modelling a
Markov chain of samples for the binary indicator variable
$\gamma_i$. However, we are not interested in the
autoregressive model itself but rather in using it to estimate
the effective sample sizes. For this purpose our approach is
found to work well, although in extreme situations
$\mbox{ESS}(\gamma_i)$ can take values which can be
counter-intuitive to the understanding of mixing. In particular,
for an MCMC sample $(\gamma_{i,m})_{m=1}^M$, which consists of
$M-1$ entries of value $0$ and one entry $1$, then
$\mbox{ESS}(\gamma_i) = M$, although one might expect a much
smaller effective sample size value. In reality, such extreme
cases are very rare though. In addition, we use the median as a summary measure to
represent the mixing properties of the MCMC chains for the
entire $\bs{\gamma}$ vector, since the median is robust to such
outliers. Also note that the effective sample size measures
$\mbox{ESS}(\gamma_i)$ are only used to compare mixing
effectiveness of various MCMC samplers which are all applied to
the same data set using the same prior specifications. Hence,
the same posterior distributions are investigated as target
distributions for the MCMC samplers, which ensures that the ESS
values of the various MCMC algorithms are comparable.

In large-scale applications such as gene expression microarray
data analysis it can happen that the majority of the
genes is never selected by an MCMC algorithm sampling from a
sparse model, i.e. that $\gamma_{i,m} = 0 \quad\forall m$ for
more than half of the variables. Then, the straightforward
$\mbox{median}_{i = 1}^p\mbox{ESS}(\gamma_i)$ is zero, because
$\mbox{ESS}(\gamma_i) = 0$ for all variables $i$, that were not
selected at all during the run of the Markov chain. This makes
comparisons of the mixing properties between samplers
impossible based on this measure. We hence prefer a weighted
mean, averaging over the median of all variables that get
selected at least once and the median of those which never get
included in the model:
\begin{eqnarray}\label{esseq}
\mbox{ESS}^*(\bs{\gamma}) &=& \frac{\#\bs{I}_\gamma}{p}\times\mbox{median}_{i \in
    \bs{I}_\gamma}\mbox{ESS}(\gamma_i) + \frac{p - \#\bs{I}_\gamma}{p}\times\mbox{median}_{i \not\in
    \bs{I}_\gamma}\mbox{ESS}(\gamma_i)\nonumber\\
    &=& \frac{\#\bs{I}_\gamma}{p}\times\mbox{median}_{i \in
    \bs{I}_\gamma}\mbox{ESS}(\gamma_i),
\end{eqnarray}
where $\bs{I}_\gamma := \{i: (\sum_{m=1}^M{\gamma_{i,m}}) > 0\}$.

There is a trade-off between the mixing performance of a Markov
chain and the computational complexity of the MCMC algorithm.
Thus, we also compare the ratios
$R$ of average effective sample sizes to CPU times $t$ required
for the $M$ MCMC iterations after burn-in
\begin{equation}
R(\bs{\gamma}) = \frac{\mbox{ESS}^*(\bs{\gamma})}{t}.
\end{equation}

Global convergence of the Markov chains to their target
distribution is monitored by plotting the traces of univariate
summary statistics, in particular model size $p_\gamma =
\sum_{i=1}^p{\gamma_i}$ and model deviance $-2 \log
p(\bs{y}|\bs{x},\bs{\beta}) = 2\sum_{j=1}^n{\log(1 + \exp(-y_j \bs{x}'_j
\bs{\beta}))}$. Also, the trace of the indicator variable vector
$\bs{\gamma}$ is plotted by indicating variables which are included
in the model as points; variables which are excluded from the
model are not shown. Finally, in simulation studies the plots
of marginal posterior probabilities $p(\gamma_i| \bs{x},\bs{y})$ can be
used to check how consistently the ``true'' model is found by
the MCMC sampler.

Since we are mostly interested in finding the most frequently
selected models and variables, we focus on regions of high
posterior probability regions, while keeping in mind that it is
likely that convergence has not yet been reached in low
probability tails of the posterior distribution. A bigger
problem here is that the posterior distribution is multi-modal
because $p >> n$, and that the chains might not have visited
all the modes. That is why good mixing and the ability of the
chains to move freely is more important here than elusive
convergence to the target distribution.

\section{Simulation studies}\label{sim}
In the following the results of two simulation studies are
presented. For both studies, 25 data sets have been simulated
according to a scheme specified below. In an initial step, a variety of possible implementations
of the neighbourhood sampler (as outlined in Section \ref{blocks}) is
applied to two selected data sets only out of all 25 sets. The
purpose of these initial runs is to determine whether there are
notable differences between the performances of the various
neighbourhood sampling implementations and which setting is doing
best. In these initial runs the threshold values $C \in \{99\%,
97.5\%, 95\%, 90\%,80\%,60\%\}$ are tested in both simulation
scenarios, corresponding to sparse estimated dependence
structures where variable pairs are only considered to be
related if their pairwise estimated absolute correlation or
partial correlation is larger than (or equal to) the $C^{th}$
percentile of all pairwise coefficients. A threshold value of
$C = 0$ means that all variables are updated in each iteration,
i.e. that the full Gibbs sampler is applied. An overview over
the MCMC samplers is given in Table \ref{samplers}. The full Gibbs sampler and the neighbourhood sampler with the
$Joint10$ updates within neighbourhoods are run for a smaller number of
MCMC iterations than all other samplers because these samplers
are extremely slow. Note that all Markov chains are started
from randomly sampled starting values for all variables,
sampled from their prior distributions.

After the initial runs, the neighbourhood samplers which performed best
are applied to all simulated data sets to compare these MCMC
neighbourhood algorithms with the vanilla samplers, i.e. the add/delete
Metropolis-Hastings and full Gibbs algorithms. The add/delete
sampler is also applied to all 25 simulated data sets, but the
full Gibbs algorithm is only run for 10 out of all 25 data sets
in both simulation scenarios because of its extreme CPU time
requirements. MCMC iteration numbers are the same as for the
initial runs listed in Table \ref{samplers}.

Because the per-iteration running time for the full Gibbs
sampler and the $Joint10$ neighbourhood sampler is exceptionally long,
these two samplers were run for a shorter total number of iterations and a shorter burn-in period than the other samplers. Global convergence in terms of trace plots for model deviance and model
size was achieved well within the chosen burn-in period for all samplers. Inference on mixing and convergence performance was
adjusted for differences in run lengths. Throughout, all
post-burn-in samples were used for posterior inference and
assessment of MCMC performance, i.e. no thinning was performed.

\begin{table}[!ht]
  \caption[MCMC samplers applied to two data sets in both simulation scenarios]{\small MCMC samplers applied in initial runs to two data sets out of all 25 sets in both simulation scenarios.}\label{samplers}
  \centering
  \small
  \begin{tabular}{|p{1.4cm}|p{1.8cm}|p{5cm}|p{2.75cm}|p{2.75cm}|}
    \hline
    Label & \multicolumn{2}{|l|}{MCMC sampler} & \multicolumn{2}{|l|}{MCMC run length N (burn-in length B)} \\
    \hline
    & \multicolumn{2}{|l|}{} & Simulation 1 & Simulation 2\\
    \hline\hline
    $AD$ & \multicolumn{2}{|l|}{Add/delete Metropolis-Hastings} & $200,000$ $(50,000)$ & $250,000$ $(50,000)$ \\
    \hline
    $Full$ & \multicolumn{2}{|l|}{Gibbs update of all $(\gamma_i)_{i=1}^p$} & $90,000$\quad $(10,000)$ & $110,000$ $(10,000)$ \\
    \hline
     & \multicolumn{2}{|l|}{Neighbourhood samplers} &  &  \\
    \hline
     & neighbour-hood type & Update within neighbourhood $\bs{I}$ & & \\
    \hline
    $Pcor$\quad$<C>$ & partial correlation & Univariate Gibbs update of all $i \in \bs{I}$ & $200,000$ $(50,000)$ & $250,000$ $(50,000)$ \\
    \hline
    $Corr$\quad$<C>$ & correlation & Univariate Gibbs update of all $i \in \bs{I}$ & $200,000$ $(50,000)$ & $250,000$ $(50,000)$ \\
    \hline
    $Random$\quad$<C>$ & random selection & Univariate Gibbs update of all $i \in \bs{I}$ & not applied & $250,000$ $(50,000)$ \\
    \hline
    $Rgibbs4$ & partial correlation & Univariate Gibbs update of subset of $\bs{I}$ of size 4  & $200,000$ $(50,000)$ & $250,000$ $(50,000)$ \\
    \hline
    $Joint4$ & partial correlation & Joint Gibbs update of subset of $\bs{I}$ of size 4 & $200,000$ $(50,000)$ & $250,000$ $(50,000)$\\
    \hline
    $Rgibbs10$ & partial correlation & Univariate Gibbs update of subset of $\bs{I}$ of size 10 & $200,000$ $(50,000)$ & $250,000$ $(50,000)$ \\
    \hline
    $Joint10$ & partial correlation & Joint Gibbs update of subset of $\bs{I}$ of size 10 & $90,000$\quad $(10,000)$ & $110,000$ $(10,000)$ \\
    \hline
  \end{tabular}
\end{table}

\subsection{Simulation scenario 1: generated covariance structure}
\subsubsection{Simulation setup}
\begin{figure}[!ht]
\begin{center}
\scalebox{0.7}{\includegraphics{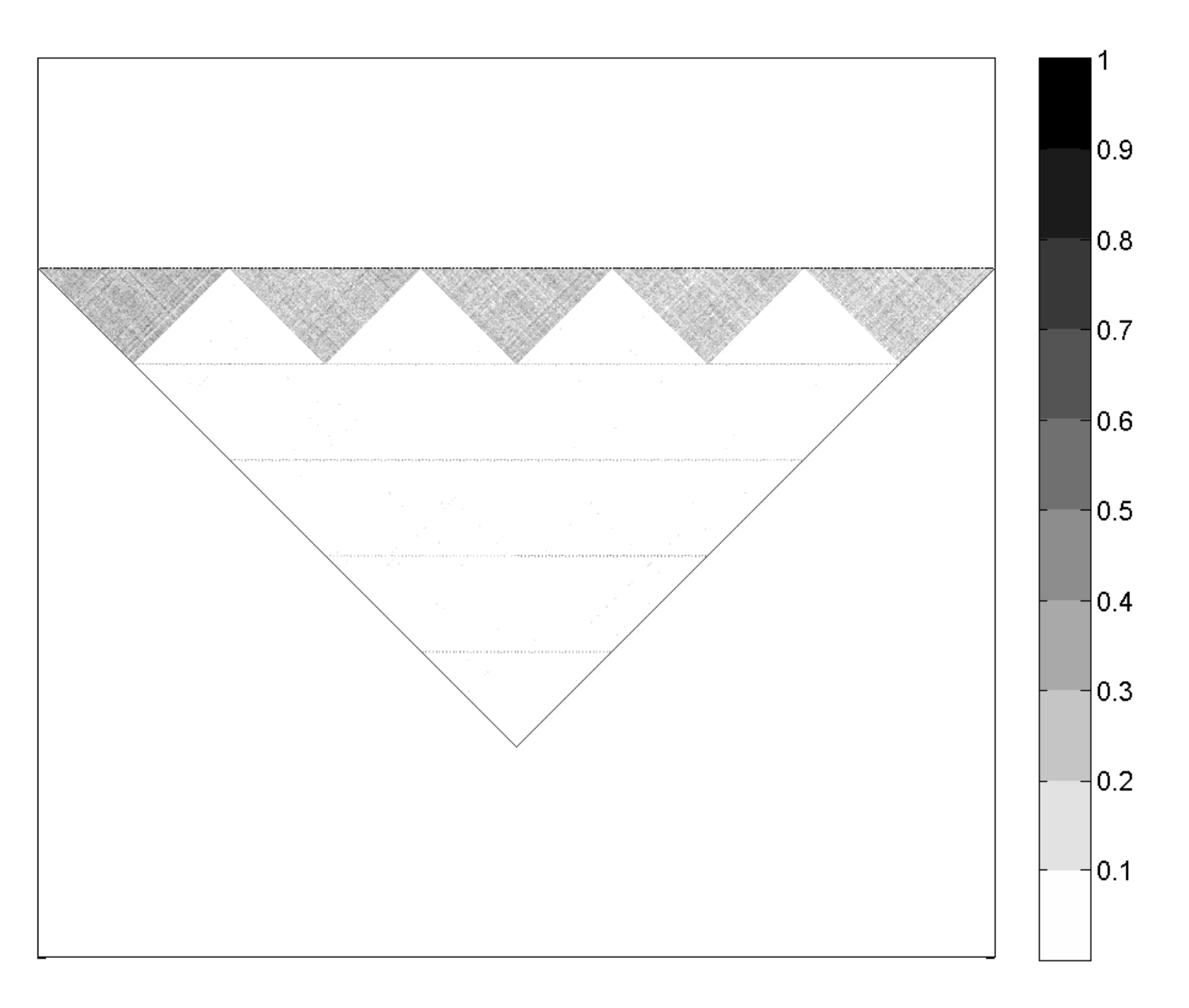}}
\caption[Squared empirical correlation structure imposed on data set 1
in simulation scenario 1]{\small \label{sim1corr} Squared empirical correlation structure imposed on data set 1
in simulation scenario 1.}
\end{center}
\end{figure}

The algorithm in Table \ref{sim1algo} is used to simulate 25
data sets $(\bs{x},\bs{y})$, so that the input data sets $\bs{x}$ have $p =
500$ variables and $n = 100$ samples, where the first $p^* = 5$ variables
$(\bs{x}_1,...,\bs{x}_5)$ are related to the binary response $\bs{y}$ via a
logistic link. The variables are simulated, in a similar manner
to example 4.2 in \citet{george93}, such that there are five
blocks of $100$ variables each, with moderately strong positive
correlations between the variables within blocks which are
induced by adding the same standard normal variable $\bs{w}$ to
$100$ independent standard normals $\bs{x}^*_1,...,\bs{x}^*_{100}$. In
addition, correlation is also introduced between blocks by
using the same variables $\bs{x}^*_1,...,\bs{x}^*_{100}$ for generating
the five blocks (but with different variables $\bs{w}$ added to
them). The correlation structure that is imposed by this
data-generating scenario is illustrated by triangular image
plot of the squared empirical correlation matrix of one
example data set in Figure \ref{sim1corr}. The variables linked to the response, i.e. $\bs{x}_1,...,\bs{x}_5$, are all in the same block. They are thus positively correlated with each other and with all other variables in their block. They are also correlated with the first five variables in all subsequent blocks, that is $\bs{x}_1$ with $\bs{x}_{101},\bs{x}_{201},\bs{x}_{301},\mbox{ and }\bs{x}_{401}$, etc. In such a scenario it is harder for a sampling algorithm to find the correct model with all five true covariates $\bs{x}_1,...,\bs{x}_5$ than if they were unrelated.
\begin{table}[!ht]
\caption[Simulation scenario 1 with generated covariance
structure]{\small\label{sim1algo}Simulation scenario 1 with
generated covariance structure.} \vspace{0.5em} \hrule
\vspace{0.5em}
For $j=1,...,n=100$ do\hfill.
\begin{enumerate}
\item $x^*_{1j},...,x^*_{qj} \mbox{ iid } \sim N(0,1)$ with $q=100$
\item For $m = 0,...,4$ do
\begin{enumerate}
\item $w_m \sim N(0,1)$
\item $x_{(m\times q + i^*)j} = x^*_{i^*j} + w_m$ ($i^*=1,...,q$)
\end{enumerate}
\item $y_j \sim \mbox{Bernoulli}\left(\frac{\exp(\bs{x}_j\bs{\beta})}{1 +
\exp(\bs{x}_j\bs{\beta})}\right)$, with vectors $\bs{\beta} = (2,2,2,2,2,0,...,0)$ and $\bs{x}_j = (x_{1j},...,x_{pj}$ have length $p=500$.
\end{enumerate}
\hrule
\end{table}

The Bayesian logistic variable selection model outlined in
Section \ref{BVS} is fitted to each of the 25 data sets
$(\bs{x},\bs{y})$.
Based on the convergence and mixing performances observed in
the first two data sets, and as shown in the following, the
$Pcor90$ sampler was chosen, i.e. the neighbourhood sampler based on the
partial correlation matrix with threshold $C = 90\%$. The prior parameter
$c^2$ in the independence prior distribution $p(\bs{\beta}) =
N(\bs{0}_{p_\gamma},c^2\bs{I}_{p_\gamma})$ was set to $c^2=5$ to guarantee a
good coverage of the range of values expected for $\bs{\beta}$. The
prior probability for $\gamma_i = 1$ was set to $\pi_i = p^*/p =
0.01$ so that the prior expected number of selected variables
is equivalent to the true number $p^* = 5$.

\subsubsection{Markov chain mixing performance}
\begin{figure}[!ht]
\begin{center}
\scalebox{0.25}{\includegraphics{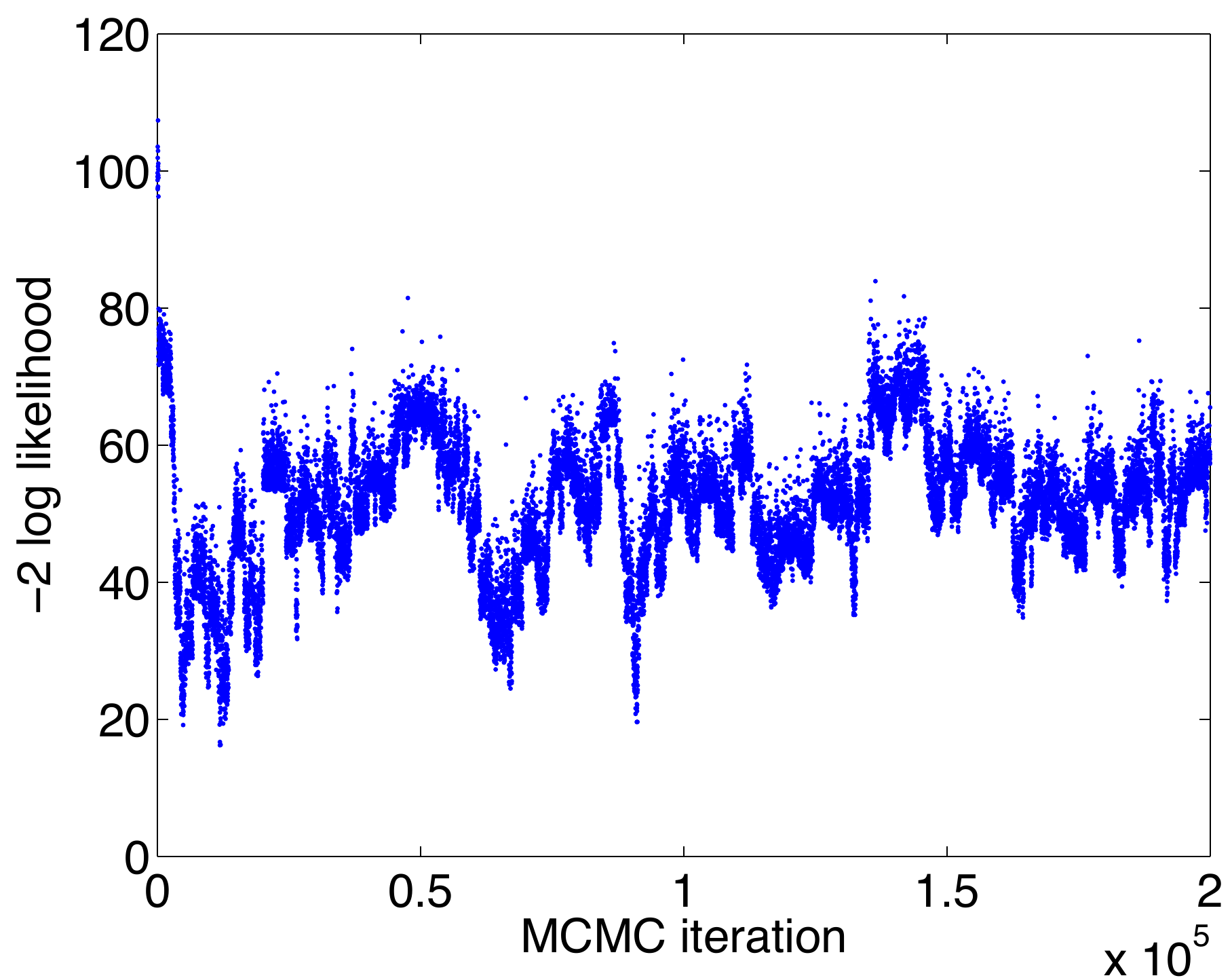}}
\scalebox{0.25}{\includegraphics{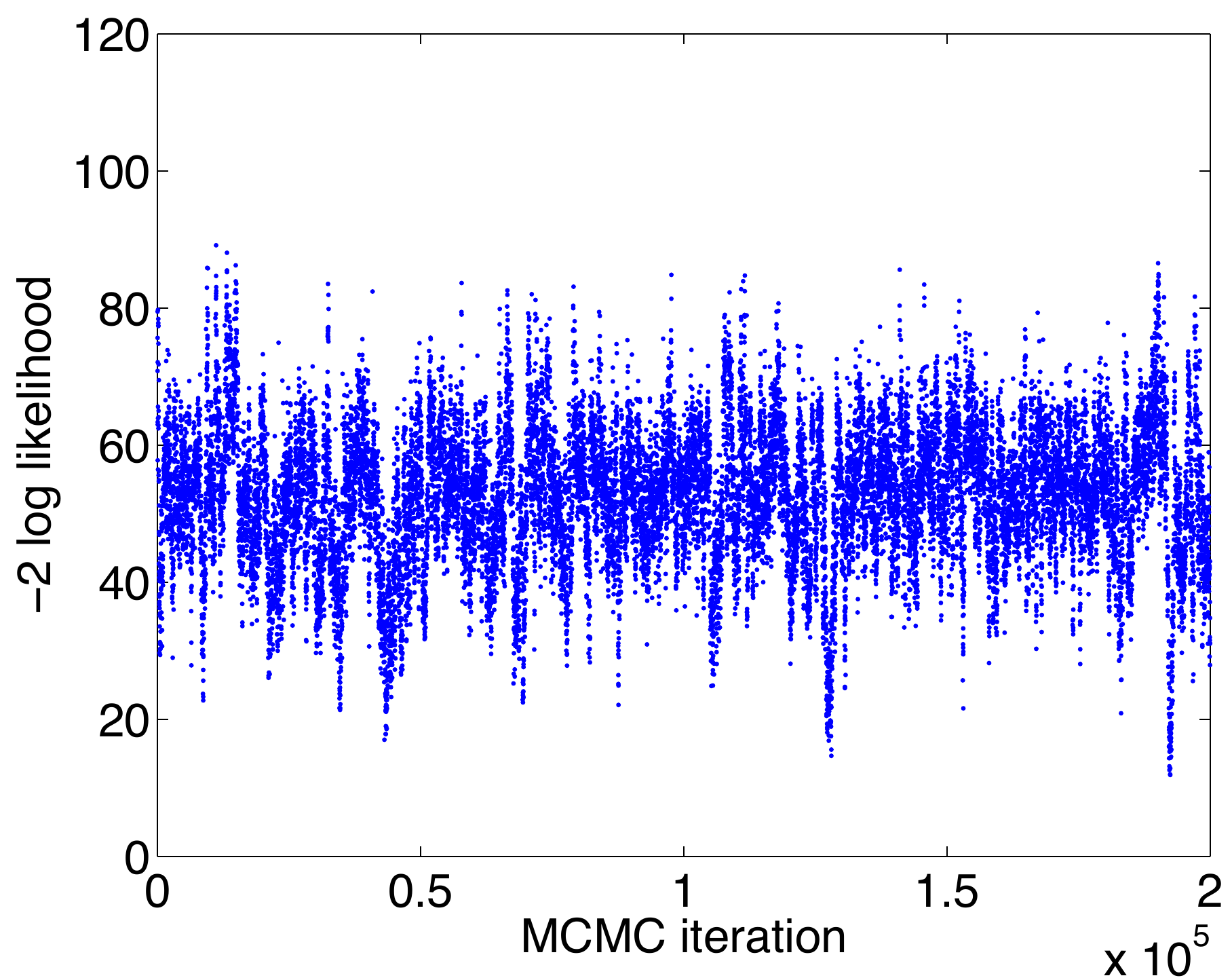}}
\scalebox{0.25}{\includegraphics{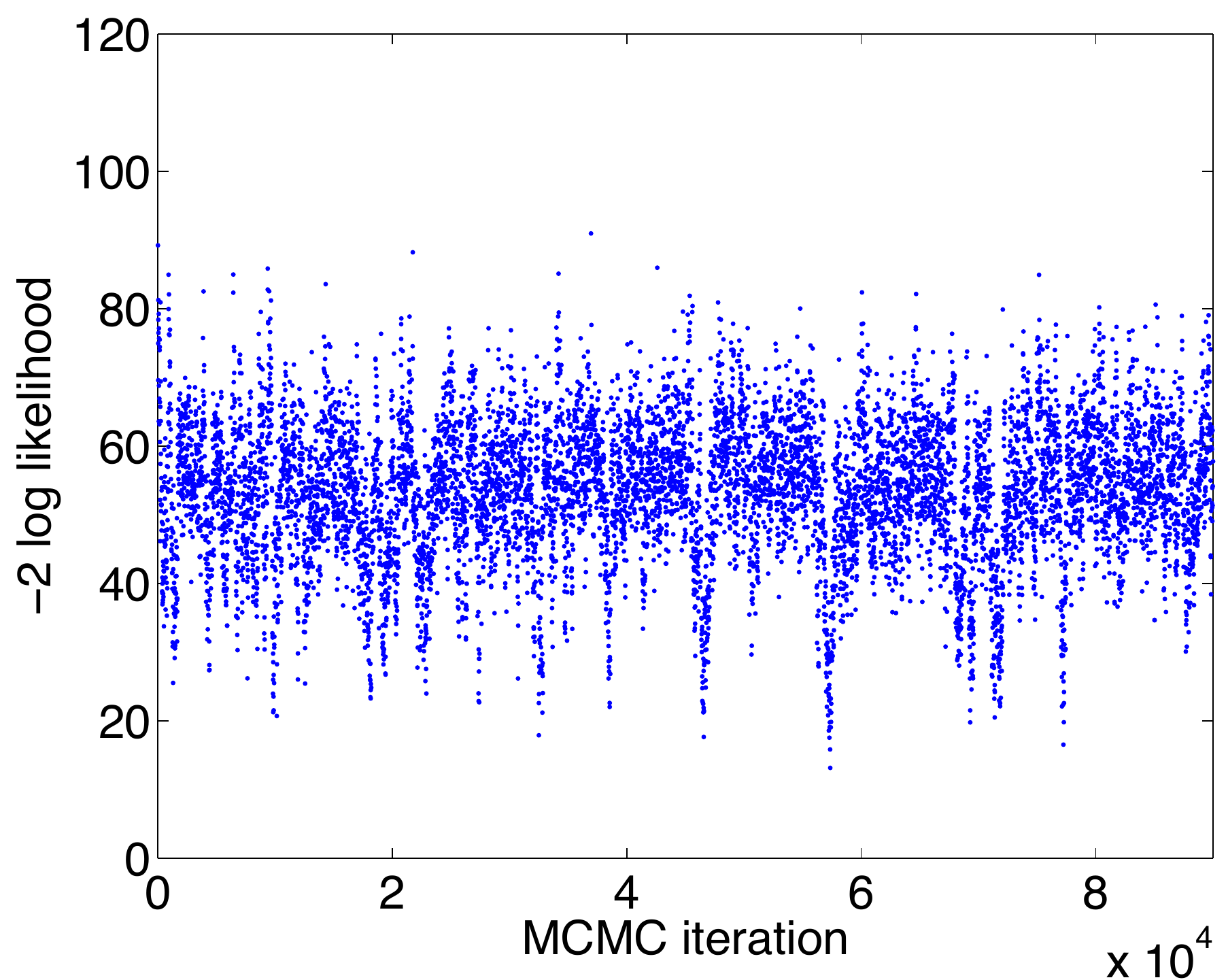}}\\
\scalebox{0.25}{\includegraphics{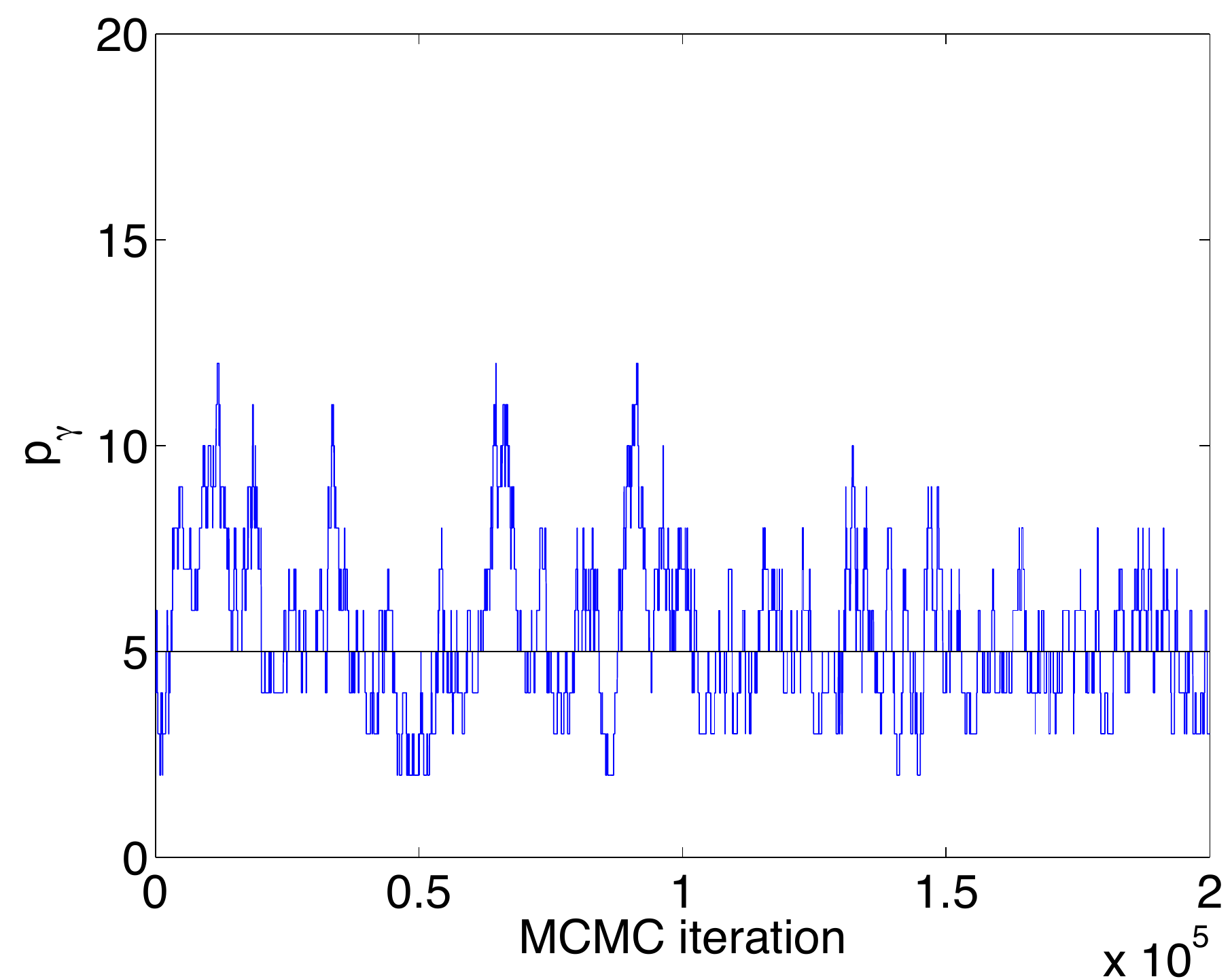}}
\scalebox{0.25}{\includegraphics{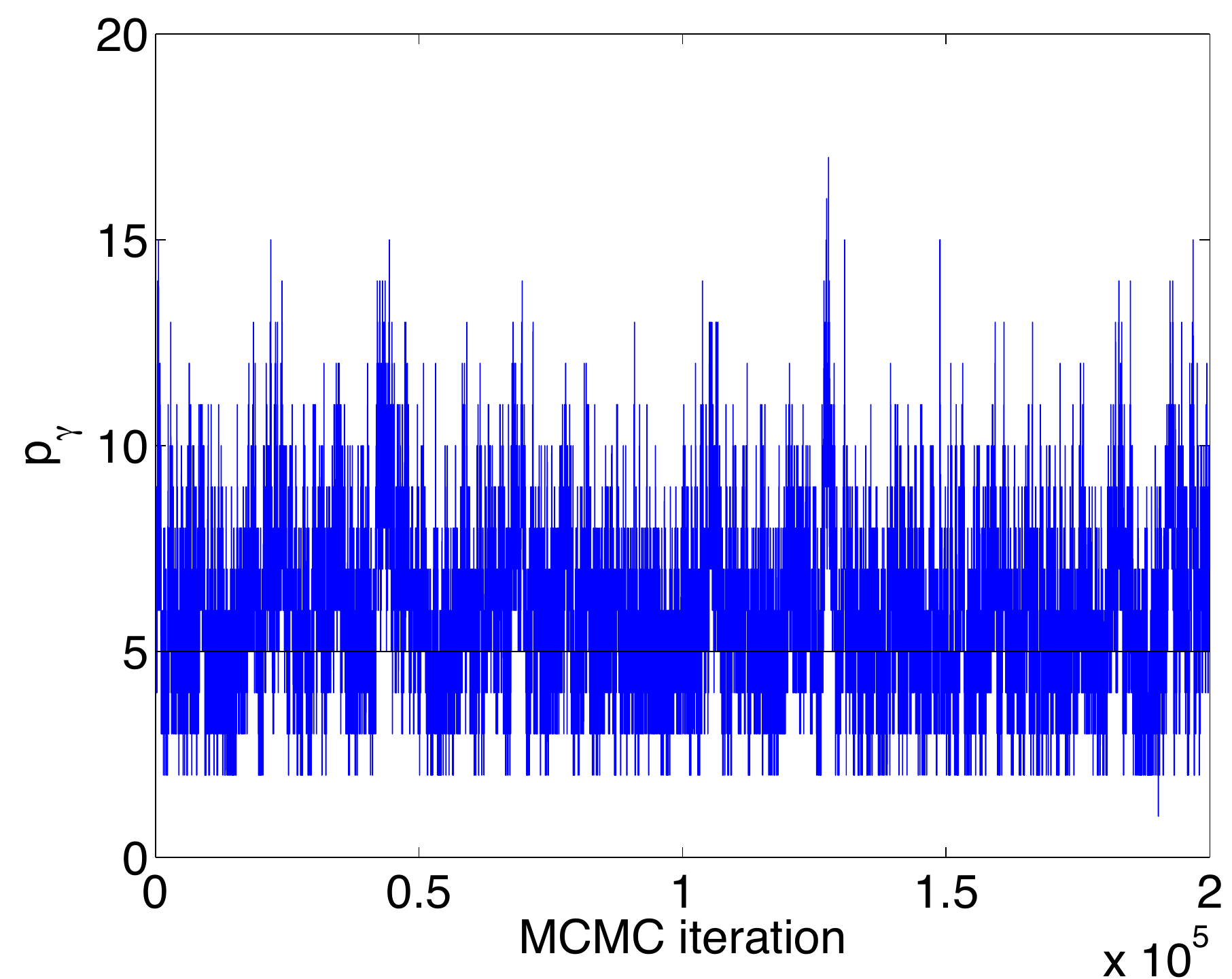}}
\scalebox{0.25}{\includegraphics{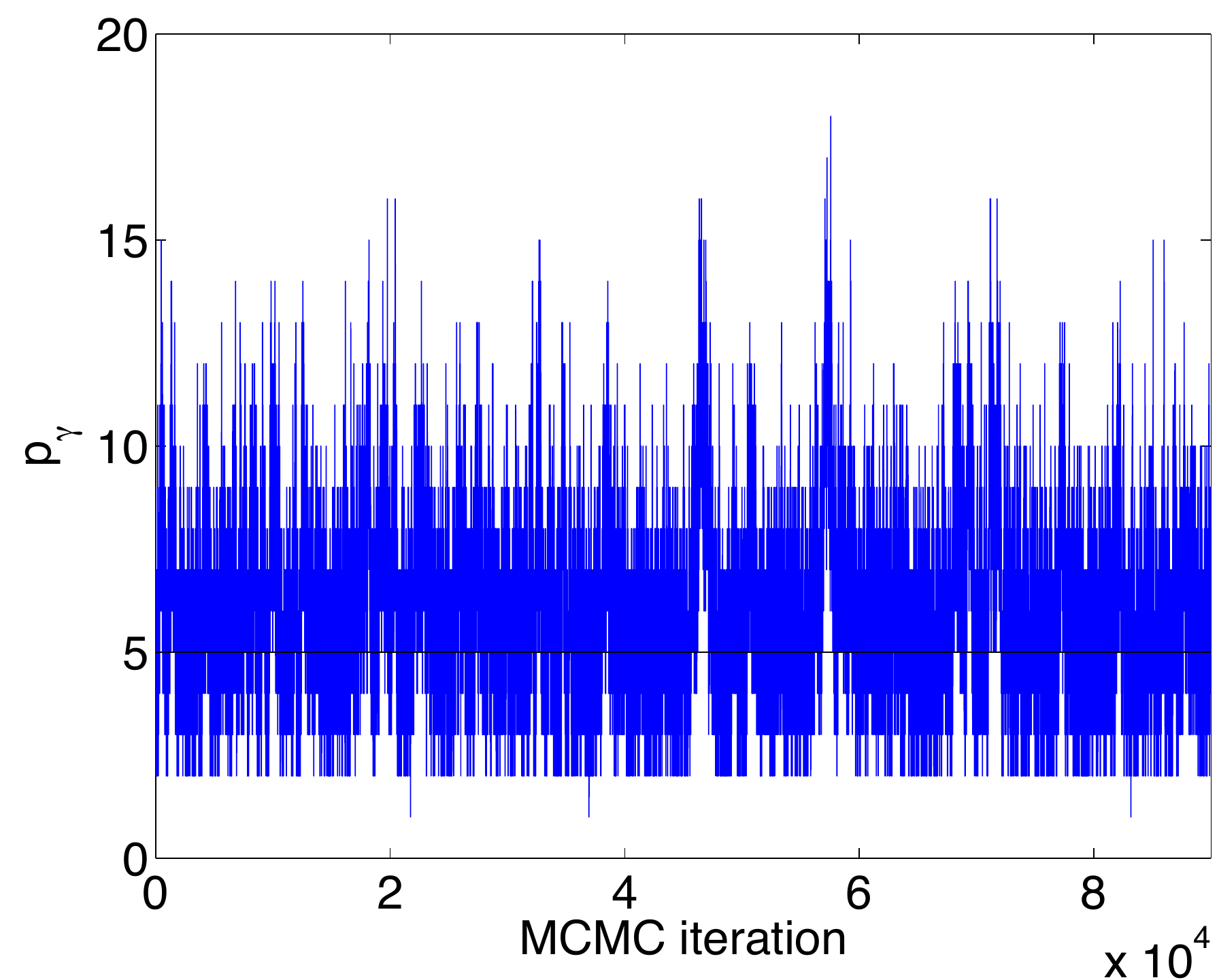}}\\
\scalebox{0.31}{\includegraphics{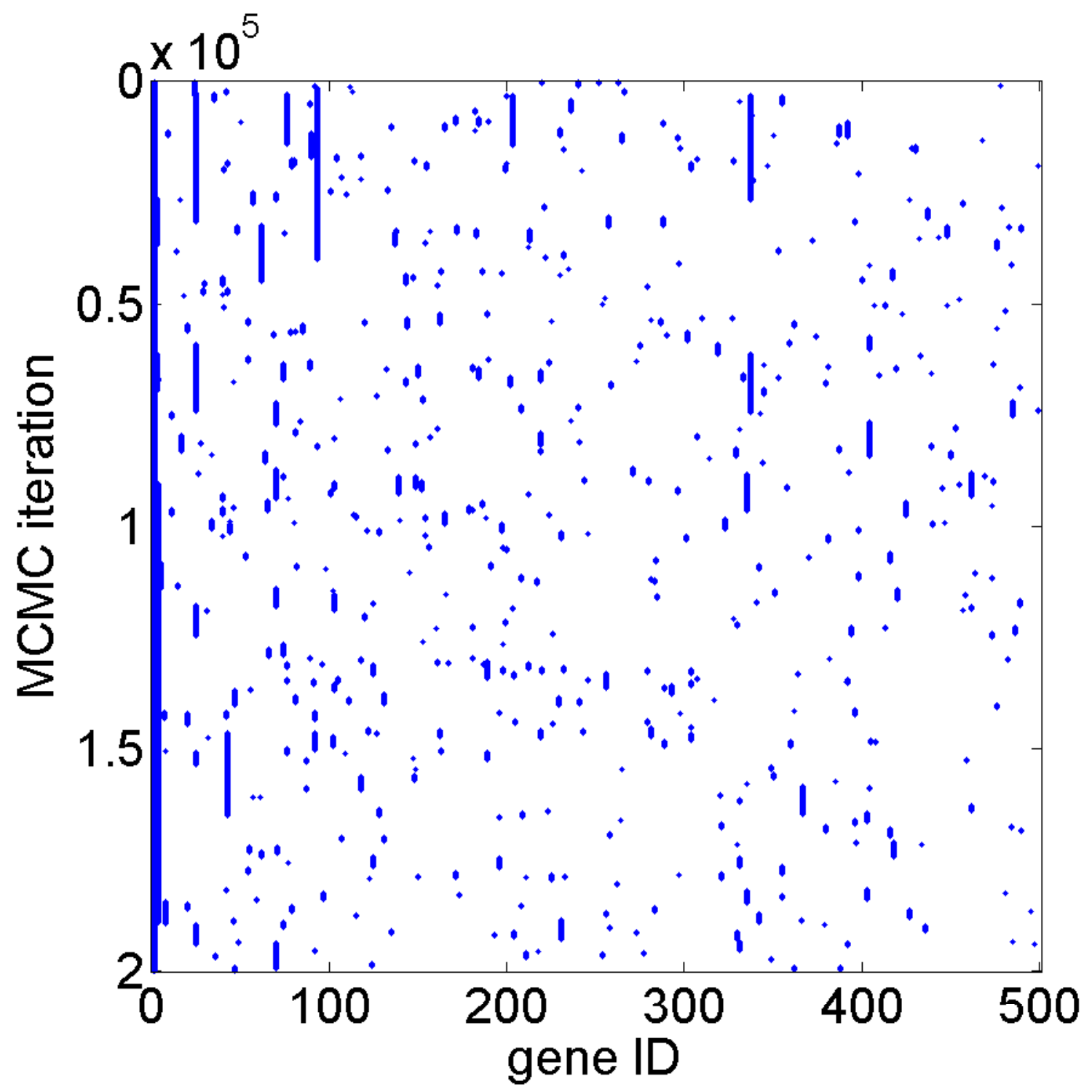}}
\scalebox{0.31}{\includegraphics{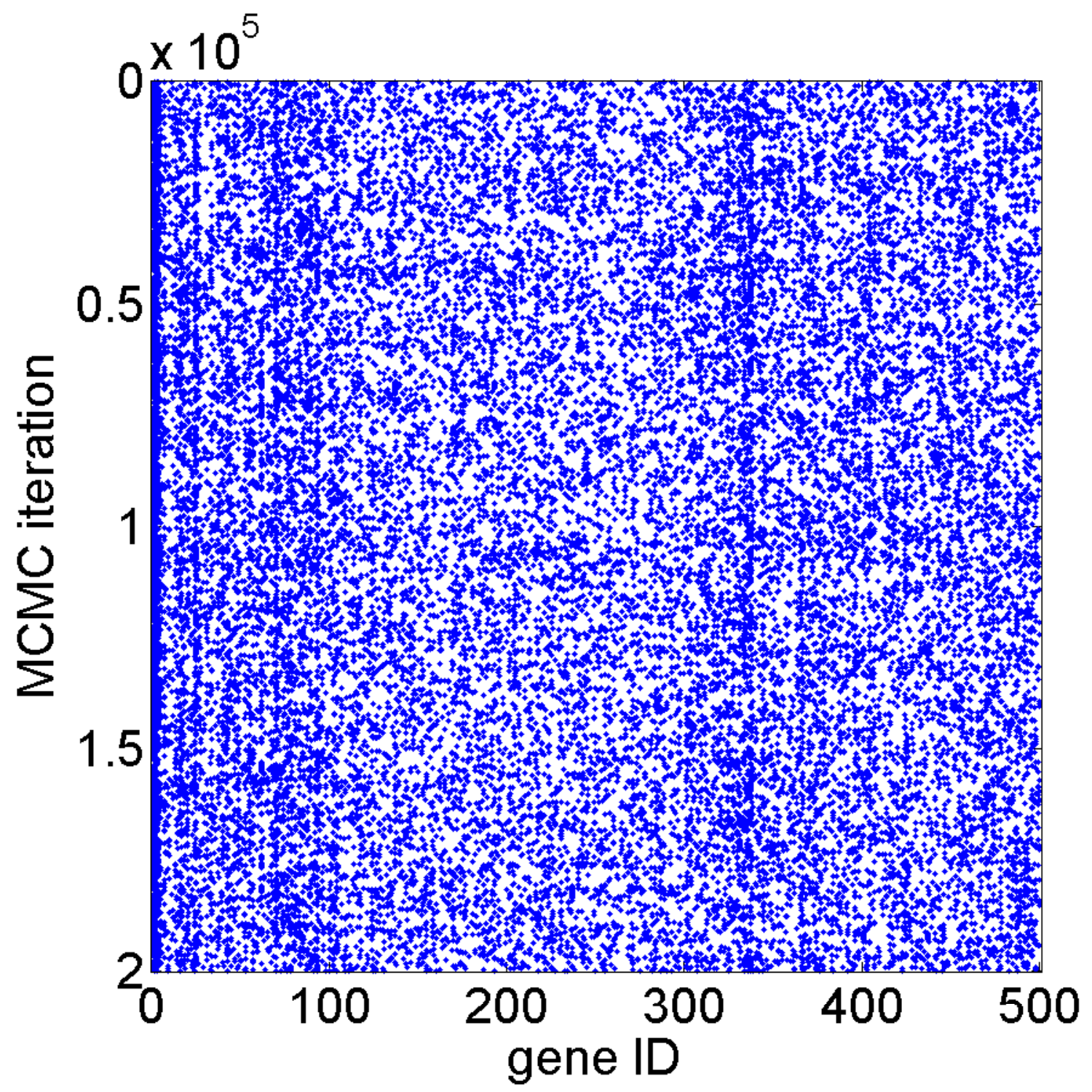}}
\scalebox{0.31}{\includegraphics{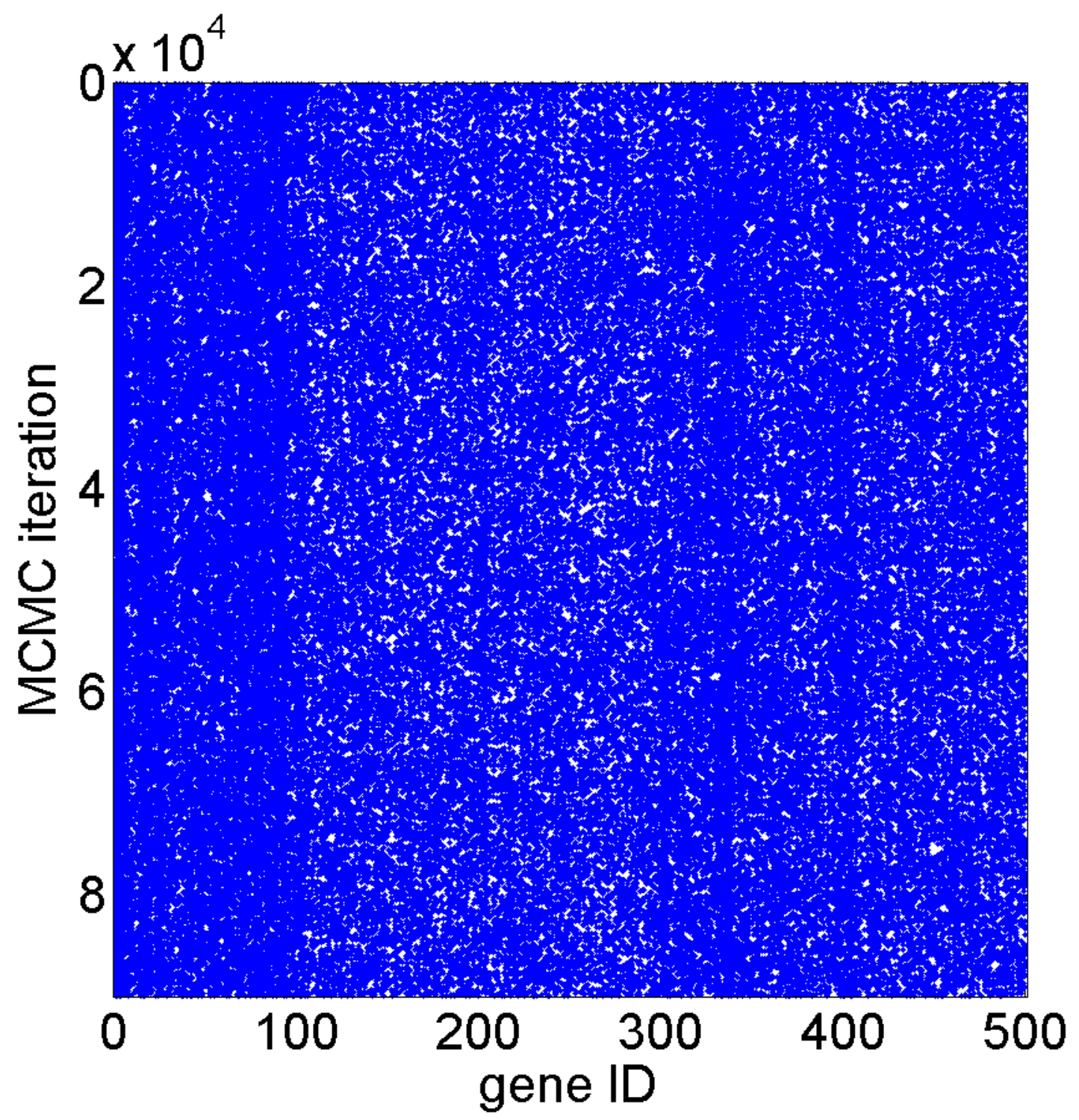}}
\caption[Trace plots of global parameters model deviance and model size $p_\gamma$,
and trace plots of $\bs{\gamma}$ vector for data set 1 in scenario 1]
{\small \label{sim1trace}Trace plots of global parameters model deviance (top)
and model size $p_\gamma$ (middle), as well as trace plots of $\bs{\gamma}$ vector (bottom) for add/delete
sampler (left), one neighbourhood sampler ($Pcor90$) (centre), and for the full
Gibbs sampler (right) for data set 1 in scenario 1.}
\end{center}
\end{figure}

Figure \ref{sim1trace} shows the traces of model deviance and model size $p_\gamma$ for the
add/delete, $Pcor90$, and full Gibbs ($Full$) samplers for simulated data set number 1. As expected, mixing is
much slower for the add/delete sampler than for the neighbourhood
($Pcor90$) and full Gibbs samplers. In particular this is also
the case for the $\bs{\gamma}$ vector, where for the add/delete
sampler the trace plot shows long ``lines'', where points are
plotted for each iteration over a long period (where
a variable stays in the model for a long time), and
equivalently long stretches of no plotted points (where variables are not included for a long time). This is
confirmed when measuring the mixing performances of the three
samplers in terms of the effective sample sizes
$\mbox{ESS}^*(\bs{\gamma})$ (see Tables \ref{sim1} and
\ref{sim1run1}). 

\begin{table}[!ht]
\caption[Mixing performance results with respect to $\bs{\gamma}$
for scenario 1 over all 25 data sets (10 data sets for $Full$
sampler, resp.)]{\small \label{sim1}Mixing performance results
with respect to $\bs{\gamma}$ for scenario 1 over all 25 data sets
(10 data sets for $Full$ sampler, respectively): median values and inter-quartile ranges.}
\begin{center} \small
\begin{tabular}{|l|r|r|r|r|r|r|}
\hline MCMC &CPU time&$\mbox{ESS}^*(\bs{\gamma})$&$R(\bs{\gamma})$&$\# \bs{I}^\sharp$&\# FP$^\dag$&\# FN$^\dag$\\
sampler &$t$ (min)&&&&&\\
\hline $AD$&38& 59 & 1.58 & 267 & 10 & 1 \\
    &(38, 38) & (55, 65) & (1.45, 1.72) & (263, 271) & (8, 14) & (1, 2)\\
\hline $Pcor90$ &168 & 3024 & 18.28 & 500  & 7 & 1 \\
     & (166, 169) & (2699, 3345) & (16.13, 20.11)  & (500, 500) & (5, 12) & (0, 1)\\
\hline $Full\ddag$&672 & 10780 & 17.63 & 500  & 8  & 1  \\
    & (664, 676) & (7791, 13700) & (16.07, 21.52) & (500, 500) & (4.5, 10.75) & (0.25, 2)\\
\hline

\multicolumn{7}{l}{\footnotesize{$^\ddag$ For $Full$ it is $M = 80,000$, compared to $M = 150,000$ for all other samplers}}\\
\multicolumn{7}{l}{\footnotesize{$^\sharp$ $\bs{I} = \{i: (\sum_{m=1}^M{\gamma_{i,m}}) > 0\}$, i.e. number of variables for which $\gamma_i = 1$ in at least one MCMC iteration}}\\
\multicolumn{7}{l}{\footnotesize{$^\dag$false positives and false negatives if cut-off at ratio
of posterior to prior probability $> 5$, i.e. if}}\\
\multicolumn{7}{l}{\footnotesize{$\hat p(\gamma_i = 1| \bs{x},\bs{y}) > 0.05$}}

\end{tabular}
\end{center}
\end{table}

\begin{figure}[!ht]
\begin{center}
\begin{center}
\includegraphics[width=0.475\textwidth]{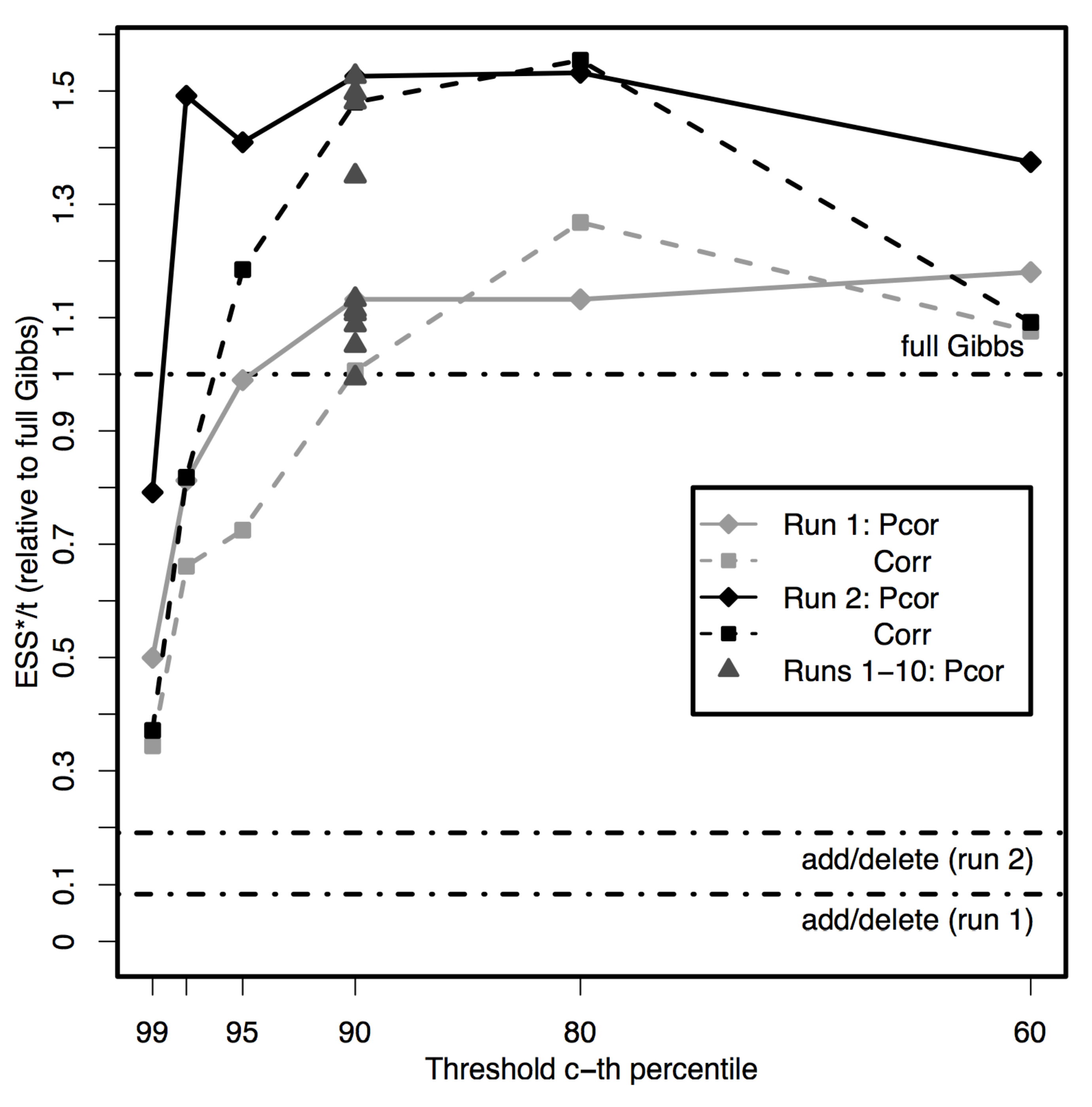}
\includegraphics[width=0.475\textwidth]{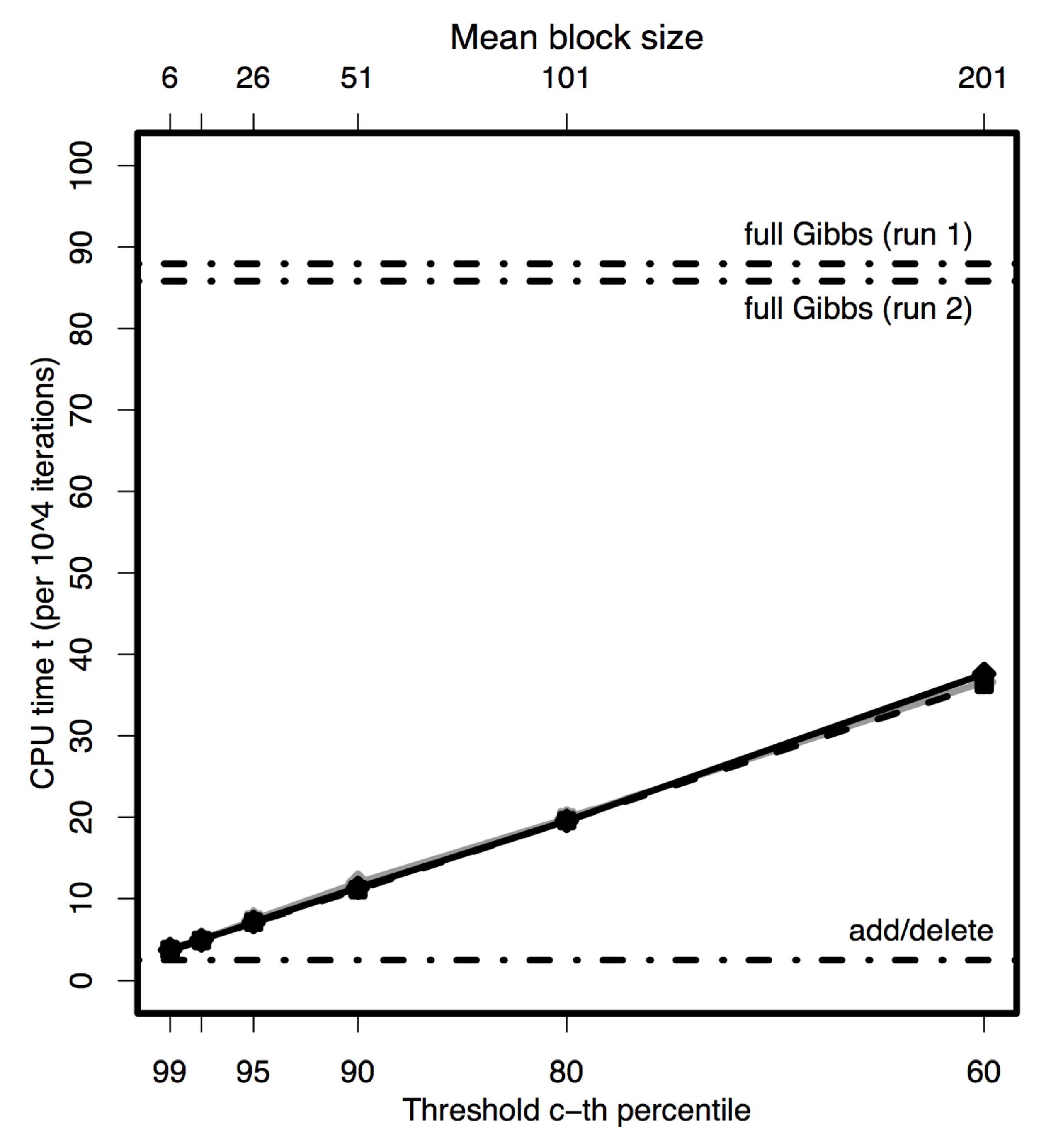}
\end{center}
\caption[Ratio of effective sample
size and CPU time, and CPU times per $10^4$ iterations, plotted
against the threshold level for data sets 1 and 2 in simulation setup 1.]{\small \label{sim1mixing}Ratio of effective sample
size and CPU time $R(\bs{\gamma}) = \mbox{ESS}^*(\bs{\gamma})/t$ (left), and CPU times per $10^4$ iterations (min) (right), plotted
against the threshold level $C$ for the neighbourhood samplers for data sets 1 and 2 in simulation setup 1.
In addition, for threshold $C = 0.9$, $R(\bs{\gamma})$ is plotted for simulated data sets 1 to 10.}
\end{center}
\end{figure}

We adjust for the computation time by computing the ratios
$R(\bs{\gamma})$ of effective sample sizes and computation times.
These ratios, relative to the ratio $R_{Full}(\bs{\gamma})$ observed
for the full Gibbs algorithm applied to the same data set, are
displayed in the left-hand side plot in Figure \ref{sim1mixing}
for all neighbourhood samplers for the first two simulated data sets.
In addition, the values $R_{Pcor90}(\bs{\gamma})$ are shown for the
$Pcor90$ samplers applied to all 10 generated data sets for
which the full Gibbs sampler was run.
The $Pcor90$ sampler is chosen for computation in all data sets and for comparison with the full Gibbs and add/delete samplers, because the comparison of a range of threshold sizes in the first two data sets indicates, that the $90\%$ percentile is within the range of threshold values $C$, for which the neighbourhood samplers are at their highest efficiency in terms of effective sample size per CPU time (see Figure \ref{sim1mixing} and Table \ref{sim1run1}).
The right-hand side of
Figure \ref{sim1mixing} shows the evolution of computation
times (per $10,000$ MCMC samples) for the neighbourhood samplers with
increasing neighbourhood sizes.

The effective sample size relative to CPU time $R(\bs{\gamma})$ is
larger for all neighbourhood samplers than for the add/delete sampler
and increases with decreasing threshold level (corresponding to
larger average neighbourhood sizes). It is also larger than the value
for the full Gibbs sampler, for all but the smallest neighbourhood
sizes. There is
no obvious difference between the neighbourhood samplers constructed
using partial correlations and those built from estimated
correlation matrices. 
All $R_{Pcor90}(\bs{\gamma})/R_{Full}(\bs{\gamma})$ ratios (for those ten
data sets for which $R_{Full}(\bs{\gamma})$ is available) are larger
than one, implicating that in this simulation scenario the
$Pcor90$ sampler leads to larger effective sample sizes
relative to CPU time requirements than the full Gibbs sampler.

The results for all 25 generated data sets are summarised in
Table \ref{sim1}. The median of the $R_{Pcor90}(\bs{\gamma})$ values
$\mbox{median}_{k=1}^{10} {R_{Pcor90}(\bs{\gamma})} = 18.28$ is slightly larger
than the median of the full Gibbs values $R_{Full}(\bs{\gamma})$,
which is equal to $\mbox{median}_{k=1}^{25}
{R_{Full}(\bs{\gamma})} = 17.63$, and although the inter-quartile
ranges overlap, we have seen from Figure \ref{sim1mixing} that
for each pairwise comparison within a generated data set it is
$R_{Pcor90}(\bs{\gamma}) > R_{Full}(\bs{\gamma})$. 

An additional indicator of Markov chain mixing, particularly in
a high-dimensional setting, is the proportion of all variables
that are visited by the Markov chain at least once. While the
add/delete algorithm only visited 265 of all 500 variables in
the application to simulated data set 1 (see Table
\ref{sim1run1}), this number is larger for all neighbourhood samplers
and increases with decreasing threshold values $C$. Comparing
the $Pcor$ and $Corr$ samplers with the smallest average neighbourhood
sizes, i.e. with the largest values of $C$, the
partial-correlation based samplers visit more variables than
the corresponding correlation based neighbourhood samplers.

\subsubsection{Posterior variable inclusion probabilities}
\begin{figure}[!ht]
\begin{minipage}{0.475\textwidth}
\begin{center}
$M=1000$
\scalebox{0.4}{\includegraphics{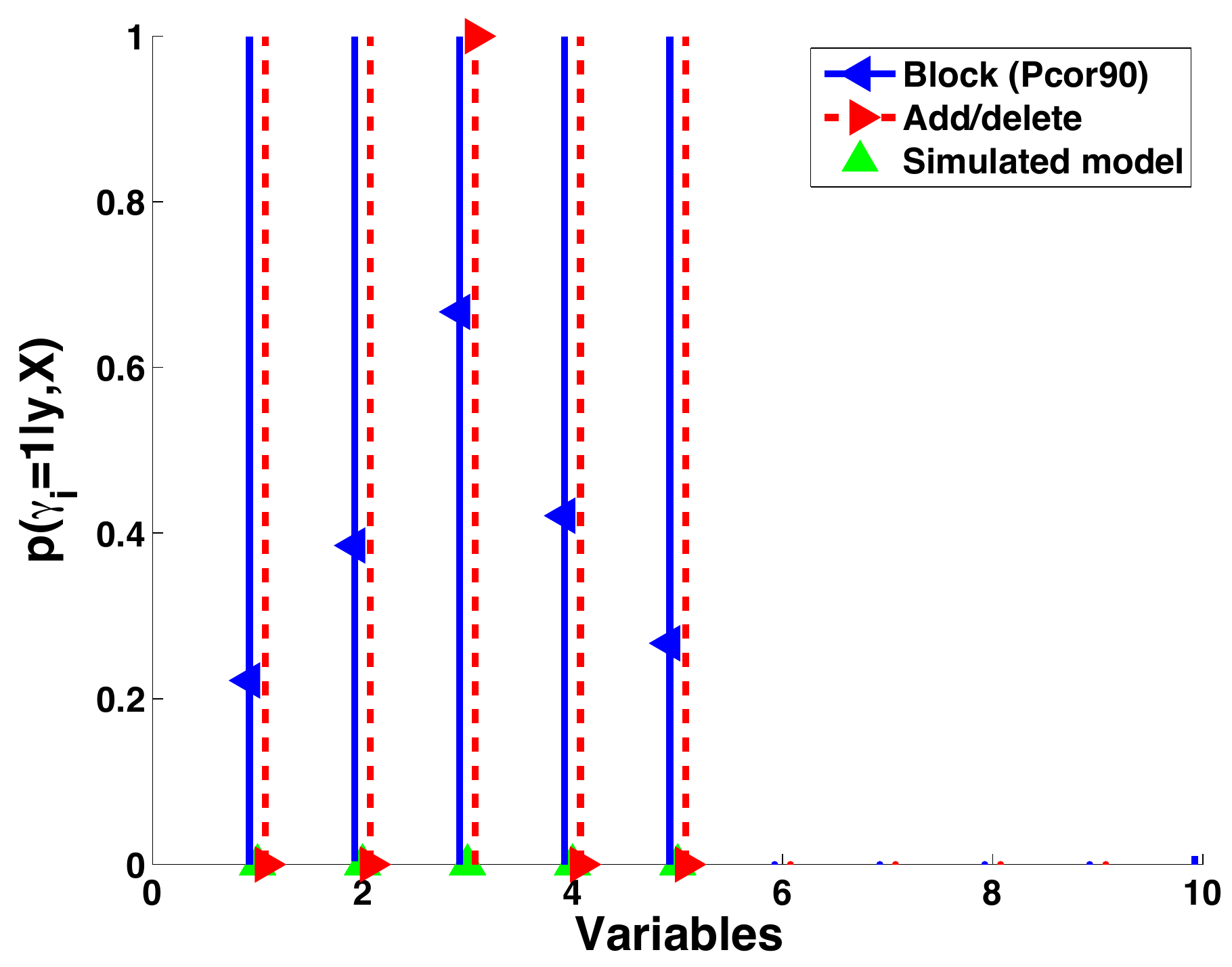}}
\end{center}
\end{minipage}\begin{minipage}{0.475\textwidth}
\begin{center}
$M=10000$
\scalebox{0.4}{\includegraphics{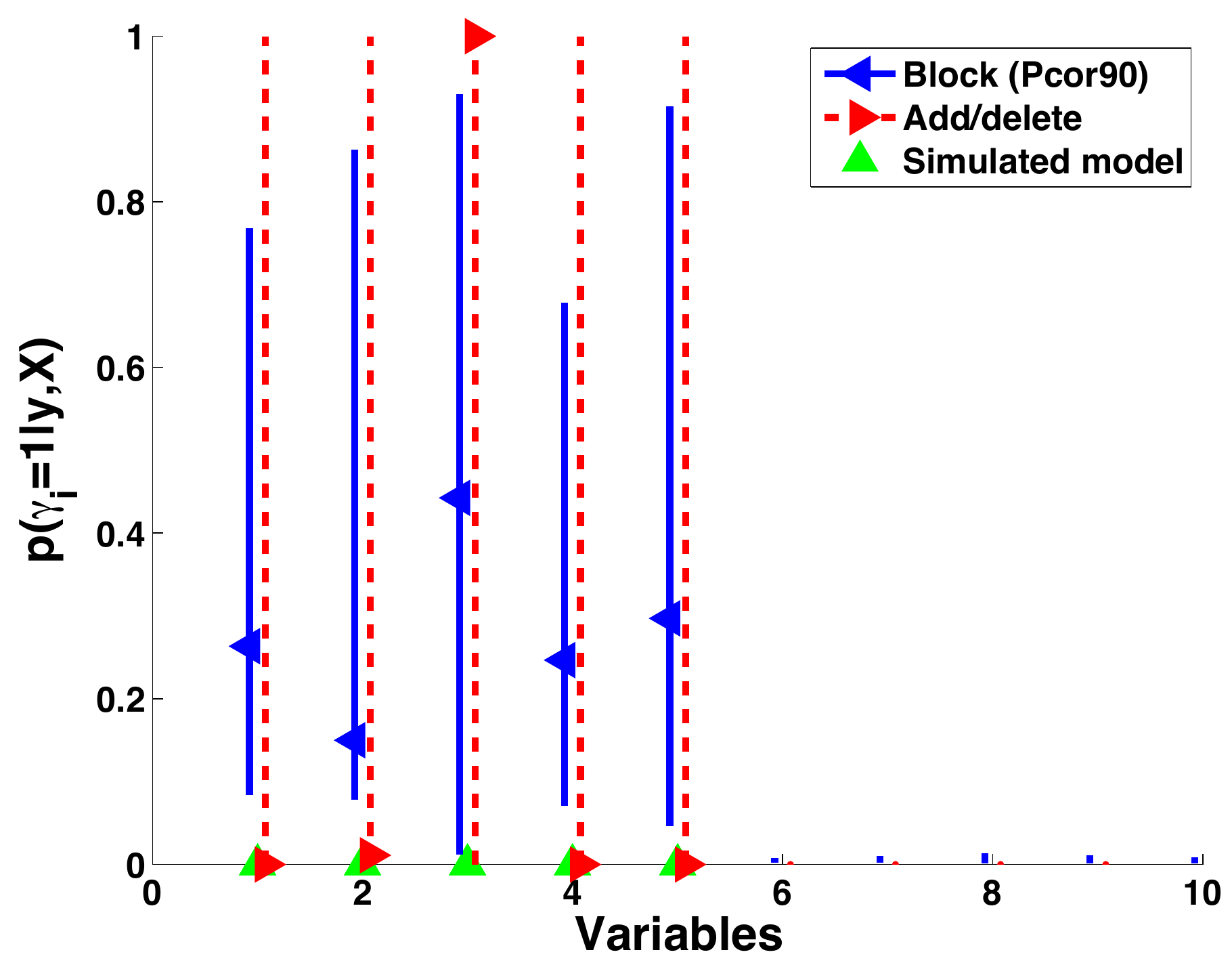}}
\end{center}
\end{minipage}

\begin{minipage}{0.475\textwidth}
\begin{center}
$M=50000$
\scalebox{0.4}{\includegraphics{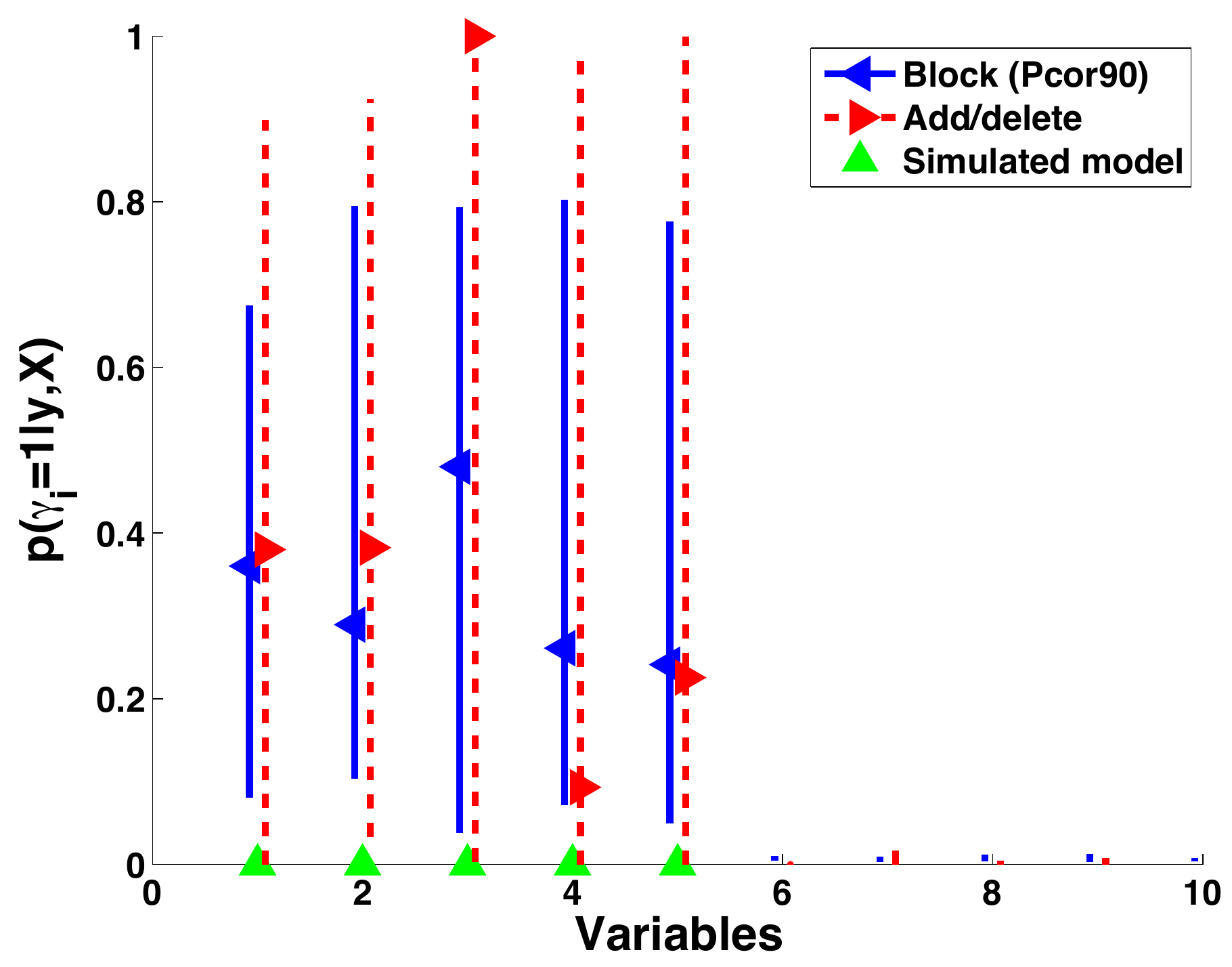}}
\end{center}
\end{minipage}\begin{minipage}{0.475\textwidth}
\begin{center}
$M=150000$
\scalebox{0.4}{\includegraphics{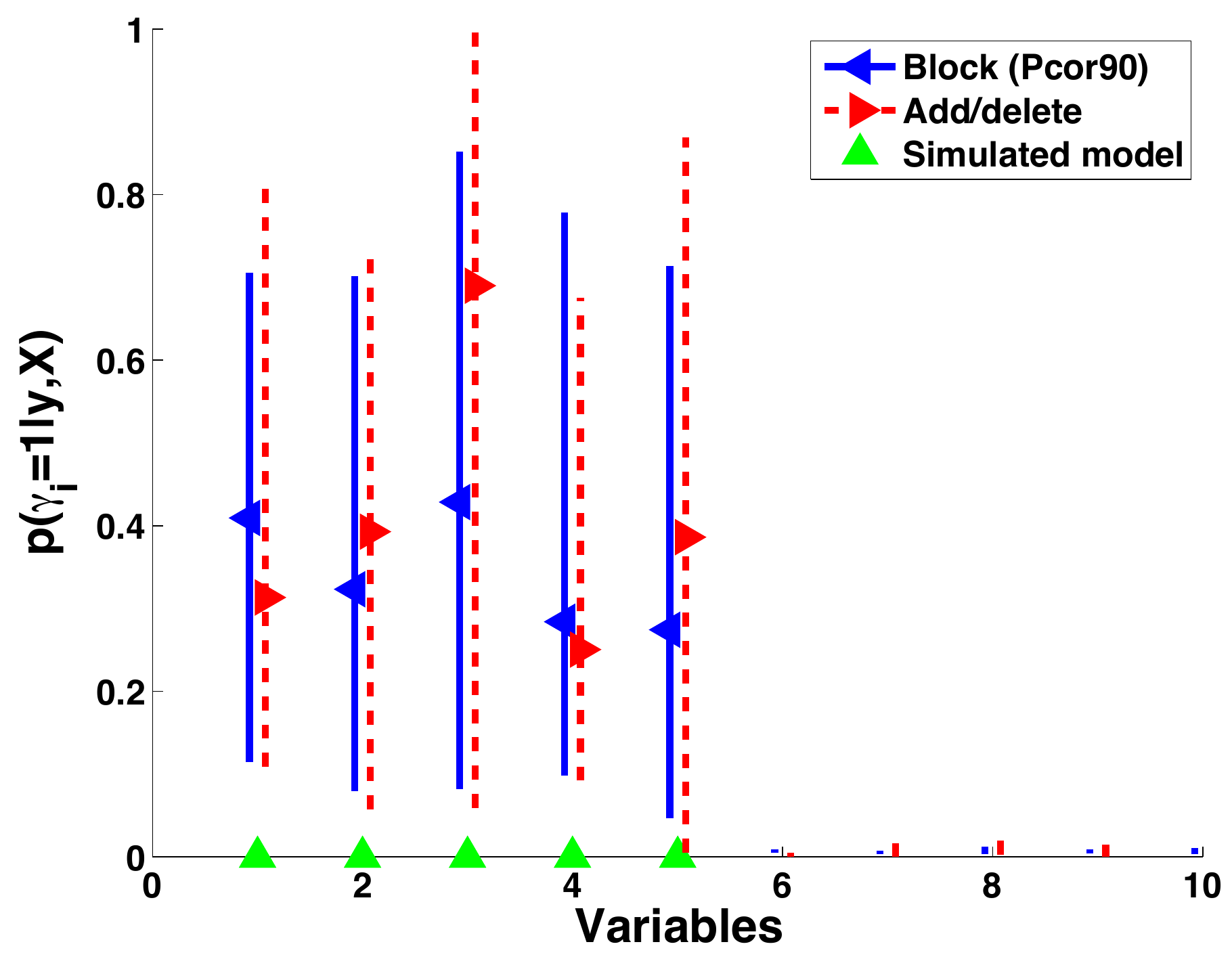}}
\end{center}
\end{minipage}
\caption[Posterior inclusion frequencies for variables
$1,...,10$ over all 25 replicates of simulation setup 1]{\small
Posterior inclusion frequencies (median and inter-quartile
ranges) for variables $1,...,10$ over all 25 replicates of
simulation setup 1 (after burn-in period).}\label{sim1post}
\end{figure}

Figure \ref{sim1post} shows the medians and inter-quartile
ranges of the MCMC estimates of the posterior variable
inclusion probabilities $\hat p(\gamma_i = 1|\bs{x},\bs{y})$ for
variables $\bs{x}_1,...,\bs{x}_{10}$, over all 25 generated data sets.
These include the $p^* = 5$ ``true'' predictors $(\bs{x}_1,...,\bs{x}_5)$
which were generated as being linked to the response variable
$\bs{y}$. In particular, the individual plots in Figure
\ref{sim1post} illustrate the evolution of the MCMC estimates
$\hat p(\gamma_i = 1|\bs{x},\bs{y})$, when the number of post-burn-in
MCMC iterations $M$ increases. The results shown for variables
$\bs{x}_6,...,\bs{x}_{10}$ are representative of the posterior
inclusion probability estimates that we find for all variables
which are simulated not to be linked to the response $\bs{y}$. As
expected, the median posterior inclusion probability estimates
of these variables are close to zero for all values of $M$ and
both samplers, with very small associated inter-quartile
ranges. 
Because in the add/delete sampler individual variables are visited and proposed for a state change so rarely, even after $M = 10,000$ post-burn-in iterations the median posterior inclusion frequencies are still either zero or very close to one for the five ``true'' predictors. Only after all $M = 150,000$ iterations do the add/delete sampler median values of the estimates start to move away from the extreme values at which they were fixed simply due to slow mixing of the Markov chain. The $Pcor90$ sampler does not have this problem,
with median values of the estimates being different from the
extremes zero and one even for $M = 1,000$ iterations. In fact,
for all $M \in \{1000, 10000, 50000, 150000\}$ the median
posterior inclusion frequencies are similar, with
inter-quartile ranges becoming narrower with increasing sample
sizes, reflecting a convergence to the true posterior variable
inclusion probabilities $p(\gamma_i = 1|\bs{x},\bs{y})$ on the level of
the individual generated data sets. Overall, the inter-quartile
ranges are narrower for the neighbourhood sampler than for the
add/delete sampler.

Note that the median values of the estimated posterior
inclusion probabilities are considerably smaller than one, and
in fact converge to values between around 0.2 and 0.4 with
increasing MCMC run lengths. In individual data sets, on
average one of the five variables has even an estimated
posterior inclusion probability smaller than 0.05. These cases
are labelled as false negative in Tables \ref{sim1} and
\ref{sim1run1}. This is linked to the fact that in individual
data sets, other variables ($\bs{x}_i$ with $i \in \{6,...,500\}$)
are sometimes found to have high posterior inclusion
probability estimates. These variables are counted as false
positive in the tables, again using a cut-off at $\hat
p(\gamma_i = 1|\bs{x},\bs{y}) > 0.05$. These results can be explained by
mixing or convergence problems of the MCMC algorithm, but also
by the presence of multi-collinearity in the input data matrix
$\bs{x}$ which necessarily arises when the number of variables $p$
is larger than the sample size $n$.

\subsubsection{Alternative Gibbs updates within the neighbourhoods}
As part of this simulation study, we apply multivariate Gibbs
sampling with $d \in \{4, 10\}$ (denoted $Joint4$ and $Joint10$) to two of the generated data
sets for both simulation scenarios. For comparison with regards to computing times we
also apply the corresponding restricted univariate Gibbs samplers
$Rgibbs4$ and $Rgibbs10$, where the maximum possible number of
variables to be updated within an MCMC iteration is also
restricted to $\min(\# \bs{I}, d)$ with $d \in \{4,10\}$. For all
these samplers, the $Pcor90$ mechanism is used to determine the
underlying neighbourhood structure, because the above comparison of
neighbourhood samplers with different threshold sizes $C$ has indicated
that the $Pcor90$ sampler performs well in
terms of the ratio of effective sample size and computation
time.

\begin{table}[!ht]
\caption[Mixing performance results with respect to $\bs{\gamma}$
for scenario 1: results for one data set]{\small
\label{sim1run1}Mixing performance results with respect to
$\bs{\gamma}$ for scenario 1: results for one data set (run 1).}
\begin{center} \small
\begin{tabular}{|p{1.75cm}|p{1.75cm}|p{1.75cm}|p{1.75cm}|p{1.75cm}|p{1.75cm}|p{1.75cm}|}
\hline MCMC &CPU time&$\mbox{ESS}^*(\bs{\gamma})$&$R(\bs{\gamma})$&$\# \bs{I}^\sharp$&\# FP$^\dag$&\# FN$^\dag$\\
sampler &$t$ (min)&&&&&\\
\hline
$AD$ & 40 & 65 & 1.65 & 265 & 7 & 1 \\
$Full^\ddag$ & 704 & 13910 & 19.77 & 500 & 7 & 1 \\
\hline
\multicolumn{7}{|l|}{Neighbourhood sampler}\\
\hline
$Pcor99$ & 56 & 558 & 9.88& 484 & 12 & 1 \\
$Pcor97.5$ & 74 & 1193 & 16.06& 497 & 10 & 1 \\
$Pcor95$ & 107 & 2094 & 19.57& 500 & 10 & 1 \\
$Pcor90$ & 170 & 3802 & 22.39& 500 & 7 & 1 \\
$Pcor80$ & 293 & 6559 & 22.39& 500 & 8 & 1 \\
$Pcor60$ & 564 & 13160 & 23.34& 500 & 7 & 1 \\
\hline
$Corr99$ & 56 & 380 & 6.80 & 411 & 14 & 1 \\
$Corr97.5$ & 74 & 967 & 13.07 & 481 & 8 & 1 \\
$Corr95$ & 107 & 1528 & 14.33 & 499 & 9 & 1 \\
$Corr90$ & 166 & 3309 & 19.89 & 500 & 9 & 1 \\
$Corr80$ & 293 & 7357 & 25.07 & 500 & 8 & 1 \\
$Corr60$ & 544 & 11580 & 21.27 & 500 & 7 & 1 \\
\hline
$Rgibbs4$ & 56 & 298 & 5.29 & 463 & 12 & 1 \\
$Joint4$ & 88 & 354 & 4.03 & 455 & 10 & 0 \\
$Rgibbs10$ & 71 & 971 & 13.64 & 499 & 7 & 1 \\
$Joint10^\ddag$ & 1423 & 620 & 0.44 & 474 & 10 & 1 \\
\hline
\multicolumn{7}{l}{\footnotesize{$^\ddag$ For $Full$ and
$Joint10$ it is $M = 80,000$, compared to $M = 150,000$ for all other samplers}}\\
\multicolumn{7}{l}{\footnotesize{$^\sharp$ $\bs{I} = \{i: (\sum_{m=1}^M{\gamma_{i,m}}) > 0\}$, i.e. number of
variables for which $\gamma_i = 1$ in at least one MCMC iteration}}\\
\multicolumn{7}{l}{\footnotesize{$^\dag$false positives and false negatives if cut-off at ratio
of posterior to prior probability $> 5$, i.e. if}}\\
\multicolumn{7}{l}{\footnotesize{$\hat p(\gamma_i = 1| \bs{x},\bs{y}) > 0.05$}}
\end{tabular}
\end{center}
\end{table}

The results for one data set of simulation scenario 1 are summarised in Table
\ref{sim1run1}. While the computation time needed for the
$Joint4$ run is with 88 minutes only about half the time needed
for the univariate Gibbs run ($Pcor90$), the time required to
run the $Joint10$ sampler explodes to nearly 24 hours for only
$N_{Joint10} = 90,000$ MCMC iterations in contrast to
$N_{Pcor90} = 200,000$ iterations. At the same time, the
effective sample sizes $\mbox{ESS}^*(\bs{\gamma})$ are only $9\%$ of
$\mbox{ESS}^*_{Pcor90}(\bs{\gamma})$ for the $Joint4$ sampler, and
only $31\%$ for the $Joint10$ algorithm when adjusting for the
differences in post-burn-in MCMC run lengths
($M_{Pcor90}=150,000$ vs. $M_{Joint10}=80,000$) by assuming a
linear relationship between $\mbox{ESS}^*(\bs{\gamma})$ and $M$.
Thus, with increasing set sizes $d$ in
multivariate-Gibbs-within-neighbourhood samplers $Joint<d>$, the
required computation time seems to increase too quickly and to outweigh
the improvement achieved in mixing as measured by
$\mbox{ESS}^*(\bs{\gamma})$. Also, the effective sample sizes of the
multivariate samplers are only modestly larger than those of
their corresponding restricted univariate Gibbs samplers
($Rgibbs4$ and $Rgibbs10$): the ratios of the effective sample
sizes are about $1.2$ for both $d=4$ and $d=10$, again when
adjusting $\mbox{ESS}^*_{Joint10}(\bs{\gamma})$ for the reduced
post-burn-in MCMC run length.

We conclude that joint moves, which update a fixed number of
variables $d$ jointly, are by themselves not a useful sampling strategy. However, it might be useful to include such updates in a
portfolio of moves, if there are covariates which are strongly
correlated. Such a flexible sampler, which could select
updating moves randomly from a portfolio of possible updates,
might benefit from occasional joint updates of strongly
correlated covariates.

\subsection{Simulation scenario 2: covariance based on gene expression data}
\subsubsection{Simulation setup}
\begin{figure}[!ht]
\begin{center}
\scalebox{0.7}{\includegraphics{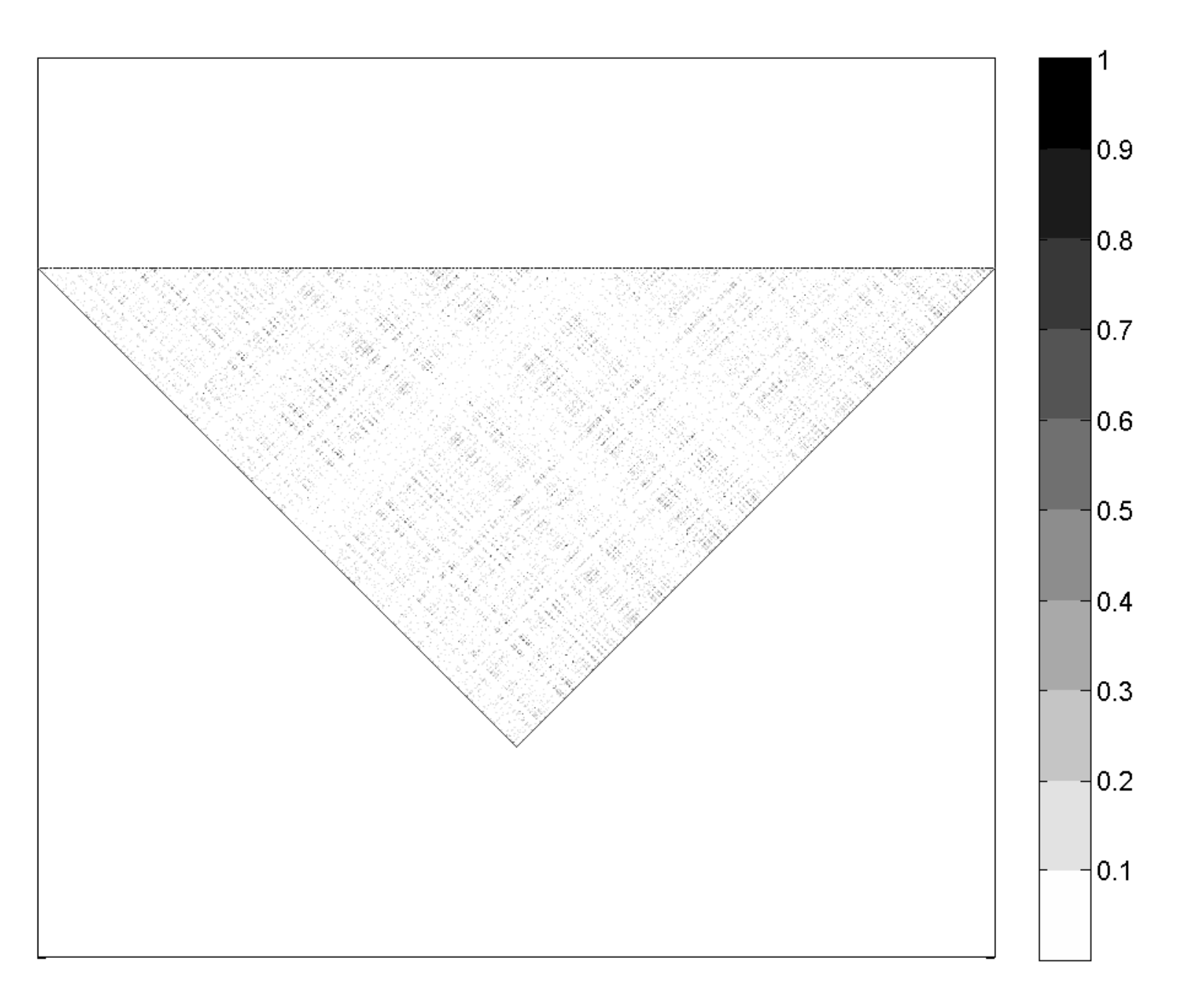}}
\caption[Squared empirical correlation structure of one data set simulated according to simulation scenario 2]{\small \label{sim2corr} Squared empirical correlation structure of one data set simulated according to simulation scenario 2.}
\end{center}
\end{figure}

In this second simulation scenario (see Table \ref{sim2algo})
we use a real gene expression data set \citep{schwartz02} to
generate the covariance structure between variables. For that
purpose, $p=500$ variables are selected at random from all
$7129$ probe sets available in the ovarian cancer gene
expression data set provided by \citet{schwartz02}. This data
set is described in more detail and analysed in Section
\ref{realdata}. It contains $n=104$ samples which are used for generating the simulated data sets. Again, 25 data
sets $(\bs{x},\bs{y})$ are generated, so that the first $p^* = 5$ variables
$(\bs{x}_1,...,\bs{x}_5)$ are related to the binary response $\bs{y}$ via a
logistic link. The natural correlation structure among the 500
randomly selected variables is illustrated by the triangular image plot of the squared empirical correlation matrix of
one of the 25 generated data sets in Figure \ref{sim2corr}. Pairwise
empirical correlations range from $-0.7$ to $0.9$. Again, the
prior parameter $c^2$ in the independence prior distribution $p(\bs{\beta}) = N(\bs{0}_{p_\gamma},c^2\bs{I}_{p_\gamma})$ is set to
$c^2=5$. The prior probability for $\gamma_i = 1$ is set to
$\pi_i = p^*/p = 0.01$.

\begin{table}[!ht]
\caption[Simulation scenario 2 based on gene expression data by
\citet{schwartz02}]{\small\label{sim2algo}Simulation scenario 2
based on gene expression data by \citet{schwartz02}.}
\vspace{0.5em} \hrule \vspace{0.5em}
\begin{enumerate}
\item $\bs{x}$ (input matrix of dimension $n\times p$): select $p=500$ gene variables at
    random from pre-processed and normalised gene
    expression microarray data set $\tilde{x}$ of dimension
    $\tilde p\times n = 7129\times 104$ by
    \citet{schwartz02} (described in Section
    \ref{realdata}), standardise all variables to have zero mean and unit
    variance.
\item For $j = 1,...,n=104$ do\\
$y_j \sim \mbox{Bernoulli}\left(\frac{\exp(\bs{x}_j\bs{\beta})}{1 +
\exp(\bs{x}_j\bs{\beta})}\right)$, with vector $\bs{\beta} = (2,2,2,2,2,0,...,0)$ of length $p=500$ and $\bs{x}_j$ ($j=1,...,n$) denoting the column vectors of $\bs{x}$.
\end{enumerate}
\hrule
\end{table}

\subsubsection{Markov chain mixing performance}
\begin{figure}[!ht]
\begin{center}
\scalebox{0.25}{\includegraphics{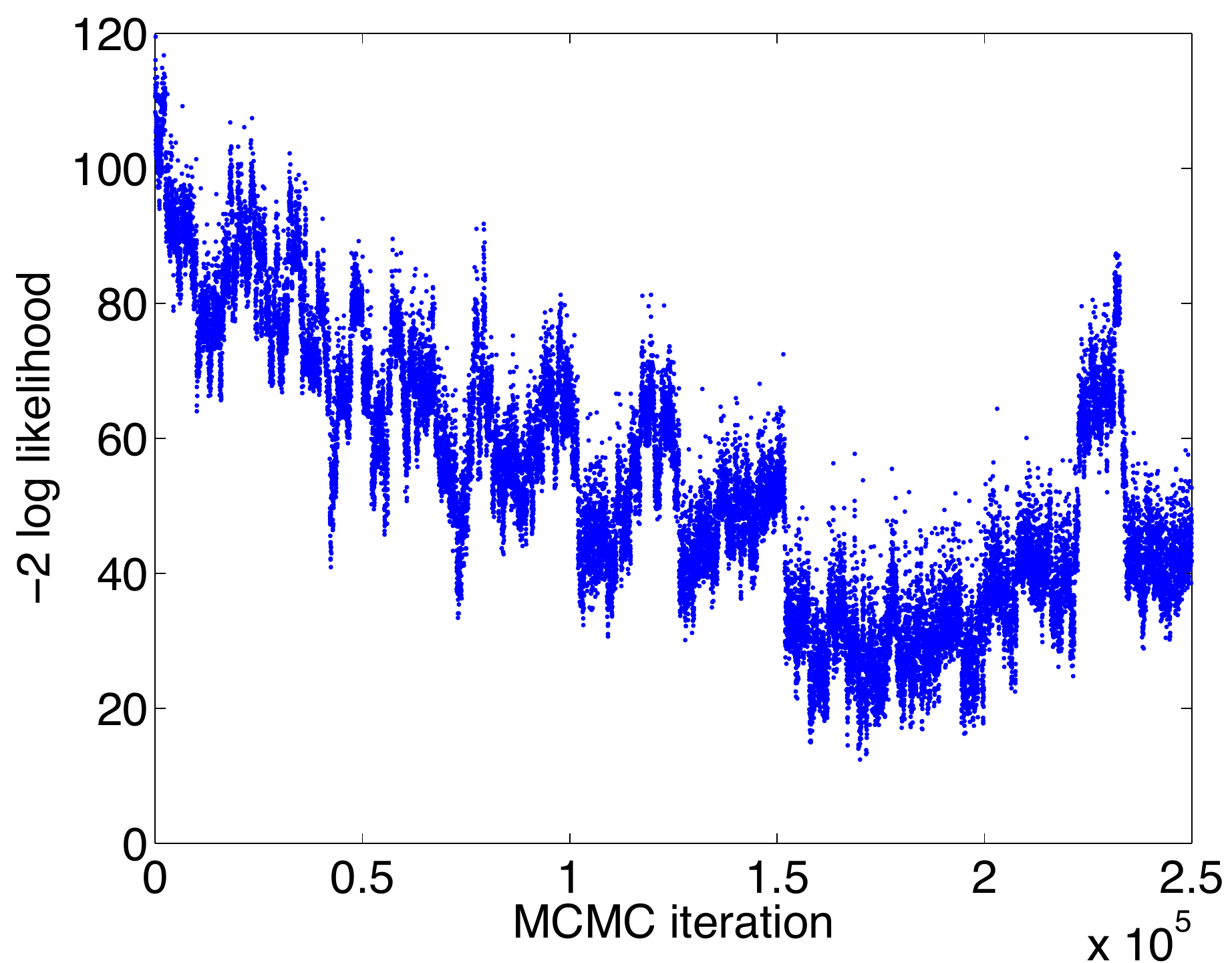}}
\scalebox{0.25}{\includegraphics{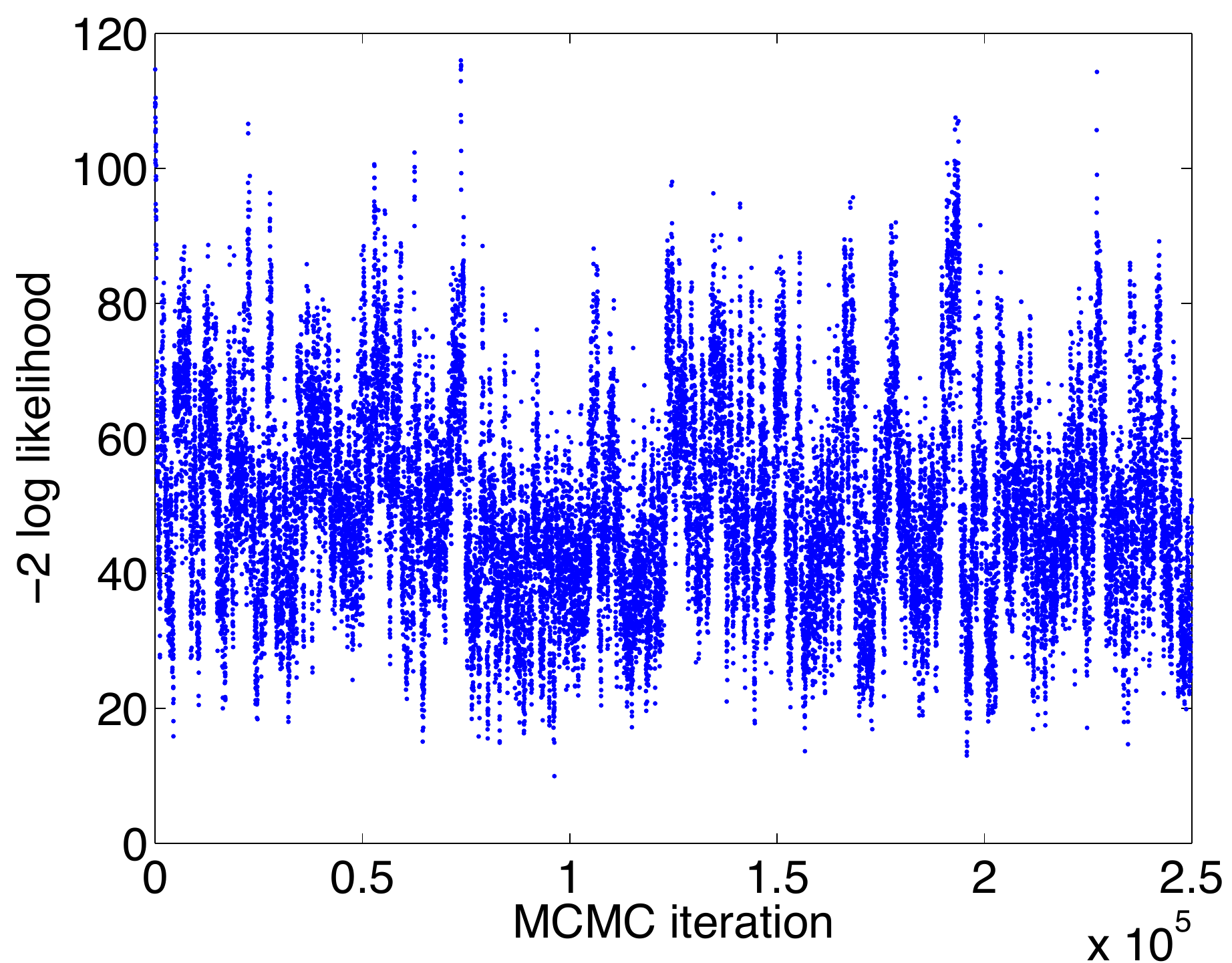}}
\scalebox{0.25}{\includegraphics{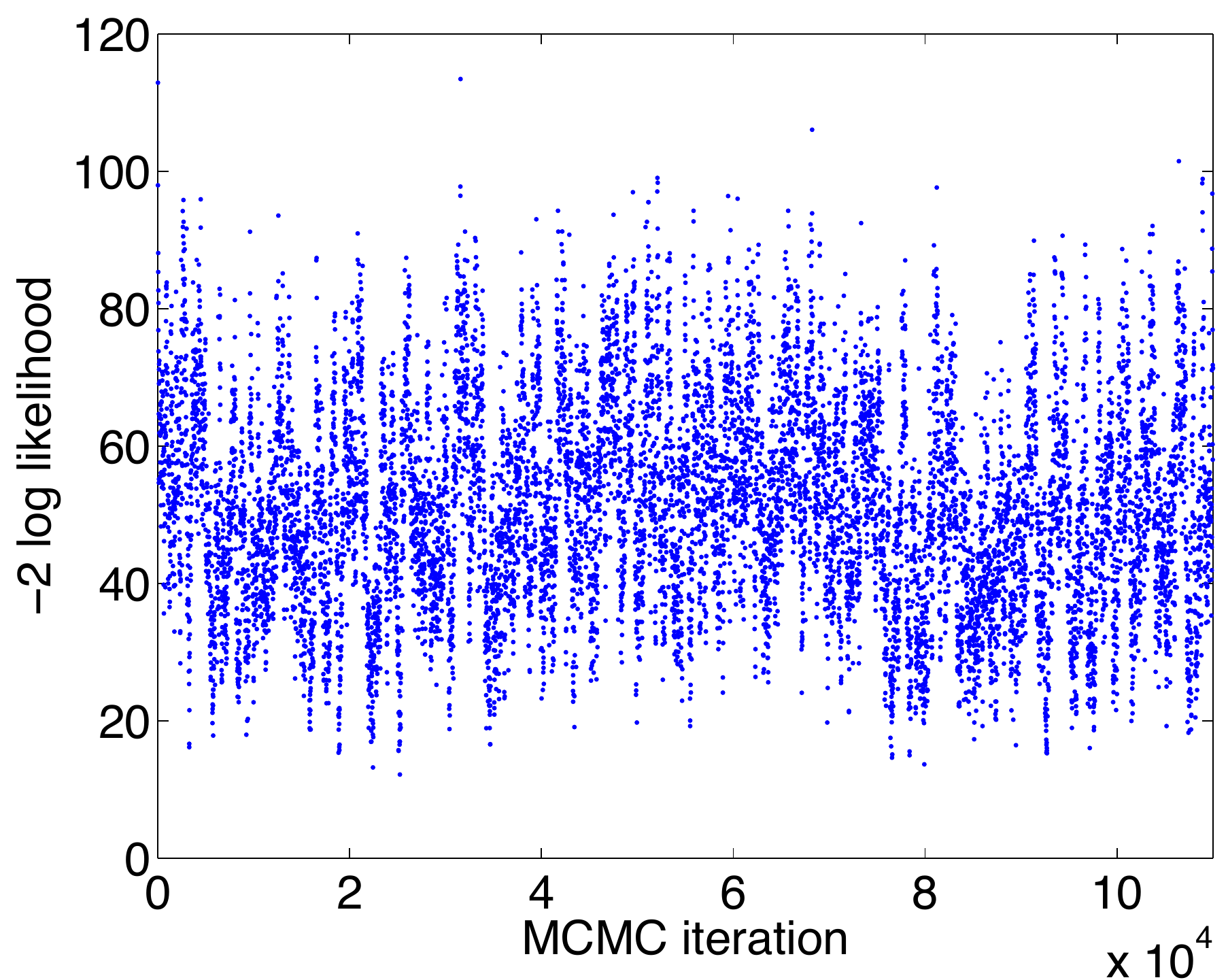}}\\
\scalebox{0.25}{\includegraphics{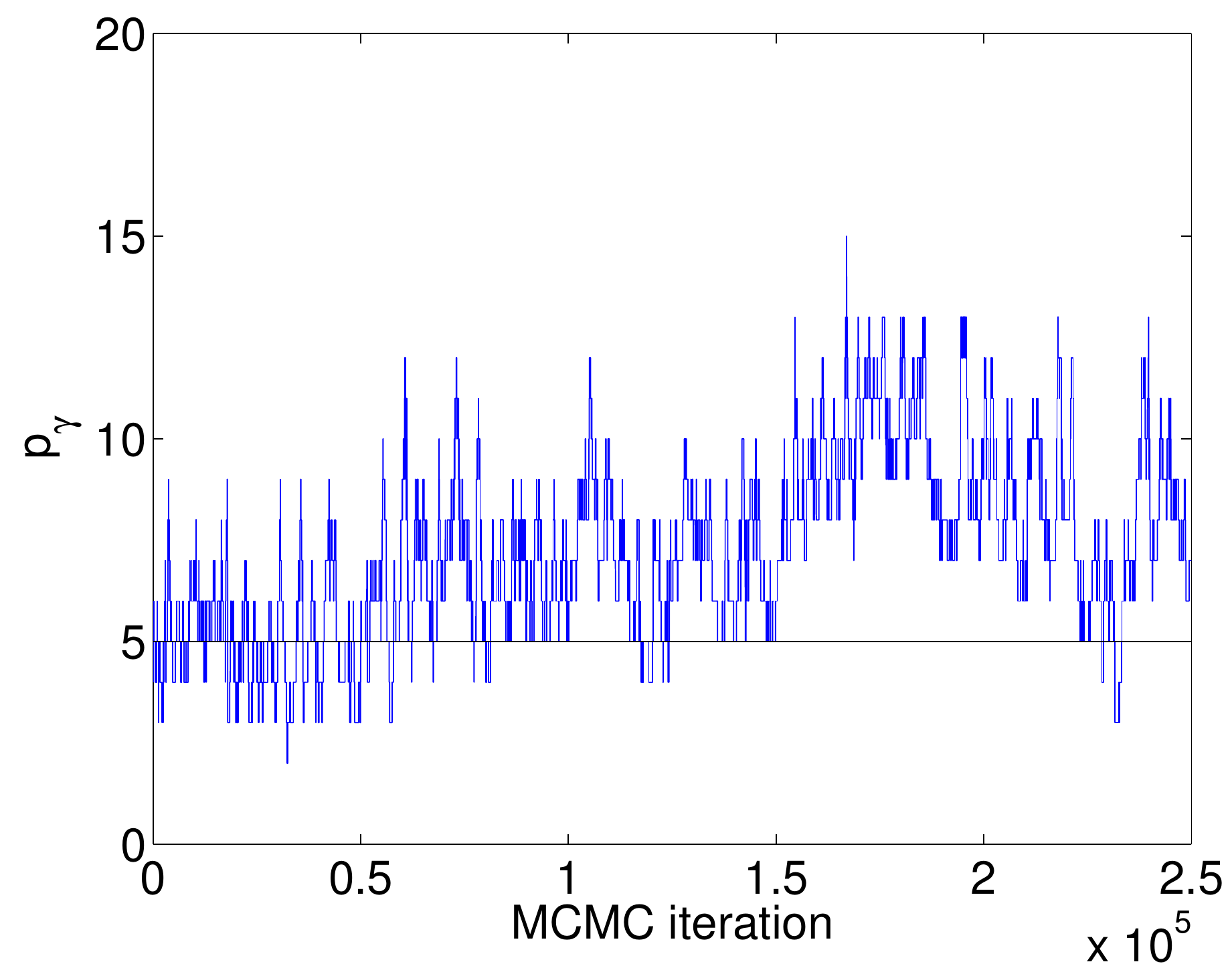}}
\scalebox{0.25}{\includegraphics{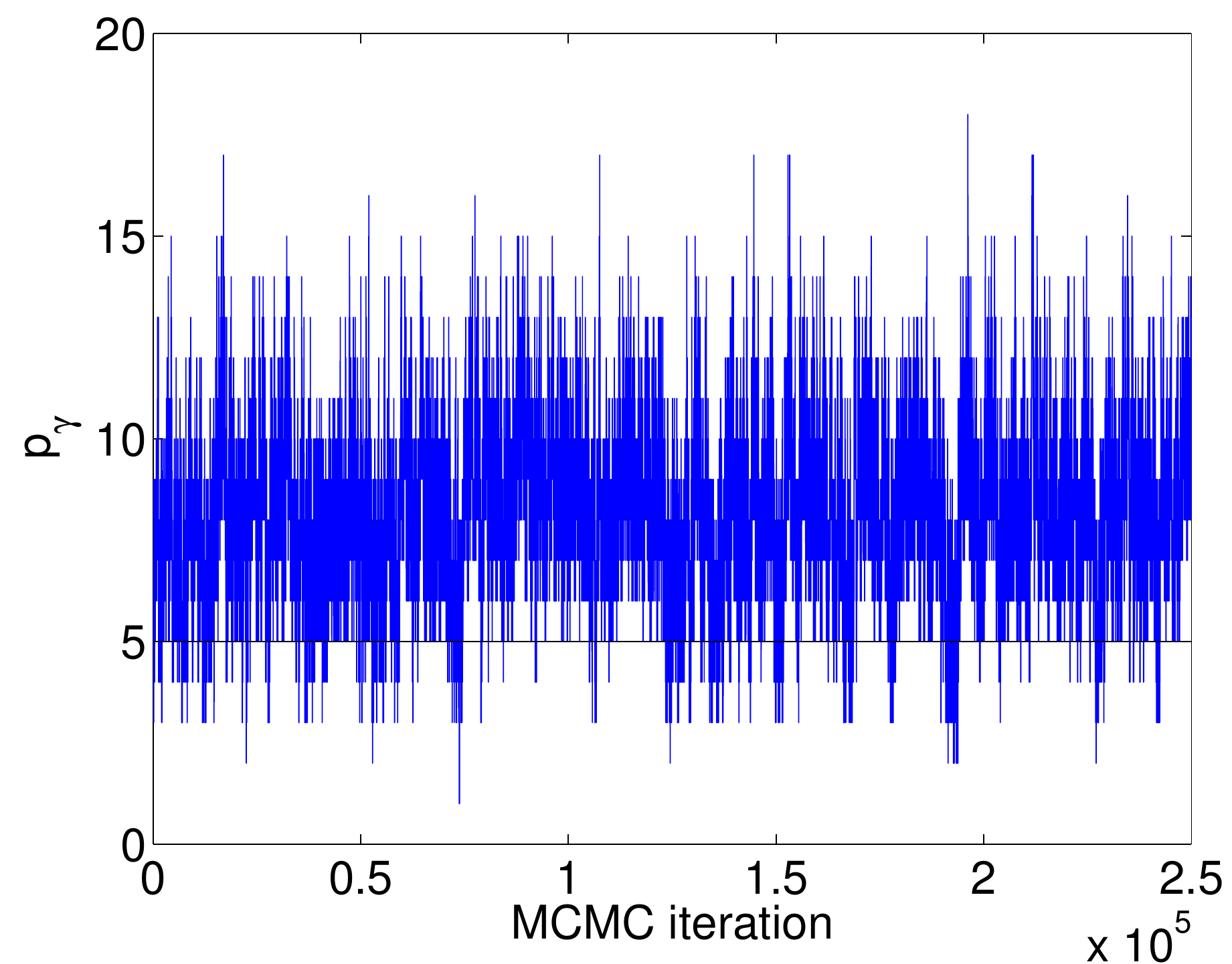}}
\scalebox{0.25}{\includegraphics{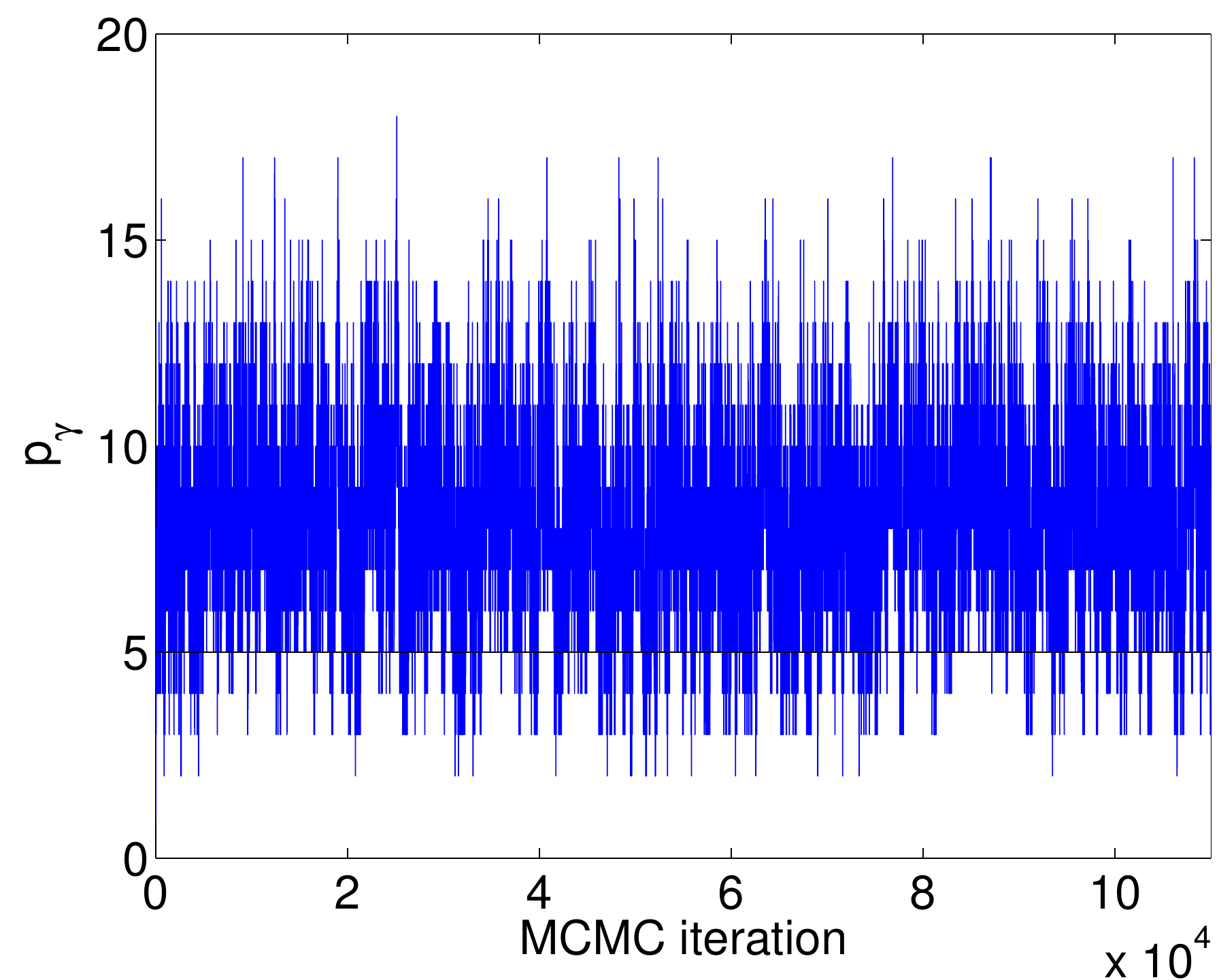}}\\
\scalebox{0.31}{\includegraphics{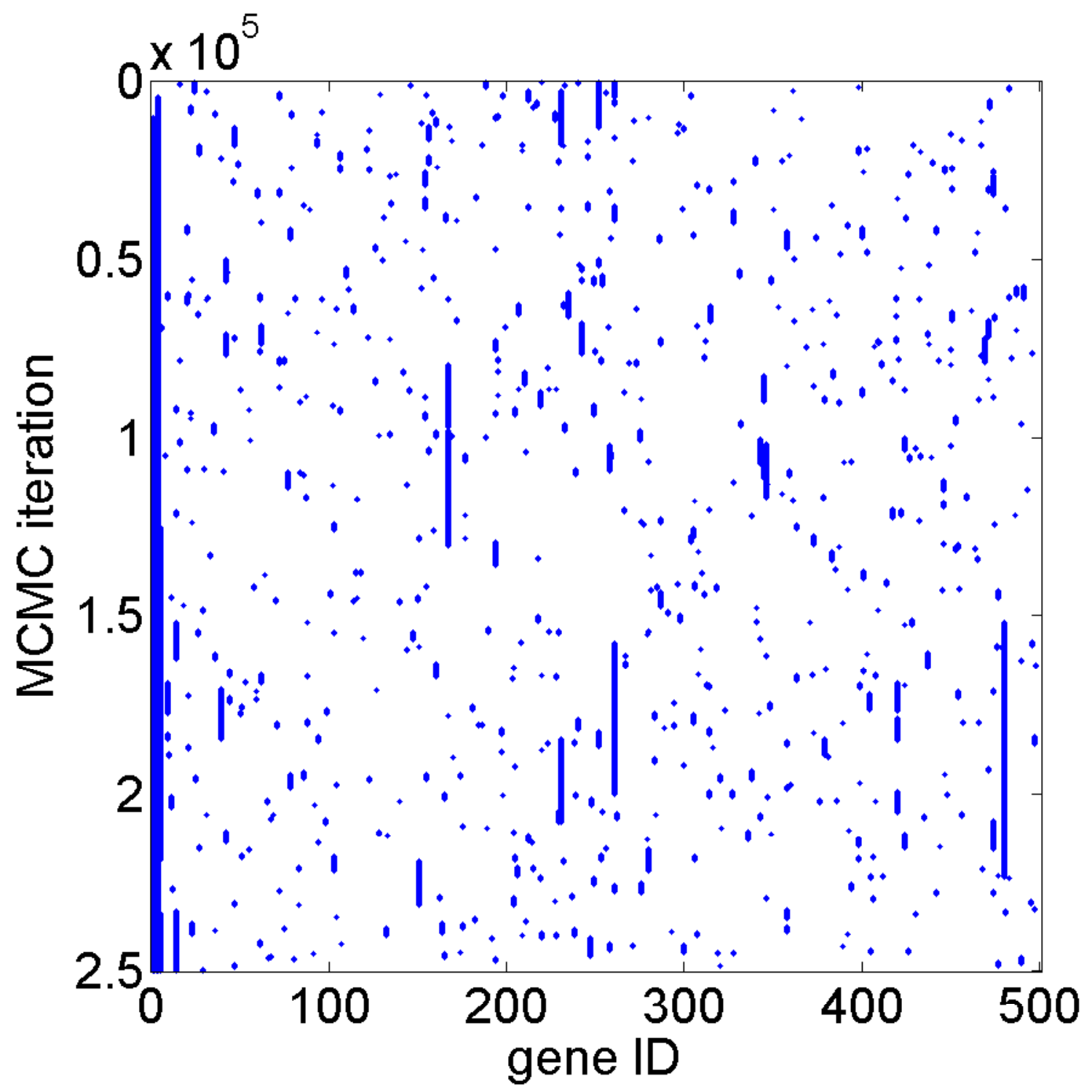}}
\scalebox{0.31}{\includegraphics{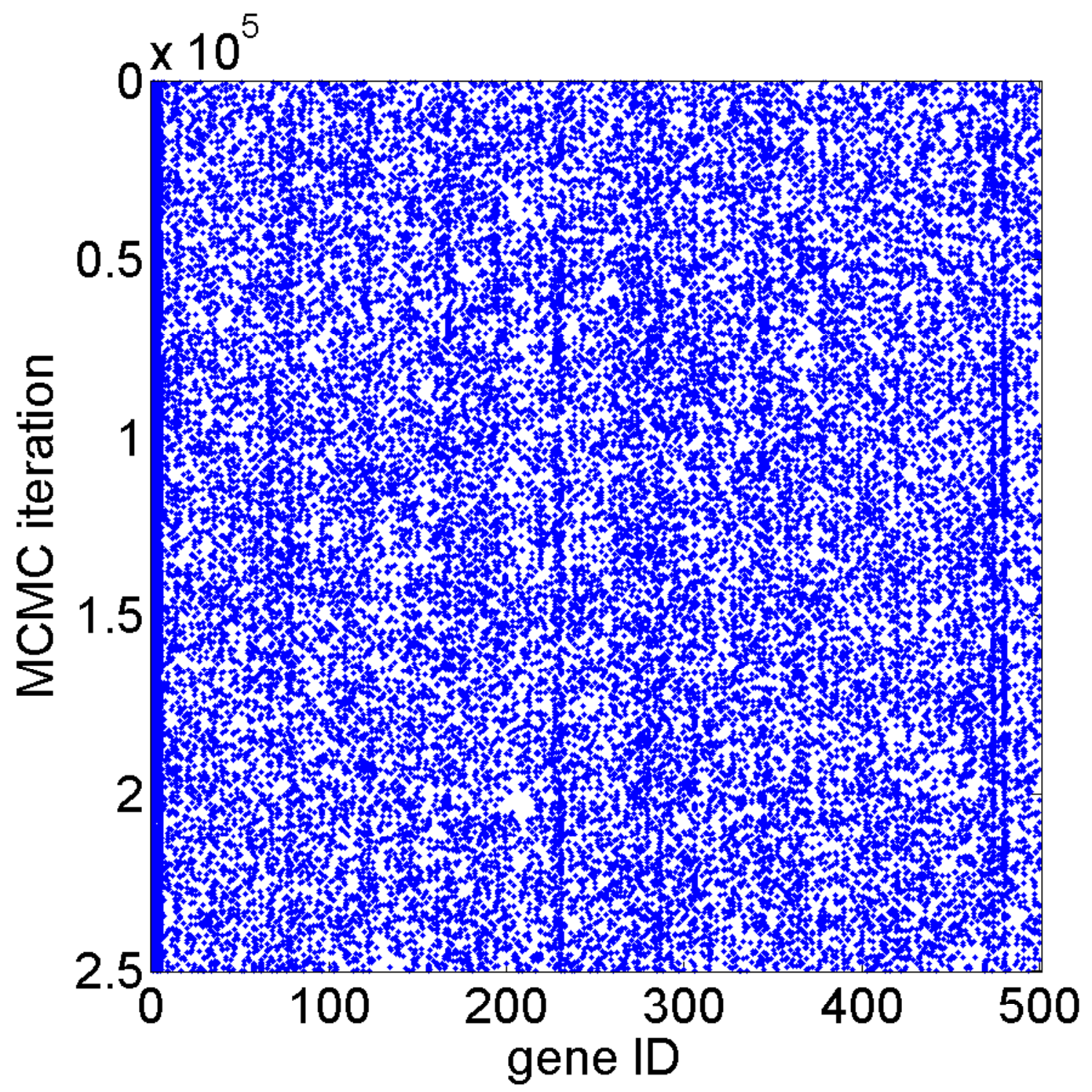}}
\scalebox{0.31}{\includegraphics{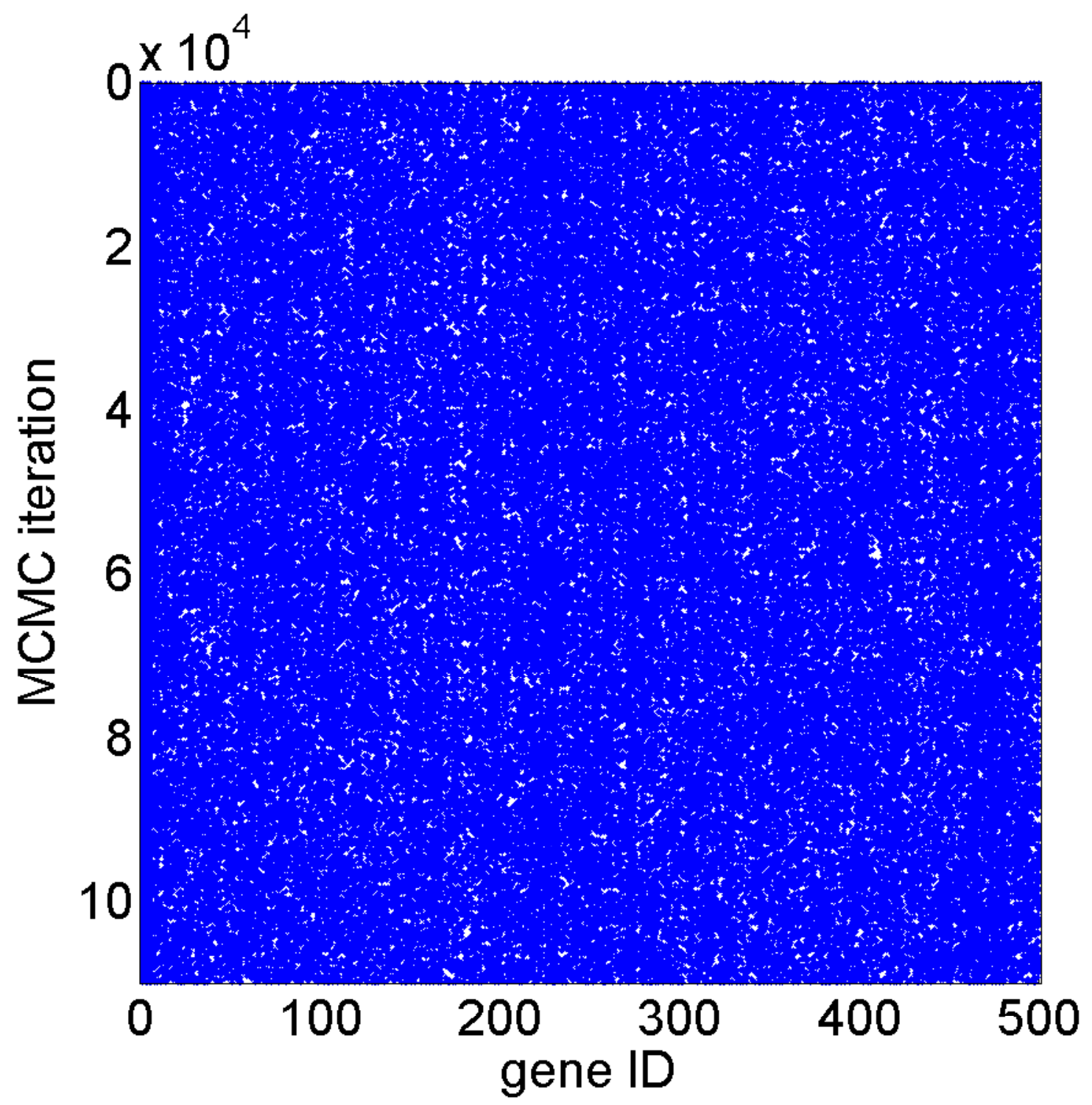}}
\caption[Trace plots of global parameters model deviance
and model size $p_\gamma$, and trace plots of $\bs{\gamma}$ vector for data set 1 in scenario 2]
{\small \label{sim2trace}Trace plots of global parameters model deviance (top)
and model size $p_\gamma$ (middle), as well as trace plots of $\bs{\gamma}$ vector (bottom) for add/delete
sampler (left), one neighbourhood sampler ($Pcor90$) (centre), and for the full
Gibbs sampler (right) for data set 1 in scenario 2.}
\end{center}
\end{figure}

We follow the structure of analysis outlined for the simulation
scenario 1 in the previous section. Figure \ref{sim2trace}
shows the trace plots of model deviance (top)
and model size (middle) and the individual traces for all
$\gamma_i$ with $i = 1,...,500$ (bottom) for the add/delete
Metropolis-Hastings sampler, the $Pcor90$ sampler and the
full Gibbs algorithm for one generated data set. The
conclusions are much the same as for simulation scenario 1,
that is mixing with respect to sampling $\bs{\gamma}$ is much slower
for the add/delete sampler than for the $Pcor90$ and full Gibbs
algorithms. In addition, there is also an obvious improvement
in mixing for the full Gibbs sampler compared to the neighbourhood
sampler, when viewing the trace plots of $\bs{\gamma}$. In terms of
the effective sample sizes $\mbox{ESS}^*(\bs{\gamma})$ (see Table
\ref{sim2run1} for the results for
data set 1), values increase about 40-fold for the neighbourhood
sampler compared to add/delete algorithm and more than 220-fold
for the full Gibbs sampler after adjustment for the reduced
post-burn-in run length $M_{Full} = 100,000$ (compared to
$M_{AD} = M_{Pcor90} = 200,000$), which is a slightly smaller
improvement than what we had observed in simulation scenario 1.

\begin{figure}[!ht]
\begin{center}
\includegraphics[width=0.475\textwidth]{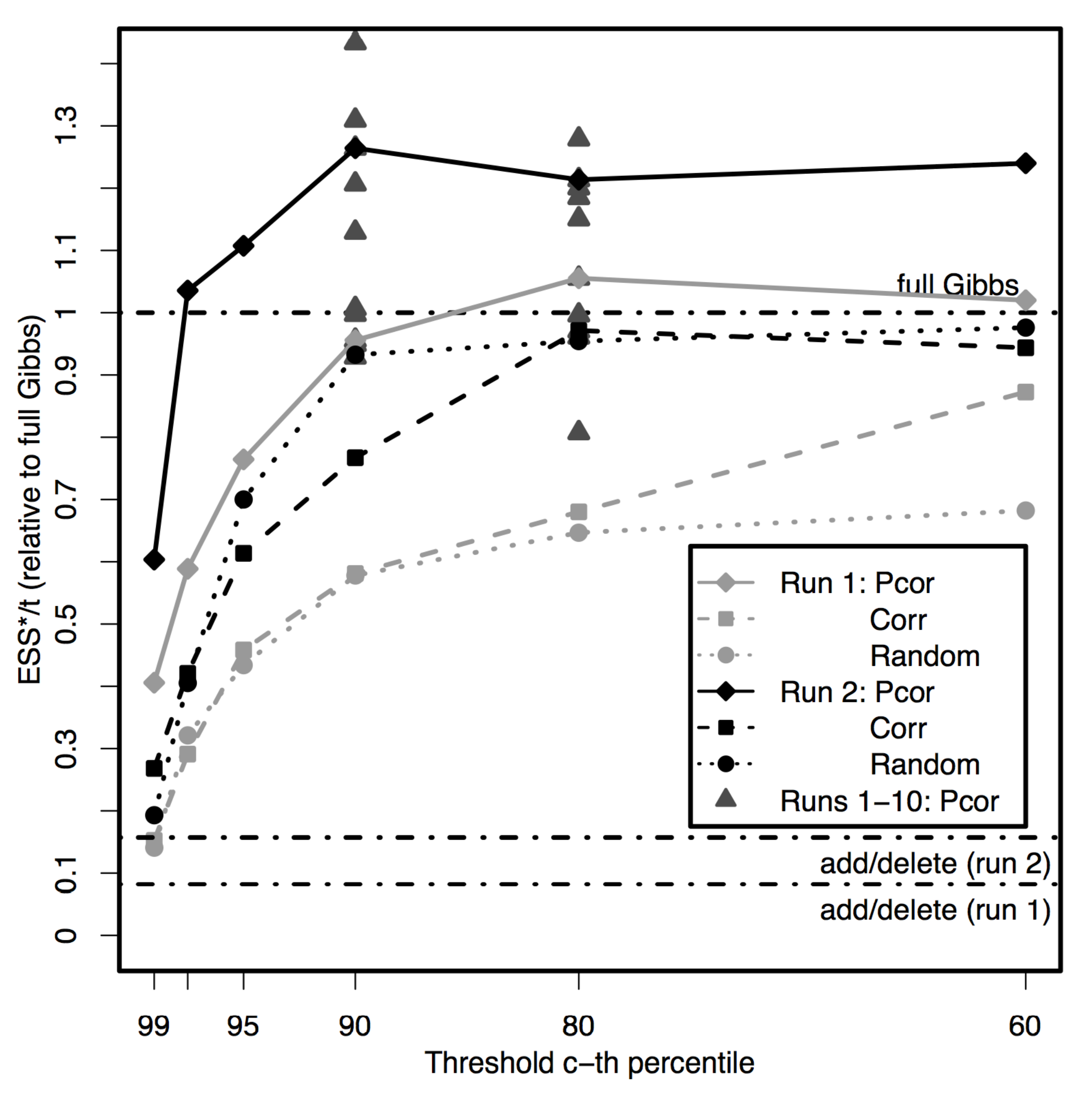}
\includegraphics[width=0.475\textwidth]{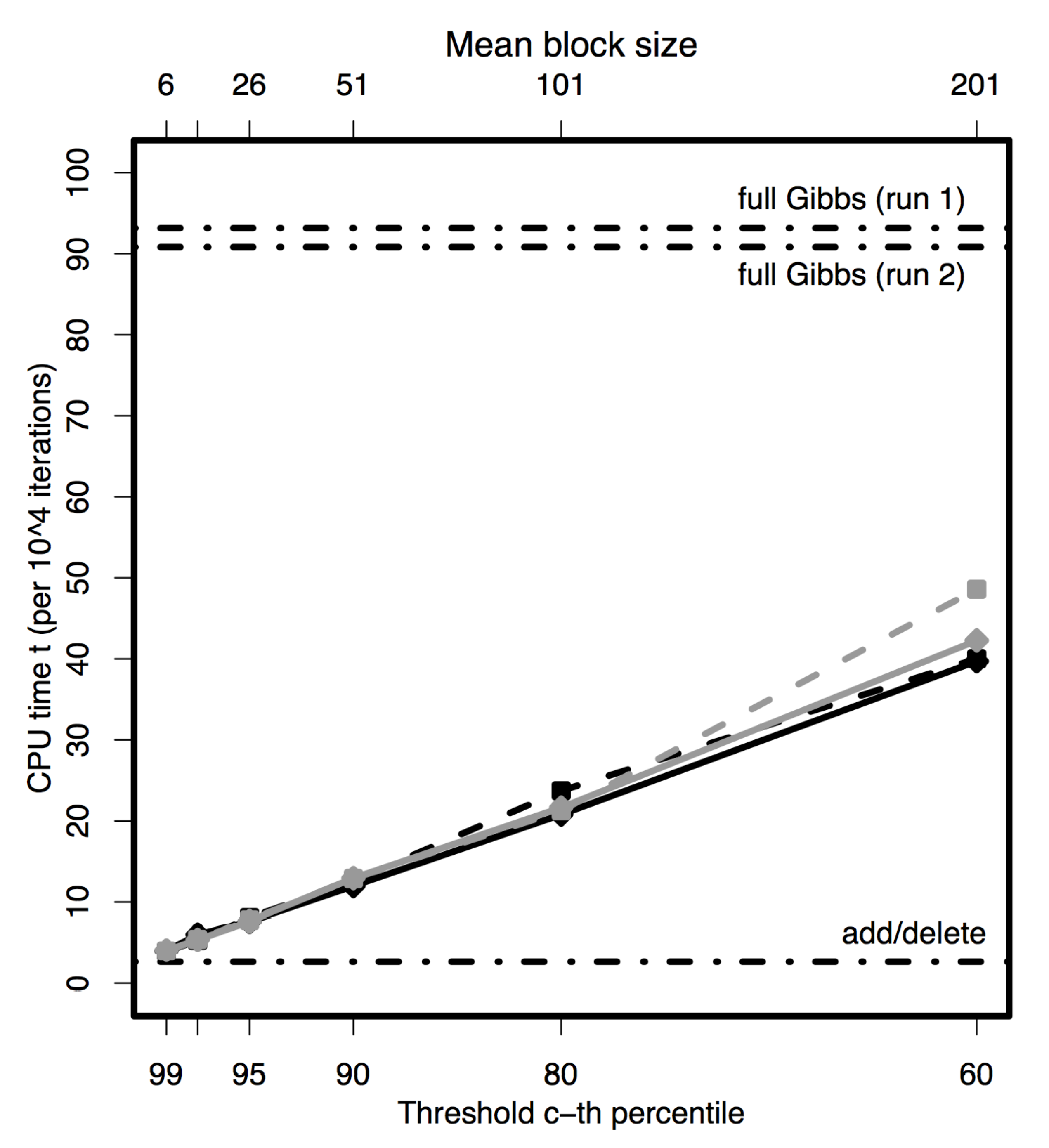}
\caption[Ratio of effective sample
size and CPU time, and CPU times per $10^4$ iterations, plotted
against the threshold level for data sets 1 and 2 in scenario 2]{\small \label{sim2mixing}Ratio of effective sample
size and CPU time $R(\bs{\gamma}) = \mbox{ESS}^*(\bs{\gamma})/t$ (left), and CPU times per $10^4$ iterations (min) (right), plotted
against the threshold level $C$ for the neighbourhood samplers for data sets 1 and 2 in scenario 2. In addition, for thresholds $C = 0.9$ and $C = 0.8$,
$R(\bs{\gamma})$ is plotted for simulated data sets 1 to 10.}
\end{center}
\end{figure}

The ratios $R(\bs{\gamma})$ of effective sample sizes and
computation times, relative to the ratio $R_{Full}(\bs{\gamma})$ for
the full Gibbs algorithm, are displayed in the left-hand side
plot in Figure \ref{sim2mixing} for all neighbourhood samplers for the
first two simulated data sets. Also, the ratios
$R_{Pcor90}(\bs{\gamma})/R_{Full}(\bs{\gamma})$ and
$R_{Pcor80}(\bs{\gamma})/R_{Full}(\bs{\gamma})$ are shown for the
$Pcor90$ and $Pcor80$ samplers applied to those 10 generated
data sets for which the full Gibbs sampler has been run. As
before, the right-hand side of Figure \ref{sim2mixing} shows
the linear evolution of computation times (per $10,000$ MCMC
samples) for the neighbourhood samplers with increasing neighbourhood sizes.

The effective sample sizes relative to CPU time $R(\bs{\gamma})$ are
larger for all neighbourhood samplers than for the add/delete sampler
and increase with decreasing threshold level (corresponding to
larger average neighbourhood sizes), until levelling off between $C =
0.9$ and $C = 0.8$. Contrary to simulation scenario 1, the
partial-correlation based neighbourhood samplers now have considerably
larger effective sample sizes and hence larger values of
$R(\bs{\gamma})$ than the samplers using correlation estimates for
neighbourhood construction. In fact, now the $Corr$ samplers do not
outperform the full Gibbs sampler in terms of $R(\bs{\gamma})$ for
the two displayed data sets, while the $Pcor$ algorithms do
result in better mixing than full Gibbs sampling if the
threshold is large enough. 
Seven out of ten $R_{Pcor90}(\bs{\gamma})/R_{Full}(\bs{\gamma})$ ratios,
for which $R_{Full}(\bs{\gamma})$ is available, are larger than one.

\begin{table}[!ht]
\caption[Mixing performance results with respect to $\bs{\gamma}$
for scenario 2 over all 25 data sets (10 data sets for $Full$
sampler, resp.)]{\small \label{sim2}Mixing performance results
with respect to $\bs{\gamma}$ for scenario 2 over all 25 data sets
(10 data sets for $Full$ sampler, respectively): median values and inter-quartile ranges.}
\begin{center} \small
\begin{tabular}{|l|r|r|r|r|r|r|}
\hline MCMC &CPU time&$\mbox{ESS}^*(\bs{\gamma})$&$R(\bs{\gamma})$&$\# \bs{I}^\sharp$&\# FP$^\dag$&\# FN$^\dag$\\
sampler &$t$ (min)&&&&&\\
\hline $AD$ & 53 & 103 & 1.96 & 316 & 10 & 1\\
    & (53, 53) & (100, 105) & (1.88, 1.99) & (303, 321) & (6, 15) & (0, 2)\\
\hline Neighbourhood & 241 & 5148 & 21.46 & 500 & 8 & 0\\
    ($Pcor90$) & (239, 243) & (4773, 6000) & (20.11, 24.89) & (500, 500) & (4, 9) & (0, 1)\\
\hline $Full^\ddag$ & 931 & 16690 & 17.94 & 500 & 9 & 0\\
    & (928, 932) & (12320, 23170) & (13.20, 25.00) & (500, 500) & (5, 9.75) & (0, 2)\\
\hline
\multicolumn{7}{l}{\footnotesize{$^\ddag$ For $Full$ it is $M = 100,000$, compared to $M = 200,000$ for all other samplers}}\\
\multicolumn{7}{l}{\footnotesize{$^\sharp$ $\bs{I} = \{i: (\sum_{m=1}^M{\gamma_{i,m}}) > 0\}$, i.e. number of
variables for which $\gamma_i = 1$ in at least one MCMC iteration}}\\
\multicolumn{7}{l}{\footnotesize{$^\dag$false positives and false negatives if cut-off at ratio
of posterior to prior probability $> 5$, i.e. if}}\\
\multicolumn{7}{l}{\footnotesize{$\hat p(\gamma_i = 1| \bs{x},\bs{y}) > 0.05$}}
\end{tabular}
\end{center}
\end{table}

The results for all 25 generated data sets are summarised in
Table \ref{sim2}. The median of the $R_{Pcor90}(\bs{\gamma})$ values $\mbox{median}_{k=1}^{25} {R_{Pcor90}(\bs{\gamma})} = 21.46$ is larger
than the median of the full Gibbs values $R_{Full}(\bs{\gamma})$ $\mbox{median}_{k=1}^{10} {R_{Full}(\bs{\gamma})} = 17.94$, although again the inter-quartile ranges overlap. 

In terms of the number of variables visited by the MCMC
algorithms, the picture is the same as for simulation scenario
1. While the add/delete algorithm only visited 323 of all 500
variables at least once in the application to simulated data
set 1 (see Table \ref{sim2run1}), this number is larger for all
neighbourhood samplers and increases with decreasing threshold values
$C$. Comparing the $Pcor$ and $Corr$ samplers with large
thresholds $C$, the partial-correlation based samplers visit
more variables than the correlation based neighbourhood samplers.

\subsubsection{Posterior variable inclusion probabilities}
\begin{figure}[!ht]
\begin{minipage}{0.475\textwidth}
\begin{center}
$M=1000$
\scalebox{0.4}{\includegraphics{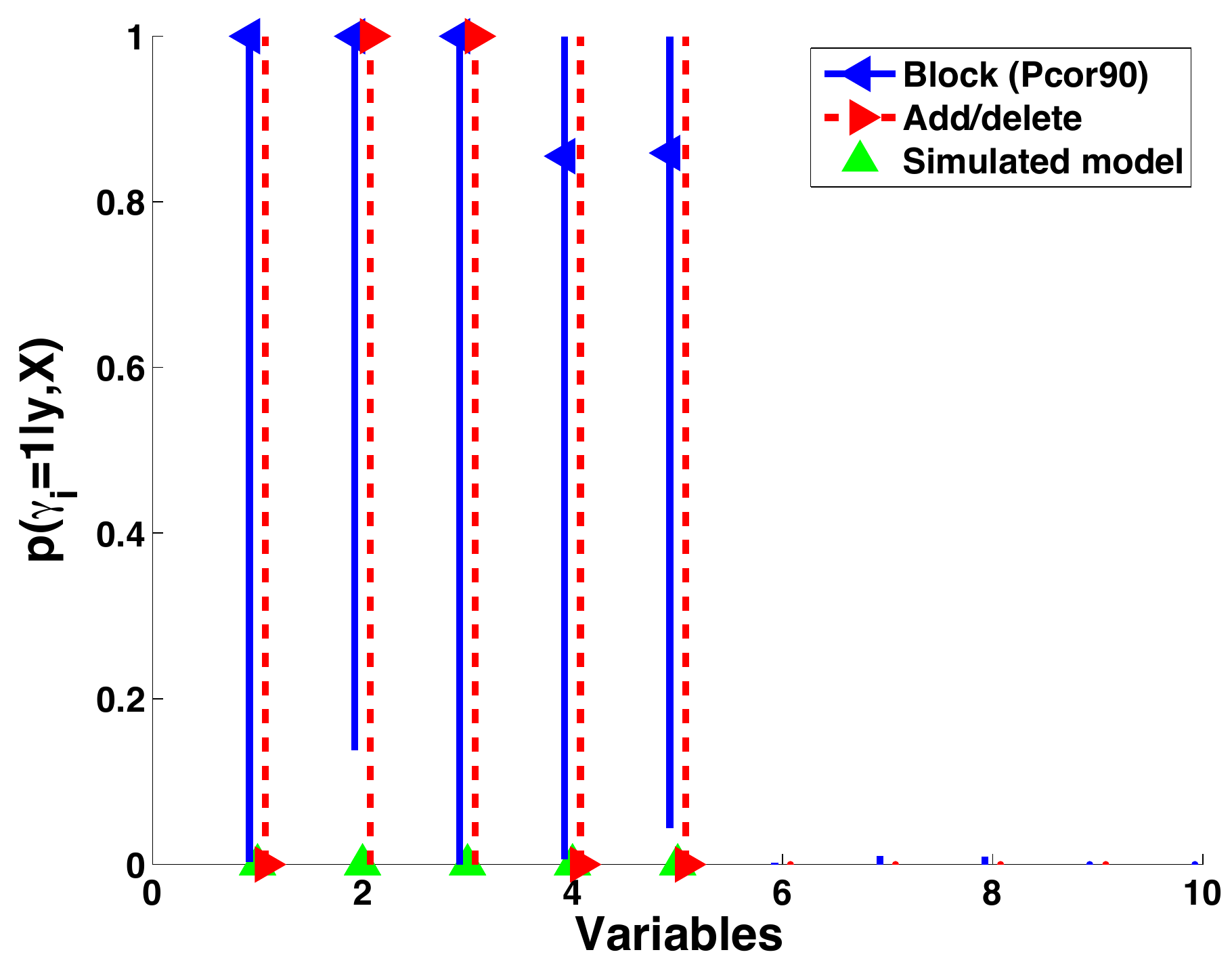}}
\end{center}
\end{minipage}\begin{minipage}{0.475\textwidth}
\begin{center}
$M=10000$
\scalebox{0.4}{\includegraphics{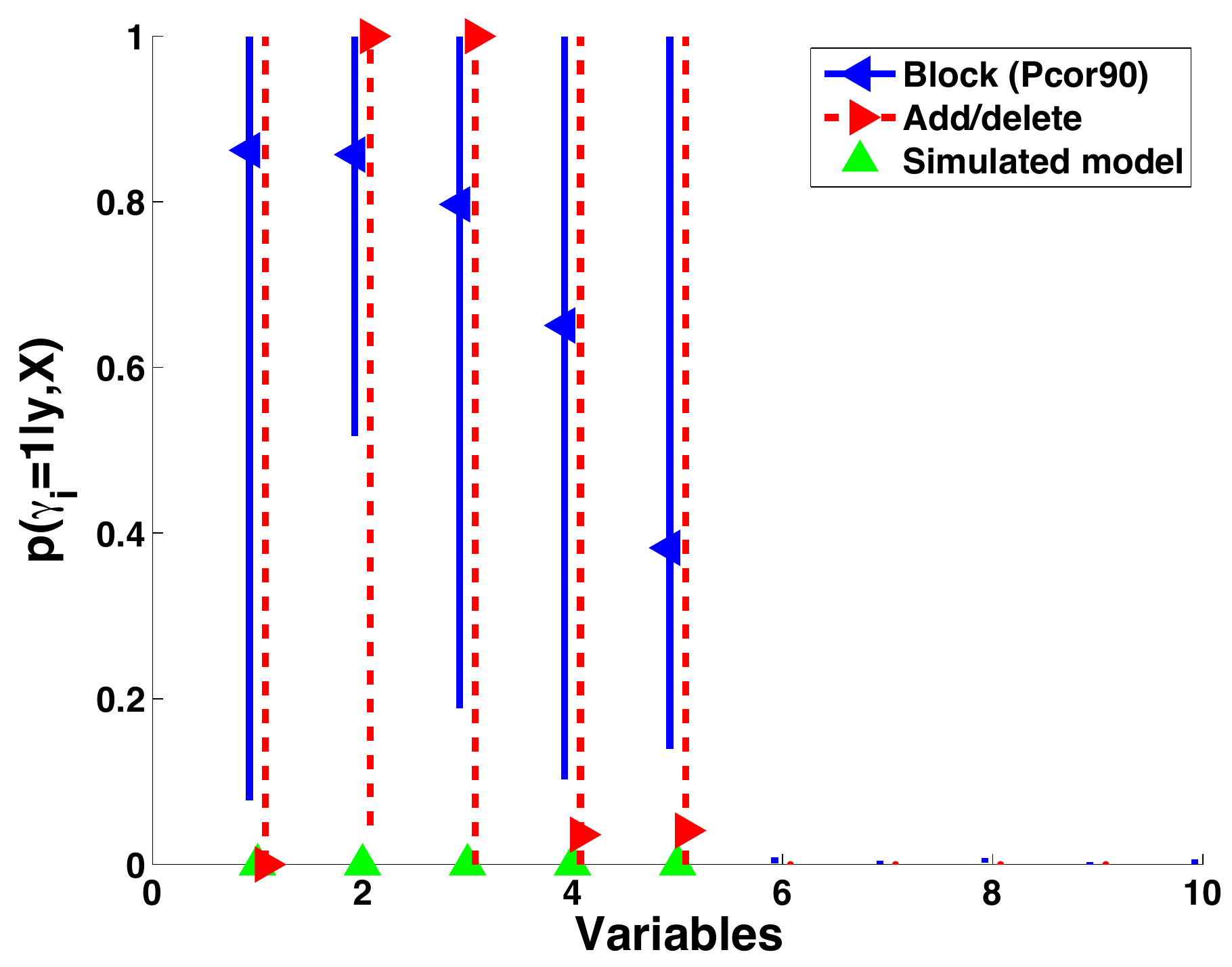}}
\end{center}
\end{minipage}

\begin{minipage}{0.475\textwidth}
\begin{center}
$M=50000$
\scalebox{0.4}{\includegraphics{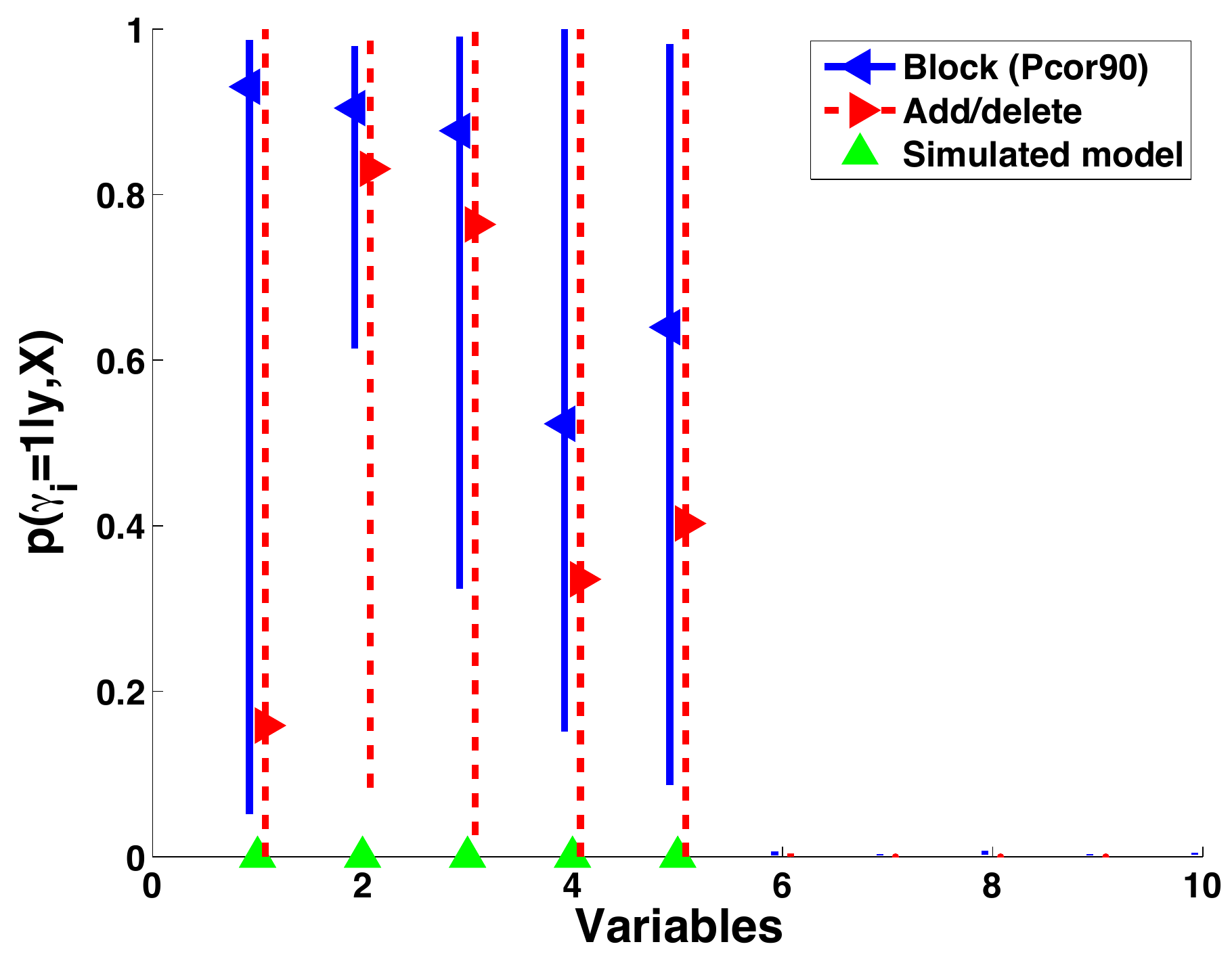}}
\end{center}
\end{minipage}\begin{minipage}{0.475\textwidth}
\begin{center}
$M=200000$
\scalebox{0.4}{\includegraphics{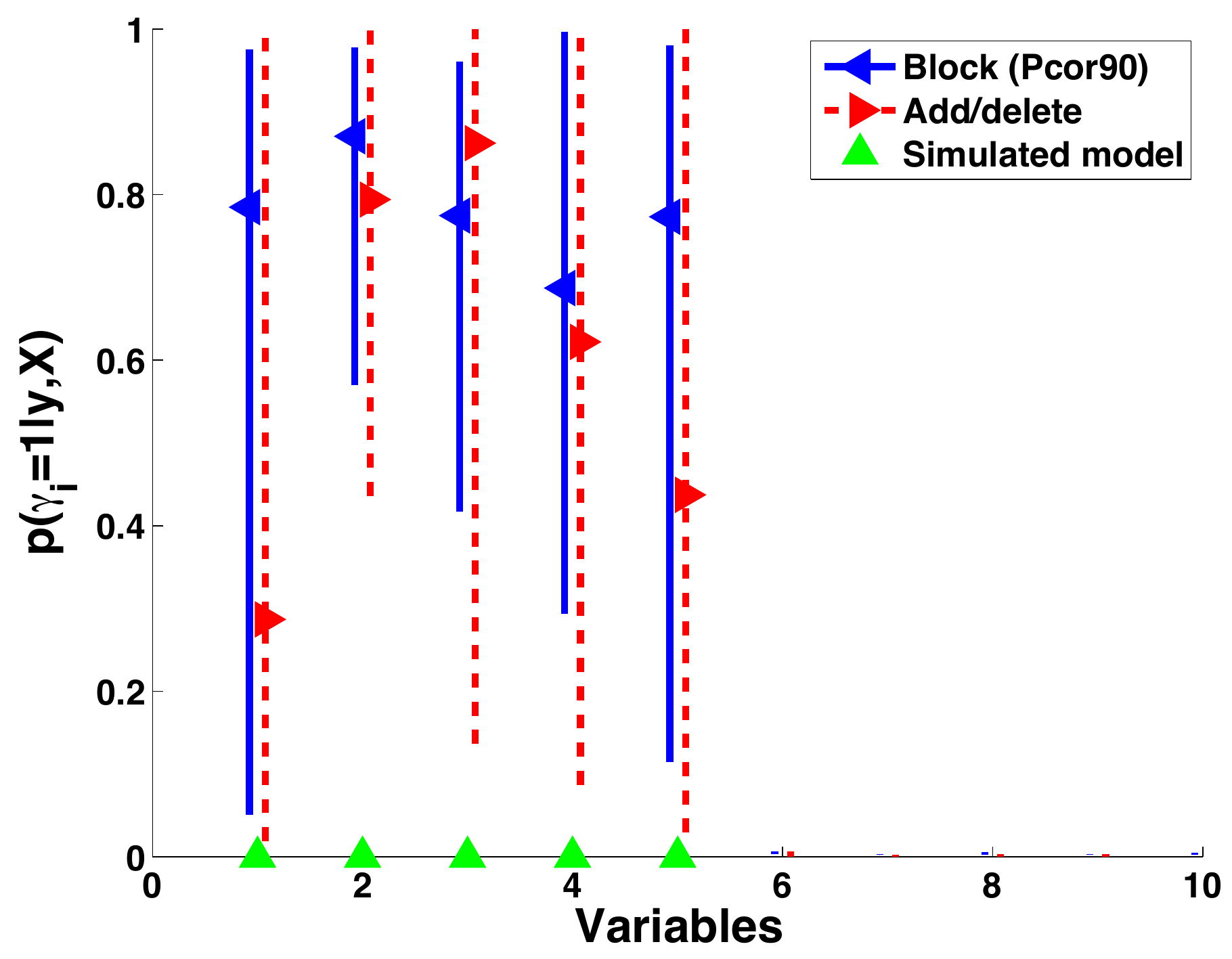}}
\end{center}
\end{minipage}
\caption[Posterior inclusion frequencies for variables
$1,...,10$ over all 25 replicates of simulation setup 2]{\small
Posterior inclusion frequencies (median and inter-quartile
ranges) for variables $1,...,10$ over all 25 replicates of
simulation setup 2 (after burn-in period).}\label{sim2post}
\end{figure}

Figure \ref{sim2post} shows the median values and
inter-quartile ranges of the MCMC estimates of the posterior
variable inclusion probabilities $\hat p(\gamma_i = 1|\bs{x},\bs{y})$ for
variables $(\bs{x}_1,...,\bs{x}_{10})$, over all 25 generated data sets.
These include the $p^* = 5$ ``true'' predictors $(\bs{x}_1,...,\bs{x}_5)$,
which were generated as being linked to the response variable
$\bs{y}$. 
The results shown for variables
$(\bs{x}_6,...,\bs{x}_{10})$ are representative for all variables, which
are simulated not to be correlated with the response $\bs{y}$, and
again, the median posterior inclusion probability estimates of
these variables are close to zero for all values of $M$ and
both samplers. 
After $M = 200,000$ post-burn-in iterations the median estimates for the five ``true'' predictors for the neighbourhood sampler converge at values of around 0.8. Overall, the inter-quartile ranges are again narrower for the
neighbourhood sampler than for the add/delete sampler, but seem to be
wider after $M = 200,000$ than in simulation scenario 1 after
$M = 150,000$ iterations. The median values of the estimated posterior inclusion
probabilities are between about
0.7 and 0.9 for the neighbourhood sampler and between 0.3 and 0.9 for the
add/delete algorithm at $M = 200,000$. In
individual data sets, the median number of false negatives is
zero for the neighbourhood and full Gibbs samplers, but one for the
add/delete algorithm (Tables \ref{sim2} and \ref{sim2run1}).
The median numbers of false positives range from 8
($Pcor90$) to 10 ($AD$), when using the same cut-off as that
used for marking the false negatives, at $\hat p(\gamma_i =
1|\bs{x},\bs{y}) > 0.05$.

\subsubsection{Alternative Gibbs samplers within the neighbourhoods}
We again apply the restricted joint Gibbs samplers $Joint4$ and $Joint10$,
and in addition the corresponding restricted univariate Gibbs
samplers $Rgibbs4$ and $Rgibbs10$. Again, the $Pcor90$
mechanism is used to determine the underlying neighbourhood structure. The results of data set 1 of simulation scenario 2 are
summarised in Table \ref{sim2run1}. While the computation time
needed for the $Joint4$ run is with 112 minutes less than half
the time needed for the univariate Gibbs run ($Pcor90$), the
time required to run the $Joint10$ sampler is 30 hours for only
$N_{Joint10} = 110,000$ MCMC iterations instead of $N_{Pcor90}
= 250,000$ iterations. At the same time, the effective sample
sizes $\mbox{ESS}^*(\bs{\gamma})$ are only $11\%$ of
$\mbox{ESS}^*_{Pcor90}(\bs{\gamma})$ for the $Joint4$ sampler, and
only $30\%$ for the $Joint10$ algorithm when adjusting for the
differences in post-burn-in MCMC run lengths
($M_{Pcor90}=200,000$ vs. $M_{Joint10}=100,000$). We draw the
same conclusions as before, i.e. that any improvement in mixing
with increasing set sizes $d$ in
multivariate-Gibbs-within-neighbourhood samplers $Joint<d>$ is
outweighed by the exponentially increasing computation time.
Also, the effective sample sizes of the multivariate samplers
are only modestly larger than those of their corresponding
restricted univariate Gibbs samplers ($Rgibbs4$ and
$Rgibbs10$), with ratios of the effective sample sizes being
about $1.1$ for both $d=4$ and $d=10$ (when adjusting
$\mbox{ESS}^*_{Joint10}(\bs{\gamma})$ for the reduced post-burn-in
MCMC run length).

\subsubsection{Random construction of neighbourhoods}
In addition to neighbourhoods constructed by means of estimated
correlation and partial correlation matrices, for comparison
reasons we here also construct neighbourhoods simply by randomly
drawing variables into neighbourhoods, matching the neighbourhood sizes with
the mean sizes observed for the partial-correlation-based
and correlation-based neighbourhood structures for the given threshold
values $C$. These $Random<C>$ neighbourhood samplers are applied to the
first two of the 25 generated data sets. The results for data
set 1 are listed in Table \ref{sim2run1}, and the curves of the
ratios $R_{Random}(\bs{\gamma})$ values relative to
$R_{Full}(\bs{\gamma})$ are shown in Figure \ref{sim2mixing}. The
effective sample sizes and consequently the ratios $R(\bs{\gamma}) =
\mbox{ESS}^*(\bs{\gamma})/t$ of the $Random<C>$ samplers are similar to
the $Corr<C>$ samplers for the two observed data sets. This
suggests that for this simulation scenario, where the
correlation structure of the data corresponds to that of a real
gene expression data set, the $Corr<C>$ samplers are not doing
better in terms of mixing relative to computation time than
neighbourhood samplers with the same mean neighbourhood sizes, where the
``neighbourhoods'' are just random selections of genes. The $Pcor<C>$
samplers, on the other hand, are consistently more efficient
than the $Random<C>$ samplers. These observations conform with the
idea that the dependence structure in gene expression data can
better be explained by a sparse partial correlation structure
than by a sparse correlation matrix, i.e. that the sparsity is
observed in terms of conditional dependence rather than
marginal dependence.

\begin{table}[!ht]
\caption[Mixing performance results with respect to $\bs{\gamma}$
for scenario 2: results for one data set]{\small
\label{sim2run1}Mixing performance results with respect to
$\bs{\gamma}$ for scenario 2: results for one data set (run 1). Diagnostic measures for Markov chain mixing with
respect to $\bs{\gamma}$.}
\begin{center}
\small
\begin{tabular}{|p{1.75cm}|p{1.75cm}|p{1.75cm}|p{1.75cm}|p{1.75cm}|p{1.75cm}|p{1.75cm}|}
\hline MCMC &CPU time&$\mbox{ESS}^*(\bs{\gamma})$&$R(\bs{\gamma})$&$\# \bs{I}^\sharp$&\# FP$^\dag$&\# FN$^\dag$\\
sampler &$t$ (min)&&&&&\\
\hline
$AD$ & 53 & 104 & 1.96 & 323 & 14 & 0 \\
\hline
$Full^\ddag$ & 932 & 11640 & 12.50 & 500 & 10 & 0 \\
\hline
\multicolumn{7}{|l|}{Neighbourhood sampler}\\
\hline
$Pcor99$ & 82 & 618 & 7.54 & 473 & 11 & 0 \\
$Pcor97.5$ & 108 & 1399 & 12.94 & 491 & 14 & 0 \\
$Pcor95$ & 153 & 2120 & 13.84 & 499 & 12 & 0 \\
$Pcor90$ & 259 & 4090 & 15.80 & 500 & 11 & 0 \\
$Pcor80$ & 433 & 6564 & 15.16 & 500 & 9 & 0 \\
$Pcor60$ & 845 & 13100 & 15.49 & 500 & 11 & 0 \\
\hline
$Corr99$ & 80 & 268 & 3.35 & 377 & 11 & 1 \\
$Corr97.5$ & 109 & 574 & 5.26 & 422 & 12 & 0 \\
$Corr95$ & 156 & 1196 & 7.67 & 463 & 12 & 0 \\
$Corr90$ & 258 & 2473 & 9.59 & 494 & 11 & 0 \\
$Corr80$ & 425 & 5163 & 12.14 & 500 & 10 & 0 \\
$Corr60$ & 972 & 11460 & 11.79 & 500 & 13 & 0 \\
\hline
$Random99$ & 77 & 187 & 2.41 & 498 & 13 & 0 \\
$Random97.5$ & 109 & 554 & 5.07 & 500 & 12 & 0 \\
$Random95$ & 153 & 1340 & 8.75 & 500 & 13 & 0 \\
$Random90$ & 247 & 2877 & 11.65 & 500 & 10 & 0 \\
$Random80$ & 457 & 5447 & 11.92 & 500 & 11 & 0 \\
$Random60$ & 846 & 10312 & 12.20 & 500 & 13 & 0 \\
\hline
$Rgibbs4$ & 75 & 412 & 5.53 & 479 & 11 & 0 \\
$Joint4$ & 112 & 448 & 3.98 & 482 & 8 & 0 \\
$Rgibbs10$ & 94 & 1099 & 11.67 & 499 & 8 & 0 \\
$Joint10^\ddag$ & 1777 & 611 & 0.34 & 485 & 12 & 0 \\
\hline
\multicolumn{7}{l}{\footnotesize{$^\ddag$ For $Full$ and
$Joint10$ it is $M = 100,000$, compared to $M = 200,000$ for all other samplers}}\\
\multicolumn{7}{l}{\footnotesize{$^\sharp$ $\bs{I} = \{i: (\sum_{m=1}^M{\gamma_{i,m}}) > 0\}$, i.e. number of
variables for which $\gamma_i = 1$ in at least one MCMC iteration}}\\
\multicolumn{7}{l}{\footnotesize{$^\dag$false positives and false negatives if cut-off at ratio
of posterior to prior probability $> 5$, i.e. if}}\\
\multicolumn{7}{l}{\footnotesize{$\hat p(\gamma_i = 1| \bs{x},\bs{y}) > 0.05$}}
\end{tabular}
\end{center}
\end{table}

\subsection{\label{sensitivity}Sensitivity analysis for prior variance parameter $c^2$}
Throughout this section, the covariance parameter $c^2$ in
$\bs{\beta} \sim N(\bs{0}_{p}, c^2 \bs{I}_p)$ was set to 5. This value was chosen
to provide a relatively flat prior across the expected range
coefficient values, and in particular to comfortably include the
``true'' regression coefficient values $\beta_1 = ... = \beta_5
= 2$, which were used to simulate the five covariates that are
linked to the response. To see, how much this choice of $c^2 =
5$ has influenced the posterior distributions, not just of
$\bs{\beta}$, but also of the main parameter of interest $\bs{\gamma}$, a
range of different values for $c^2$ has been applied in this
section. Both the add/delete and $Pcor90$ samplers have
been applied to one of the data sets generated according to
simulation scenario 2.

In several previous publications, where the probit model was
used for variable selection in binary regression rather than
the logistic model \citep[e.g.][]{brown98_2,lee03,tadesse05},
the authors had warned that the posterior inference about the
covariate indicator variable $\bs{\gamma}$ can be influenced by the
choice of the prior covariance parameter $c^2$. For g-priors,
i.e. $\bs{\beta} \sim N(\bs{0}_{p}, c^2 (\bs{x}'\bs{x})^{-1})$, the suggestion by
\citet{smith96} to use large values of $c^2$ ranging between 10
and 100 is often followed. For the independence prior, which is
used here, \citet{brown02} and \citet{sha04} suggest values
which are small relative to typically expected regression
coefficient values $\bs{\beta}$ and are chosen in order to allow for
good inference about $\bs{\gamma}$ (rather than $\bs{\beta}$). In
particular, \citet{sha04} argue for using a value $c^2$ which
implies a ratio of prior to posterior precision of between 0.1
and 0.005. The prior precision is $1/c^2$ for all variables
$\bs{x}_i$ and the posterior precision is $1/c^2 + e_i$ where the
vector $(e_i)_{i=1,...,n-1}$ of eigenvalues of the precision
matrix is equal to the inverse of the vector of non-zero
eigenvalues of the empirical covariance matrix. Consequently,
the range of $c^2$ is given by
\begin{equation}\label{sha2004}
c(\bar{e},0.1) < c^2 < c(\bar{e},0.005),
\end{equation}
where $c(e,p) = (1-p)/(p e)$ and $\bar{e}$ denotes the mean
eigenvalue. This criterion is proposed for
data matrices where the condition number, i.e. the ratio of
maximum and minimum eigenvalues, is not too large.

We use the simulation scenario 2 to compare the sensitivity of both, probit and logistic,
BVS regression models with regards to the influence of $c^2$ on
Markov chain mixing and convergence behaviour, and posterior
inference about $\bs{\gamma}$ and $\bs{\beta}$. The probit regression
model is implemented in the auxiliary variable formulation
described by \citet{albert93} as given in equation
(\ref{albert1993}). We apply the MCMC sampling algorithms for Bayesian probit regression detailed in
\citet{holmes06}.

\begin{figure}[!ht]
\begin{center}
\scalebox{0.27}{\includegraphics{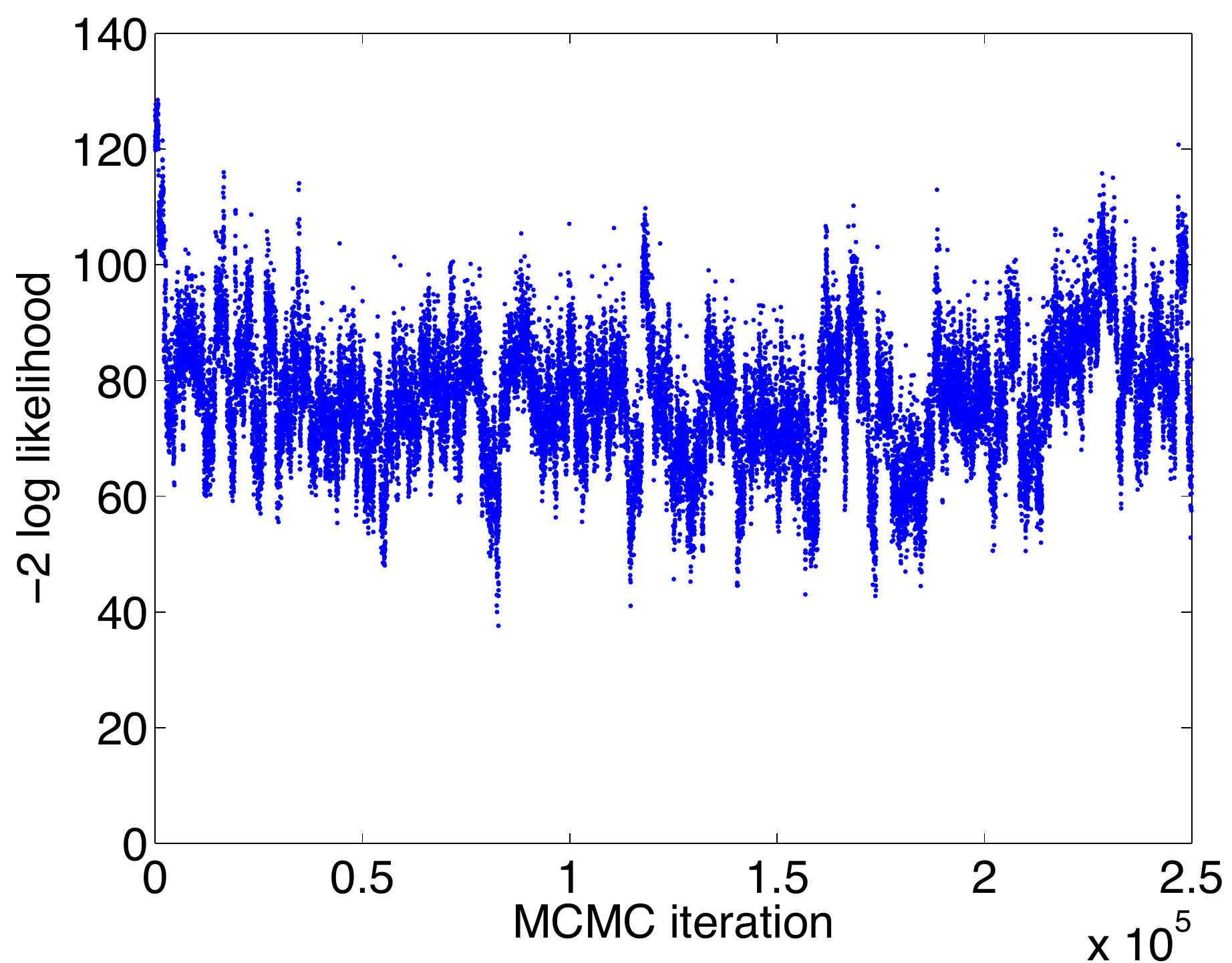}}
\scalebox{0.27}{\includegraphics{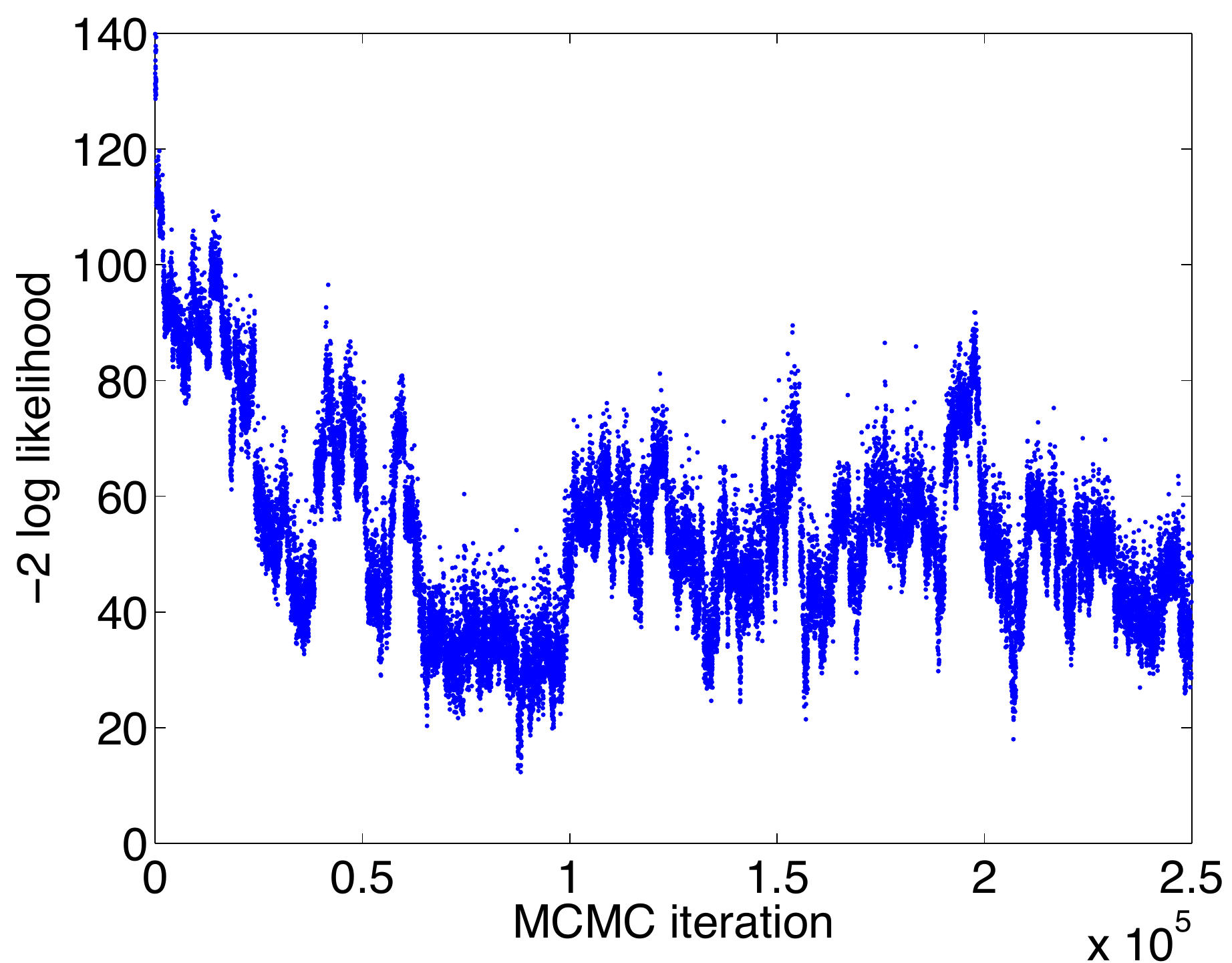}}
\scalebox{0.27}{\includegraphics{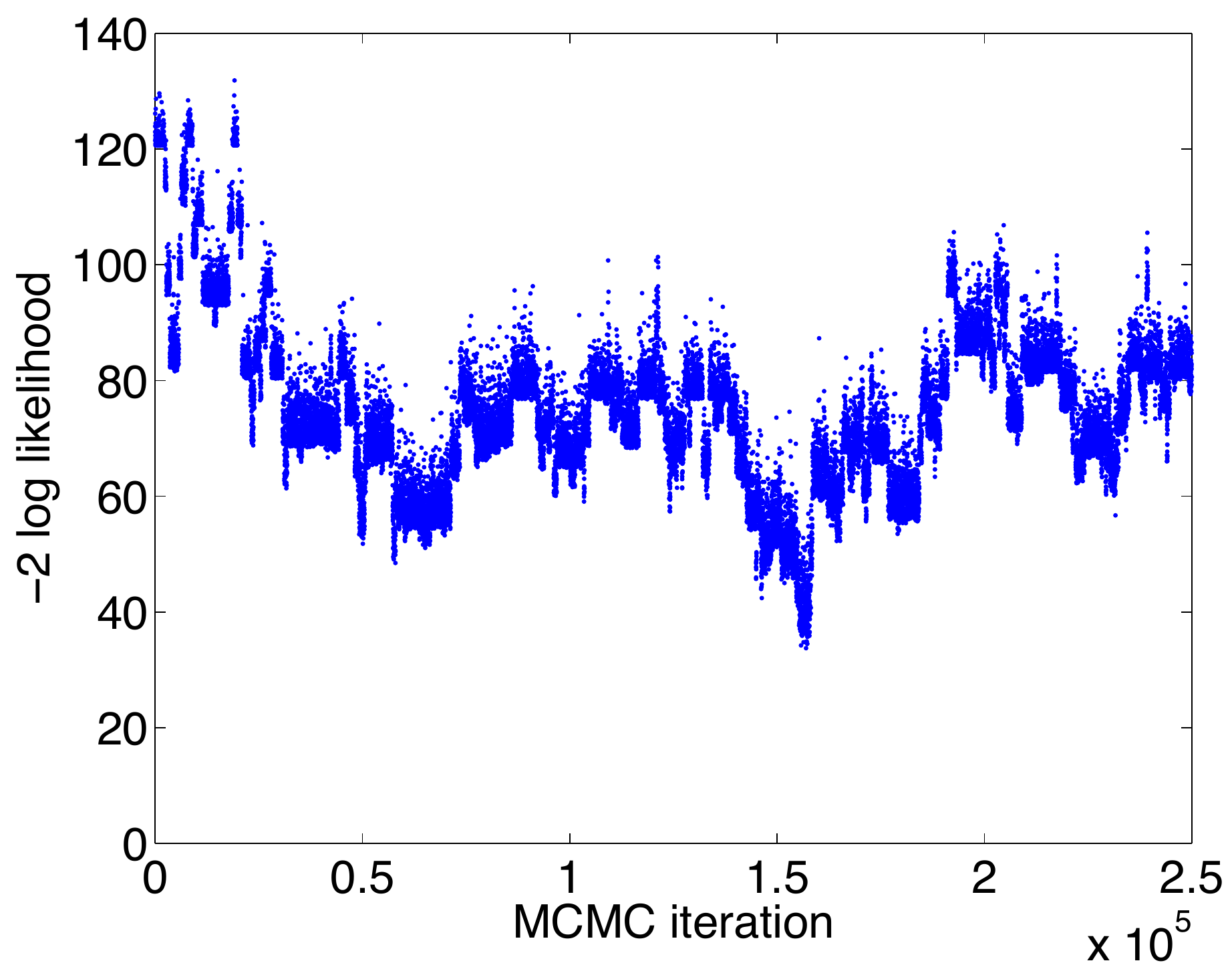}}\\
\scalebox{0.27}{\includegraphics{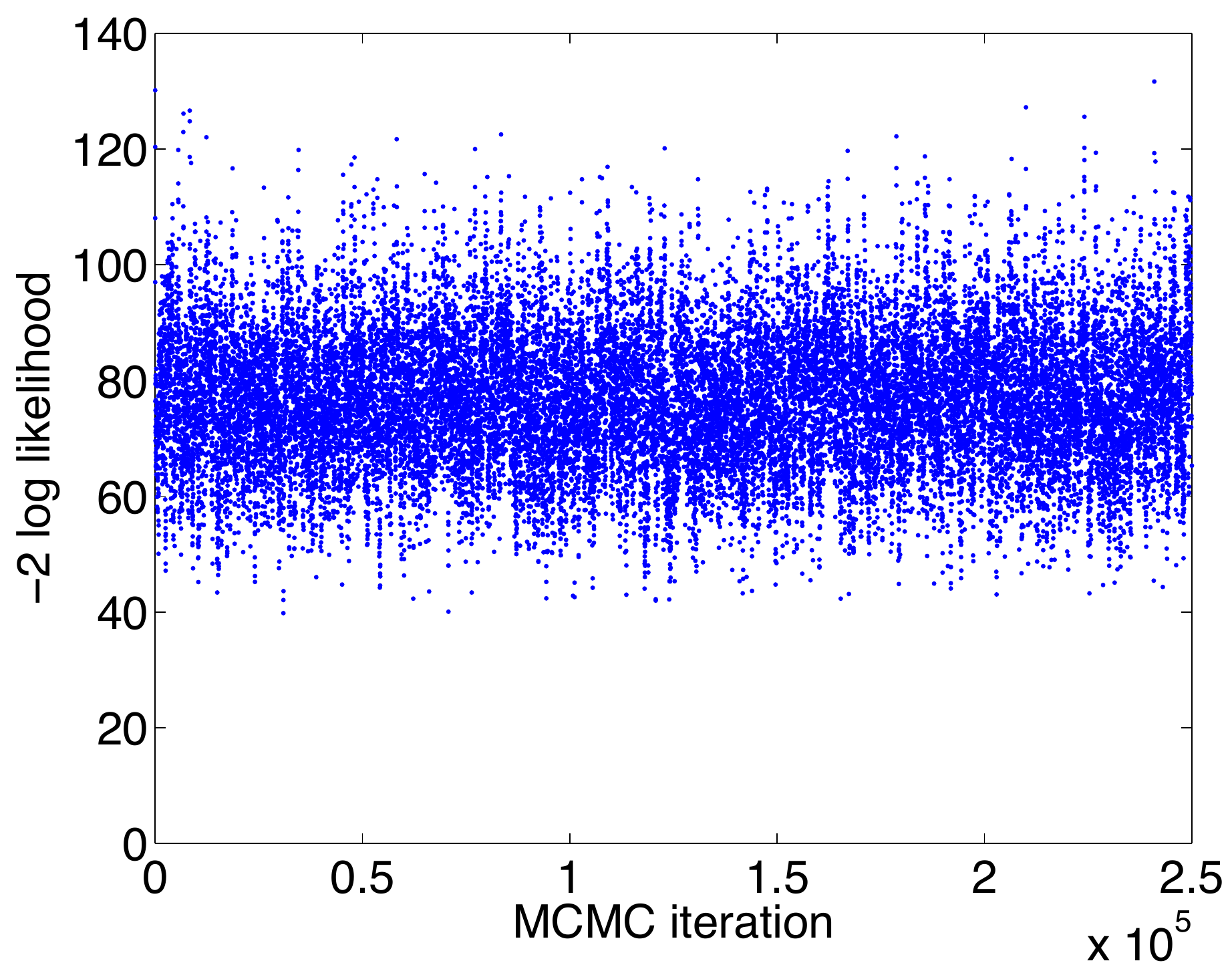}}
\scalebox{0.27}{\includegraphics{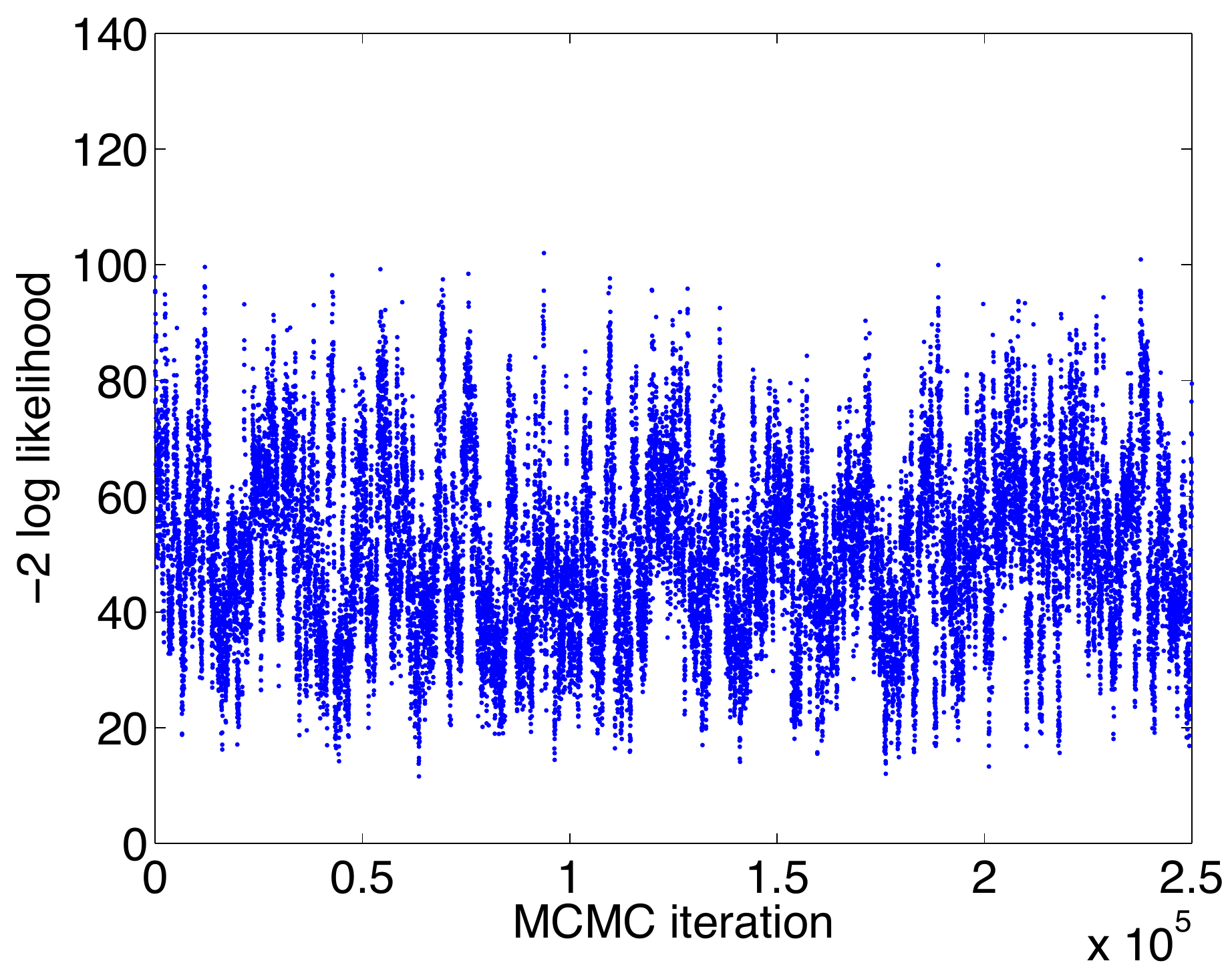}}
\scalebox{0.27}{\includegraphics{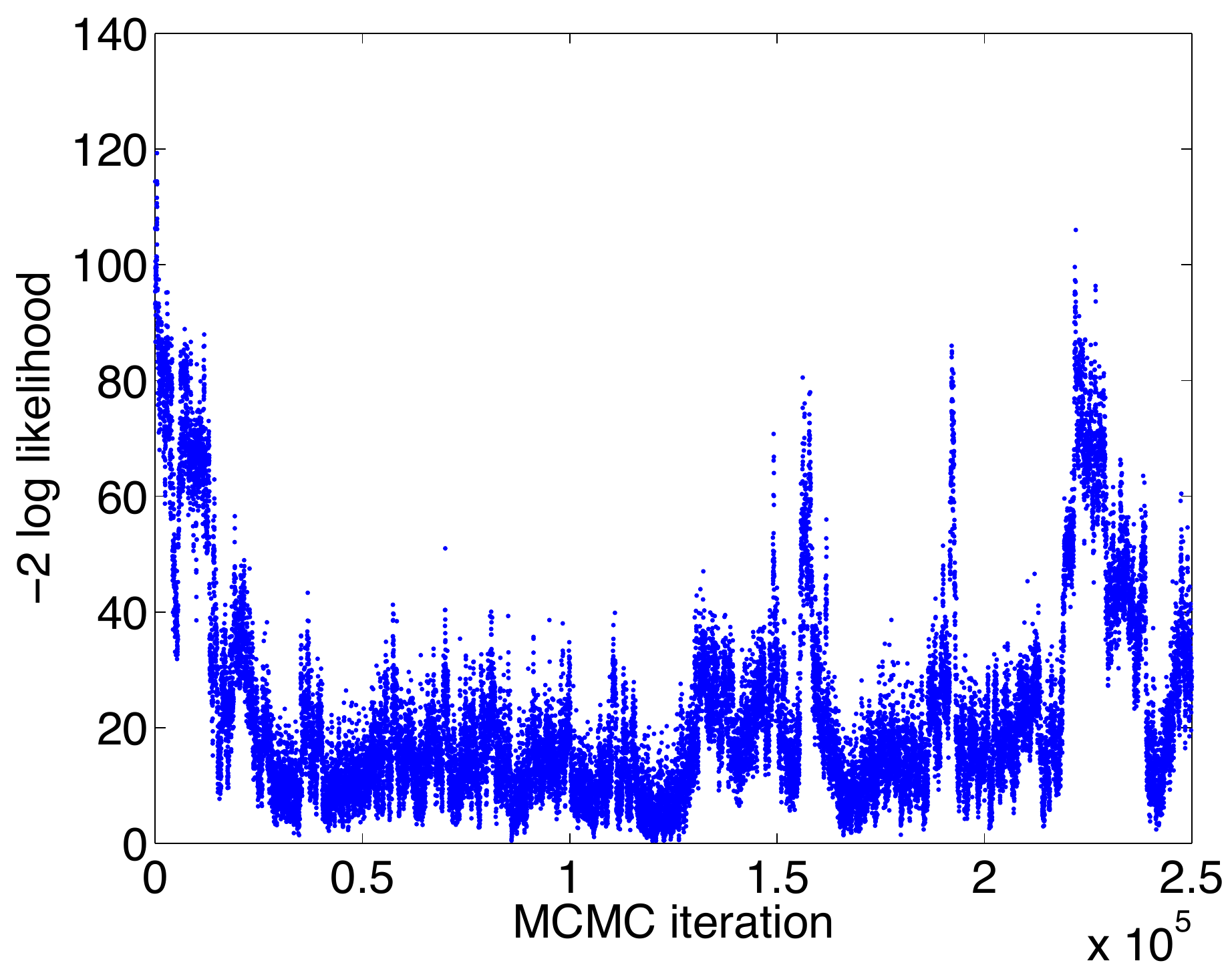}}
\caption[Logistic BVS model: trace plots of deviance for
add/delete samplers and neighbourhood samplers with prior covariance parameters
$c^2 = 0.5$, $c^2 = 5$, and
$c^2 = 50$]{\small \label{tracelogit}Logistic BVS model: trace plots of model deviance for
add/delete samplers (top) and neighbourhood samplers ($Pcor90$)
(bottom) with prior covariance parameters
$c^2 = 0.5$ (left), $c^2 = 5$ (centre, standard value chosen
throughout simulation studies in previous section), and
$c^2 = 50$ (right).}
\end{center}
\end{figure}

Starting with the logistic variable selection model, the trace
plots of the model deviances in Figure \ref{tracelogit} are
used to visually monitor MCMC convergence and mixing. In terms
of model deviance, the Markov chains mix better and converge
faster, if the prior covariance parameter $c^2$ is chosen to be
small, for both add/delete and neighbourhood sampling algorithms. This
is also true for chain mixing at the level of the individual
covariate indicators $\bs{\gamma}$ as indicated by the effective
sample sizes $\mbox{ESS}^*$ and the numbers of variables $\# \bs{I}$
visited by the chains at least once (see Table \ref{sensitive}). This behaviour is not unexpected,
since decreasing the size of $c^2$ restricts the posterior
parameter space so that it is easier for Markov chains to cover
the entire posterior distribution and find the regions of high
density quickly. Note that in the logistic variable selection
model, for all choices of $c^2 \in \{0.5, 5, 50\}$ Markov
chains generated by the $Pcor90$ sampler mix better than
the corresponding add/delete Metropolis-Hastings Markov chains.

In addition to monitoring the mixing and convergence properties
of the Markov chains, we also look at the posterior mean estimates
of $\bs{\beta}_\gamma$ and $\bs{\gamma}$. 
Remember that the ``true'' underlying vector of regression coefficients
used to simulate the data set is $\bs{\beta} = (2,2,2,2,2,0,...,0)$
and the ``true'' value of $\bs{\beta}_\gamma$ for the model defined
by $\bs{\gamma} = (1,1,1,1,1,0,...,0)$ would be $\bs{\beta}_\gamma =
(2,2,2,2,2)$. So in Table \ref{sensitive}, the estimates
$\bs{\hat\beta}_\gamma$ from the posterior distribution are
summarised in terms of the ranges (minimum and maximum values)
of the variables $\beta_1,...,\beta_5$ on the one hand, and of
$\beta_6,...,\beta_{500}$ on the other hand. While we expect
the estimates of the former to be close to the value two, the
latter should vary around zero. Indeed, the estimates
$\hat\beta_{\gamma i}$ for $i = 6,...,500$ vary around zero,
with the ranges becoming larger with increasing values of
$c^2$. The values of $\hat\beta_{\gamma i}$ for $i = 1,...,5$
also depend on the choice of $c^2$, with those posterior
estimates obtained with the prior covariance paramater $c^2 =
5$ being closest to the expected value 2 (although being
slightly too large with ranges of $(2.13,3.03)$ for the
add/delete sampler and $(2.22,3.16)$ for the neighbourhood sampler).
Note that the $\hat\beta_{\gamma i}$ are the marginal estimates computed by averaging over all MCMC iterations after burn-in $m=1,...,M$, where $\gamma_{i,m} = 1$, not taking into account, which other variables are included in the model at each iteration. Hence, the estimates are not conditioned to the ``true'' model, where $\bs{\gamma} = (1,1,1,1,1,0,...,0)$. 
More important for the variable selection problem is
posterior inference about the covariate indicator vector
$\bs{\gamma}$. The results for $\bs{\gamma}$ are summarised in terms of
false positives and false negatives, defined using the
threshold $\hat p(\gamma_i = 1| \bs{x},\bs{y}) > 0.05$ as before in this manuscript. The add/delete sampler with $c^2 = 50$ is the only
MCMC run, where not all five ``true'' predictors are detected
at that level, with variable 3 never even having been visited
by the Markov chain. There is no obvious difference in the
numbers of false positives selected by the samplers at other values of $c^2$. In summary, the logistic variable
selection model is quite robust to the choice of the prior
covariance parameter $c^2$ in terms the covariate indicator
vector $\bs{\gamma}$. This allows to use the samplers for inference
about variable selection and model selection without need for
fine-tuning $c^2$.

\begin{figure}[p]
\begin{center}
\scalebox{0.3}{\includegraphics{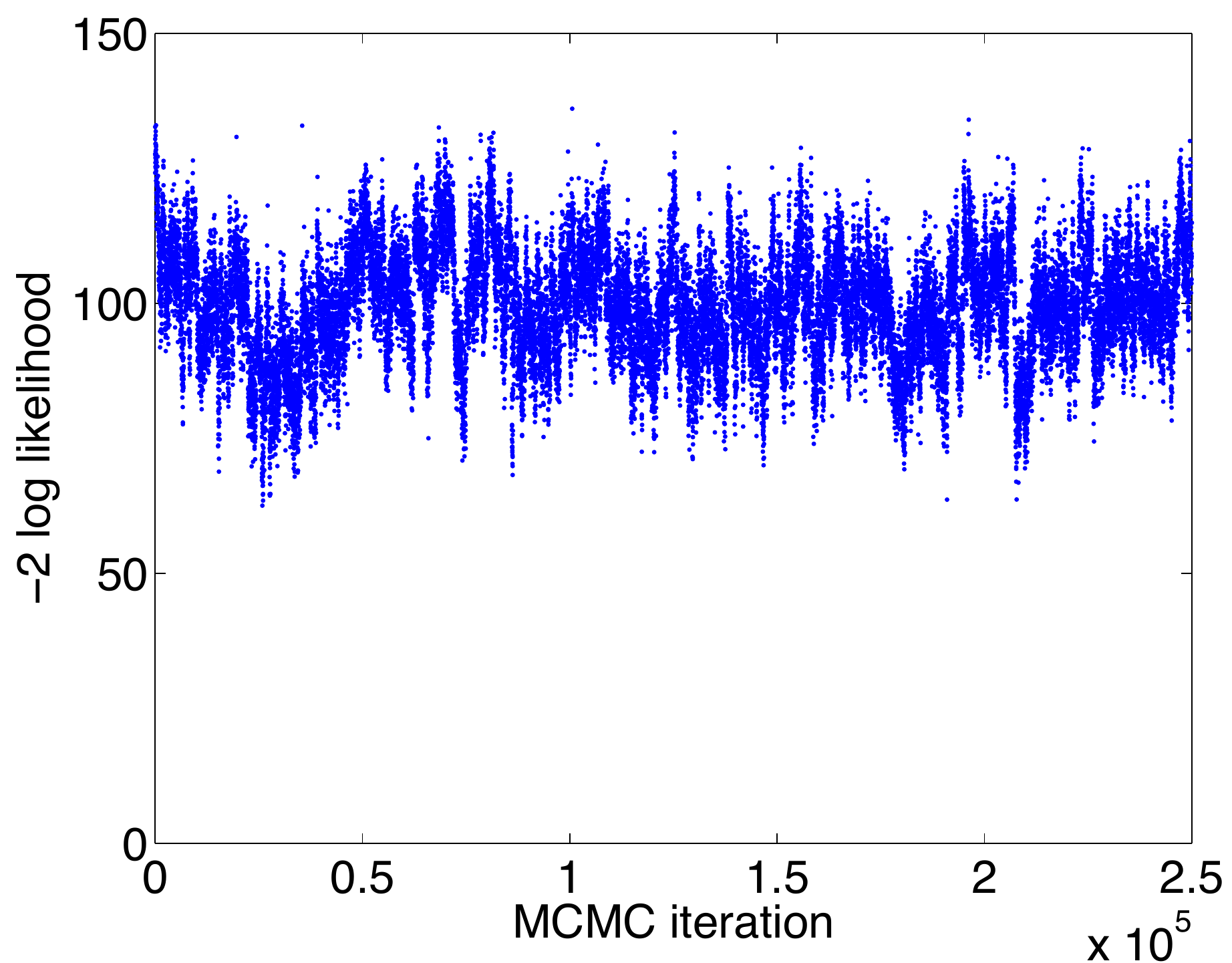}}
\scalebox{0.3}{\includegraphics{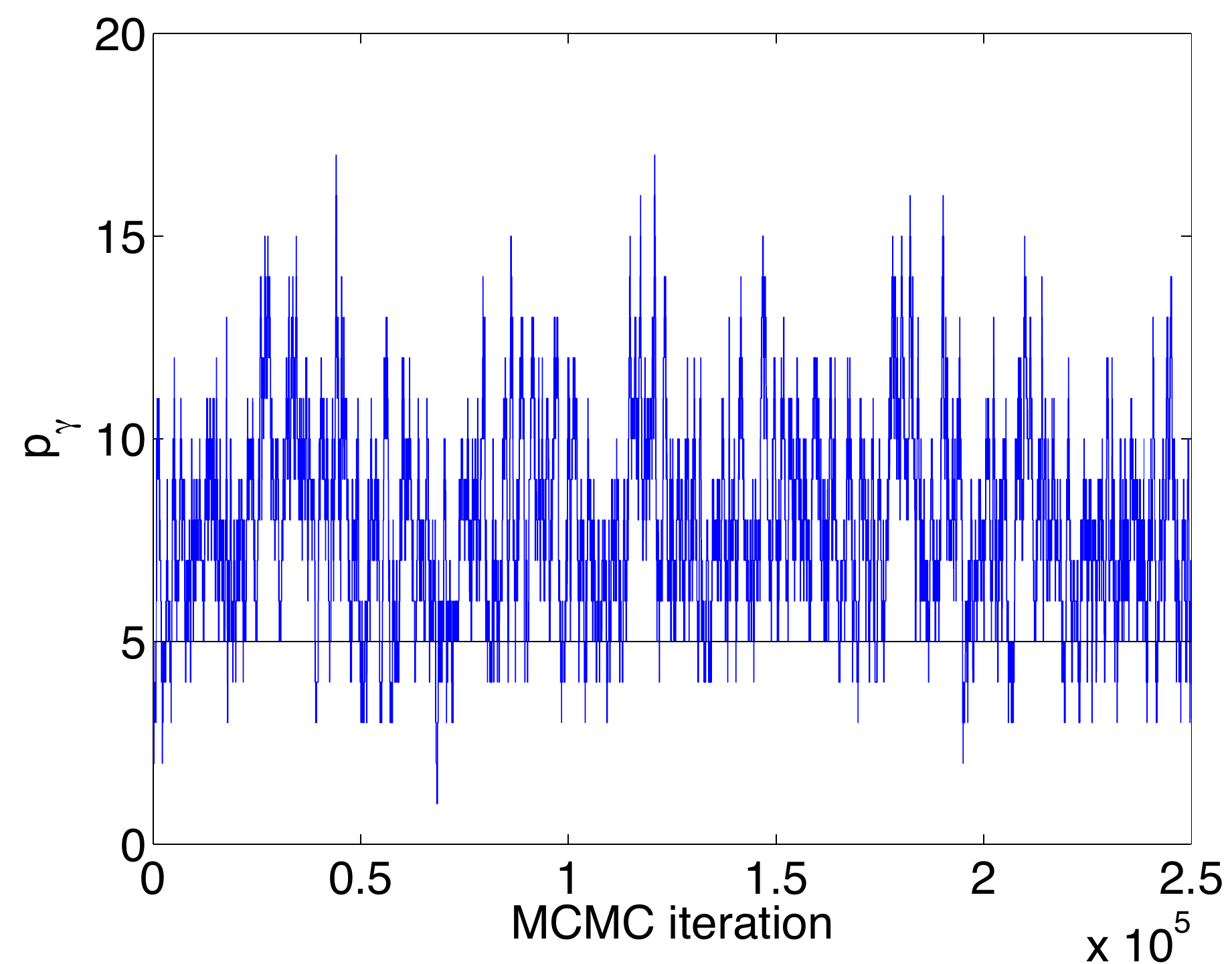}}\\
\scalebox{0.3}{\includegraphics{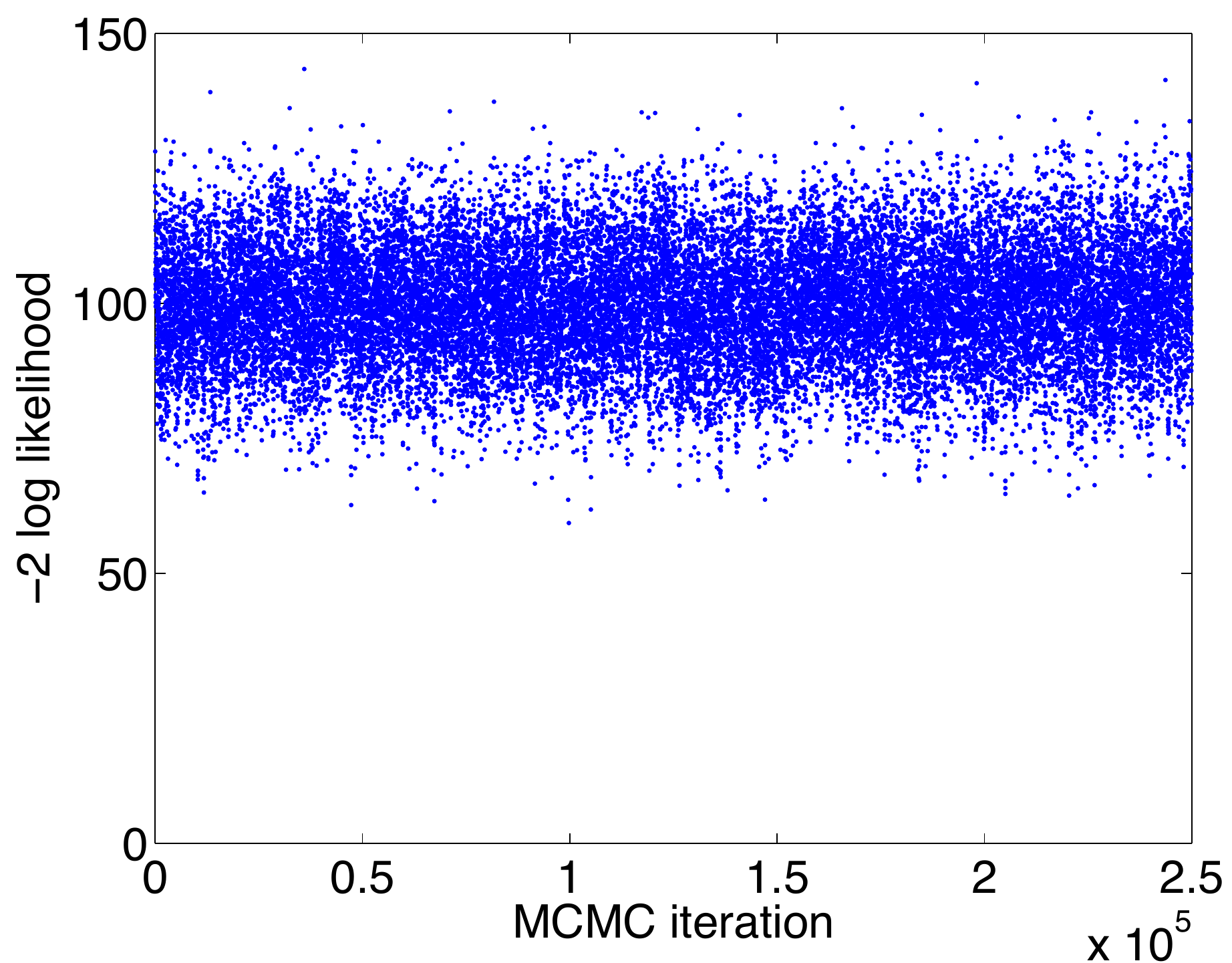}}
\scalebox{0.3}{\includegraphics{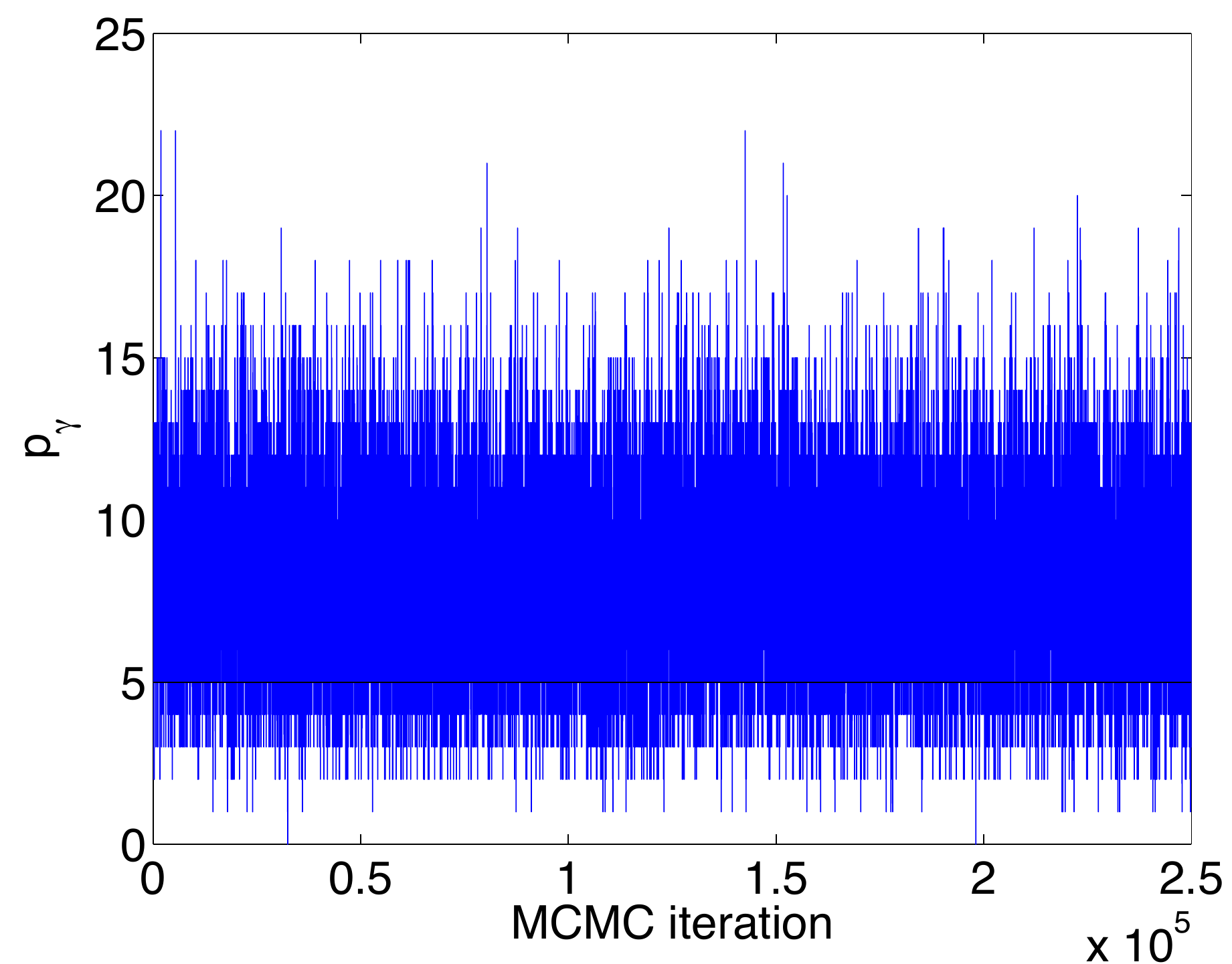}}\\
\scalebox{0.3}{\includegraphics{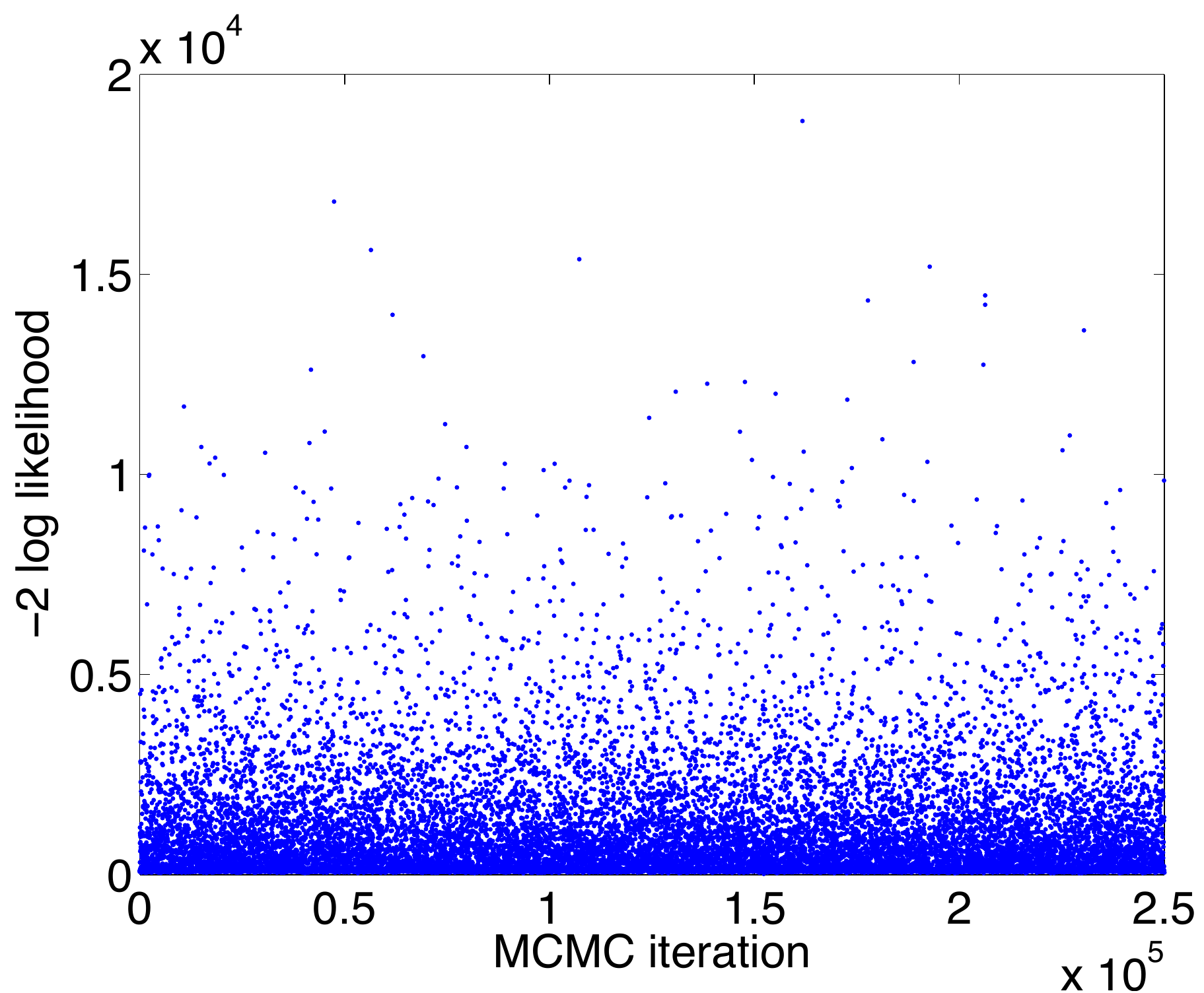}}
\scalebox{0.3}{\includegraphics{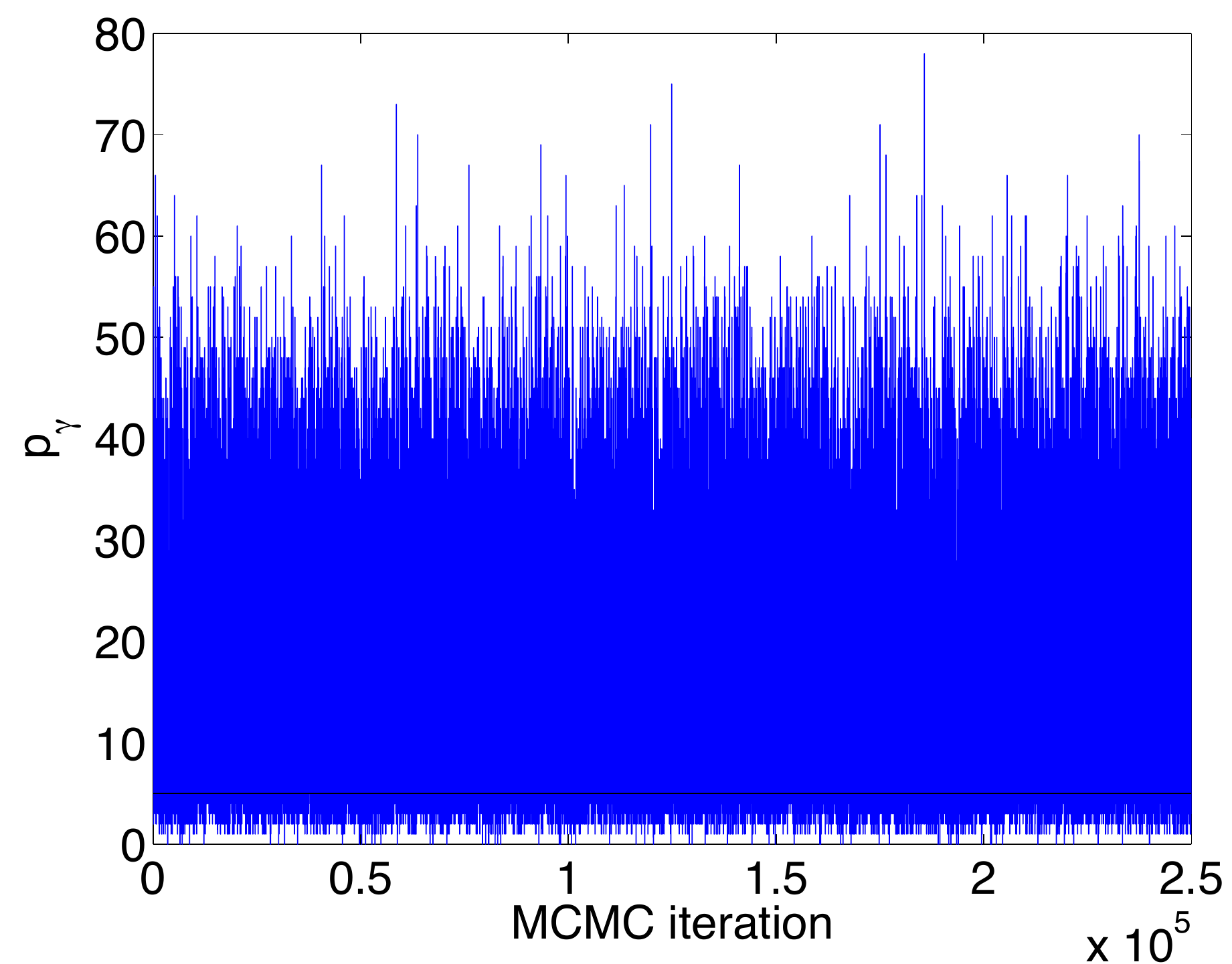}}\\
\scalebox{0.3}{\includegraphics{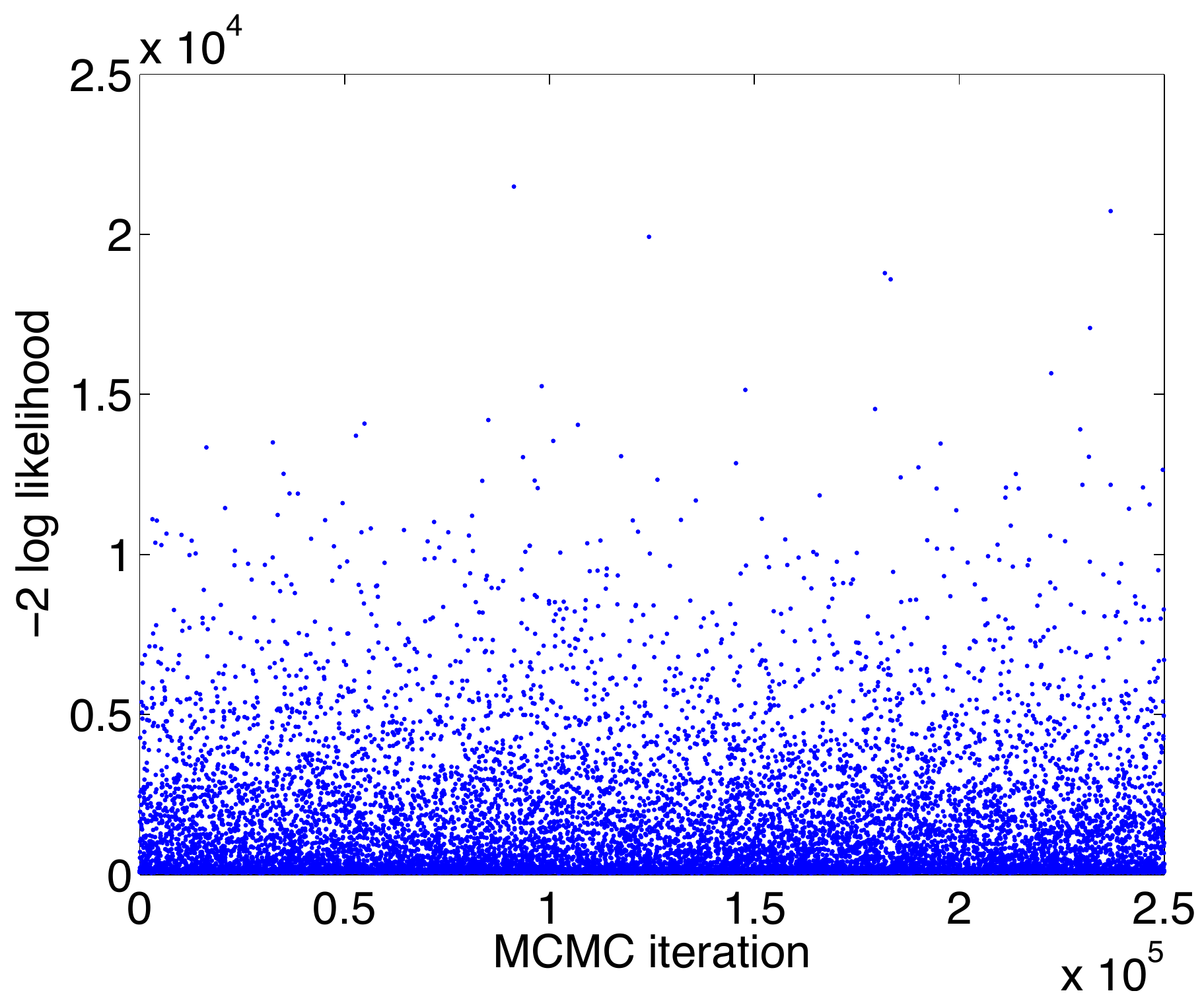}}
\scalebox{0.3}{\includegraphics{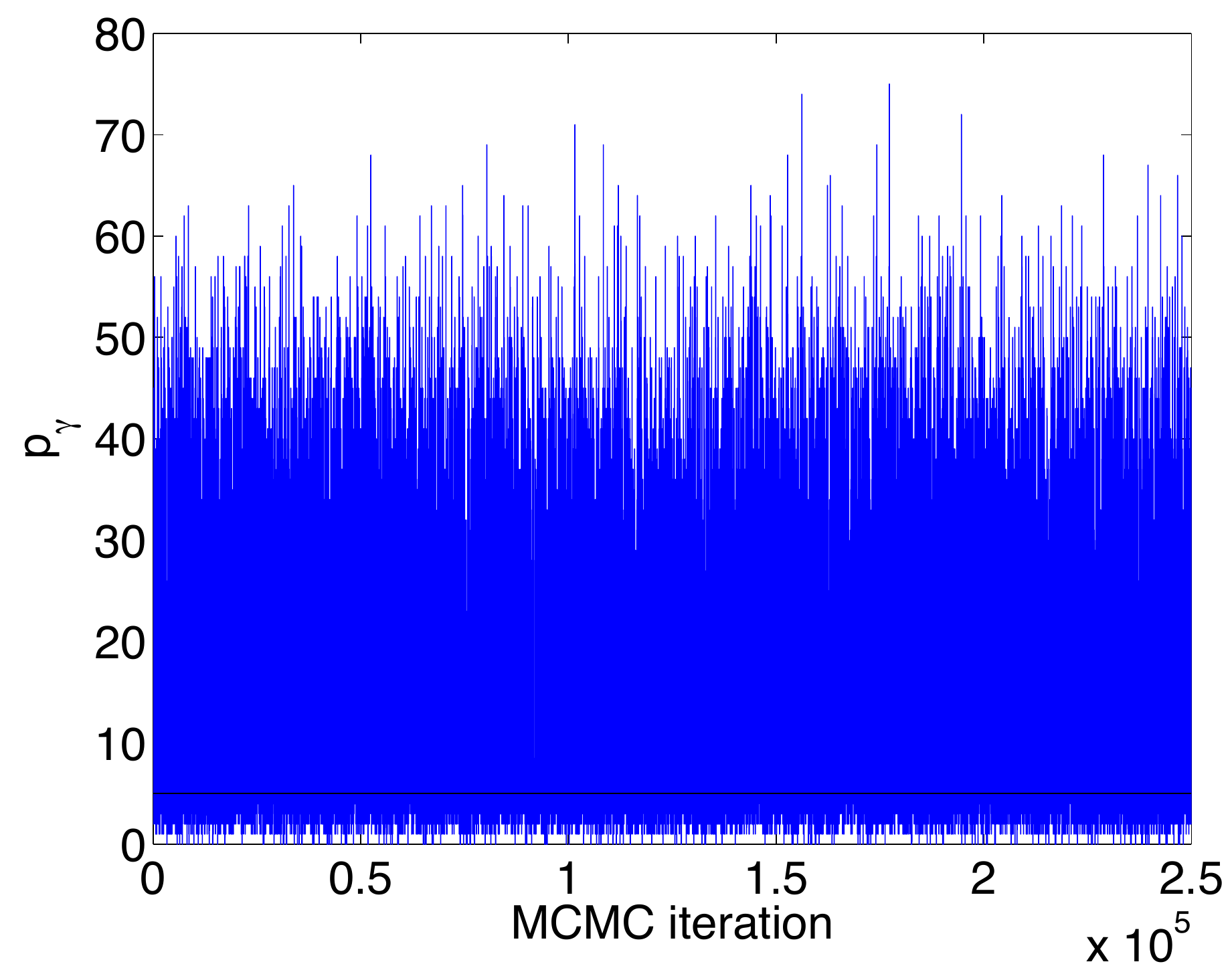}}\\
\scalebox{0.3}{\includegraphics{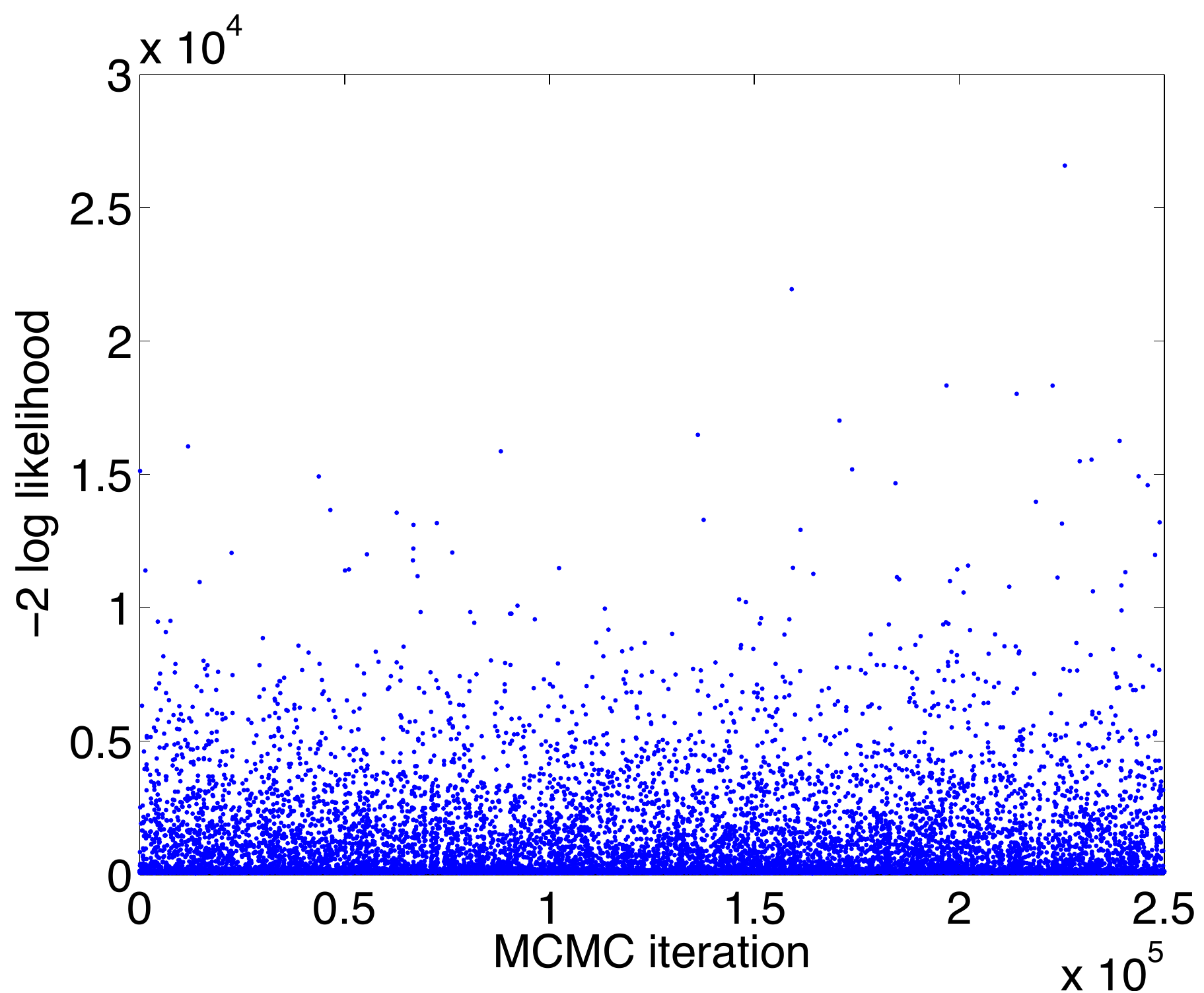}}
\scalebox{0.3}{\includegraphics{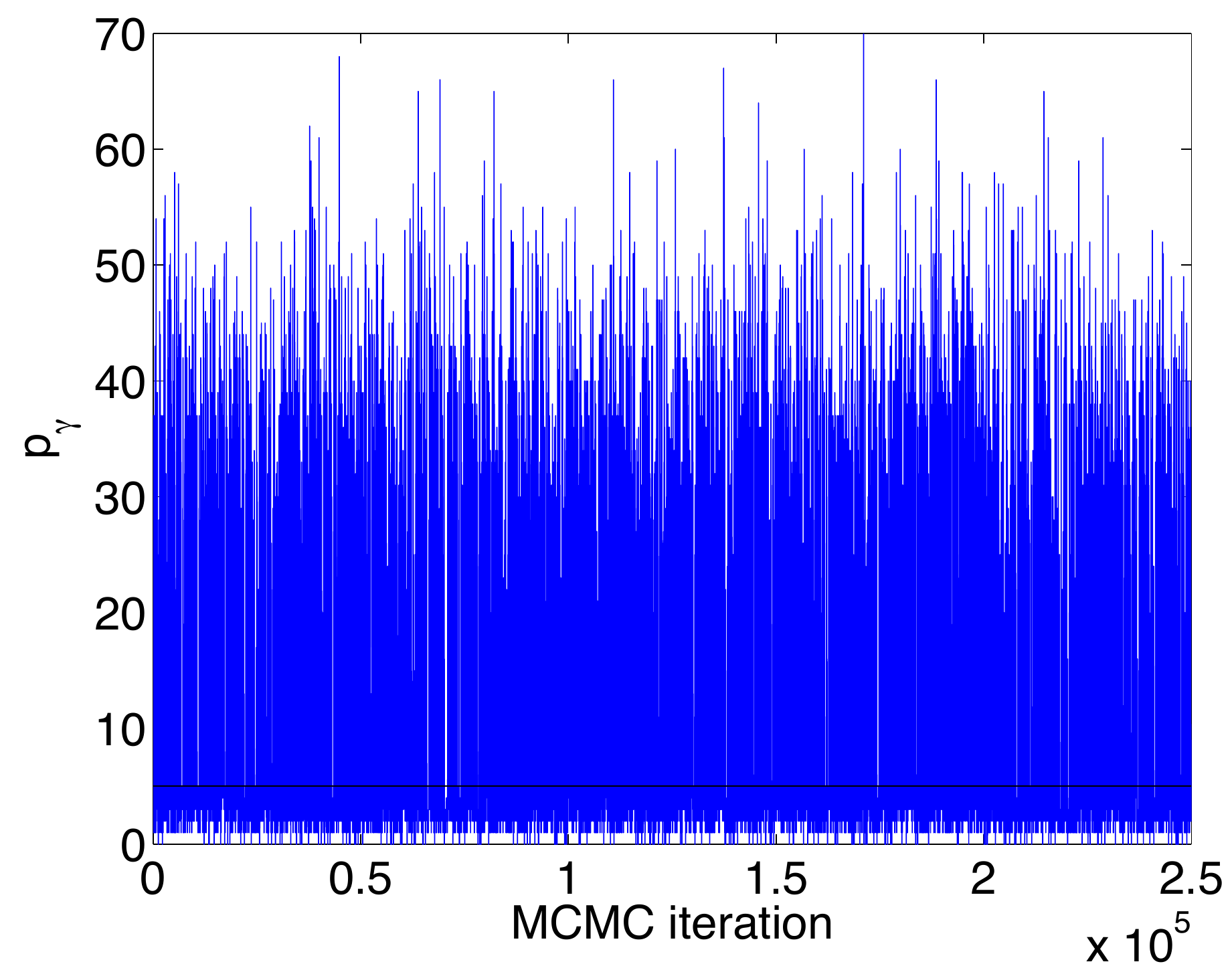}}\\
\caption[Probit BVS model: trace plots of model deviances and model sizes for
add/delete sampler with $c^2 = 0.05$ and neighbourhood samplers with
$c^2 = \{0.05,0.5,5,50\}$]
{\small \label{traceprobit}Probit BVS model: trace plots of model deviances (left) and model sizes (right) for
add/delete sampler with prior covariance parameter $c^2 = 0.05$ and neighbourhood samplers ($Pcor90$) with
$c^2 = \{0.05,0.5,5,50\}$ (from top to bottom).}
\end{center}
\end{figure}

The probit variable selection model is much more sensitive to
the choice of $c^2$, especially the add/delete algorithm, which
does not even converge if $c^2$ is chosen too large (see Table
\ref{sensitive}). Instead, the samplers start to include more
and more variables until the number of variables in the model
became larger than the sample size $n=104$. Consequently, the
samplers slow down significantly, due to the necessity to
invert large matrices of size $k\times k$ with $k =
\min(p_\gamma, n)$ in every iteration. At that point, the
sampling process was stopped manually due to convergence
problems. This problem is related to the fact, that in sparse situations
with small variable inclusion probability $\pi = p^*/p$,
the acceptance probability for deleting variables tends to zero
with $\pi \to 0$ in the add/delete Metropolis-Hastings
algorithm. This in turn means that the algorithm proposes to
add variables much more often than to delete variables, leading
to the sampler running off to include more and more variables.
In our simulation runs the convergence problem could only be avoided by choosing a
very small prior covariance parameter value of $c^2 = 0.05$,
which then resulted in very small posterior mean estimates
$\hat\beta_{\gamma i}$, but good posterior inference on the
probabilities $\hat p(\gamma_i = 1|\bs{x},\bs{y})
> 0.05$ for variable inclusion (see Table \ref{sensitive}).
Incidentally, $c^2 = 0.05$ does not fit within the range of
values suggested by \citet{sha04}, as the values in equation
(\ref{sha2004}) correspond to the range $3.33 < c^2 < 73.70$
for the data set used in this example. Contrary to the add/delete algorithm, the Gibbs sampler using $Pcor90$ neighbourhoods does not break down, if a large $c^2$ is chosen. However, the trace plots of the model size $p_\gamma$ (Figure
\ref{traceprobit}) illustrate that large models are often
visited with many more variables being included than in the
logistic variable selection models with the same value of
$c^2$. In addition, the model deviance trace plots shown in
Figure \ref{traceprobit} indicate that the sampler frequently
moves into regions of low posterior probability, and the posterior mean
estimates of $\bs{\beta}_\gamma$ run off to extreme values with magnitudes up to
$10^{12}$ for $c^2 = 50$ (Table \ref{sensitive}). However, the
posterior inference about variable inclusion probabilities
$p(\gamma_i = 1|\bs{x},\bs{y})$ is still quite robust, as reflected by
the number of false negatives and false positives (Table
\ref{sensitive}). Most samplers still find all five ``true''
predictor variables, but the number of false positives is
slightly larger than was observed for the corresponding
logistic models for some values of $c^2$, in particular for
$c^2 = 0.5$. One could circumvent this problem of having to fine-tune $c^2$
in a probit BVS model by introducing a hyper-prior distribution
for $c^2$. For the g-prior $c^2 (\bs{x}'\bs{x})^{-1}$ possible hyper-prior implementations $p(c^2)$  have for example been presented by
\citet{bottolo10}. 

Finally, in terms of the binomial prior distribution
$p(\gamma_i) = \pi_i$ ($i = 1,...,p$), it should be mentioned that our strategy to choose
the prior so that all $\pi_i$ correspond to the expected fraction of
true predictors among all variables, might not be
the best strategy, if the main interest lies in finding the
``true'' predictors rather than the overall ``true'' model. In
that situation, choosing a binomial prior probability $\pi_i$,
which is larger than the expected proportion $p^*/p$, would
mean that the models which are visited by the Markov chain will
tend to be larger than the expected size $p^*$, which will
increase the chance that all ``true'' variables of interest
will be included in that model. 

\begin{table}[!ht]
\caption[Results of sensitivity analysis regarding the choice
of $c^2$]{\small \label{sensitive}Results of sensitivity
analysis regarding the choice of $c^2$. MCMC samplers are
evaluated on data set 1 in simulation scenario 2.}
\begin{center} \small
\begin{tabular}{|r|r|r|r|r|r|r|r|}
\hline $c^2$ & abort due to&  $(\min \hat\beta_{\gamma i},\max \hat\beta_{\gamma i})$ & $(\min \hat\beta_{\gamma i},\max \hat\beta_{\gamma i})^\ddag$  &  \# FP$^\dag$ &\# FN$^\dag$ & $\mbox{ESS}^*(\bs{\gamma})$ & $\# \bs{I}^\sharp$\\
  & convergence & for $i = 1,...,5$ & for $i = 6,...,500$ & & & &\\
  & problems?& & & & & &\\
\hline\hline
\multicolumn{8}{|c|}{\textbf{Logistic BVS model}}\\
\multicolumn{8}{|l|}{Add/delete sampler}\\
\hline
0.5& no & (0.92, 1.52) & (-0.85, 0.87) & 13 & 0 & 177 & 437\\
5& no & (2.13, 3.03) & (-2.34, 1.88) & 15 & 0 & 84 & 330\\
50& no & (1.61, 3.51)$^\flat$ & (-1.42, 1.88) & 10 & 2 & 46 & 147\\
\hline
\multicolumn{8}{|l|}{Neighbourhood sampler ($Pcor90$, univariate Gibbs within neighbourhoods)}\\
\hline
0.5& no & (0.89, 1.49) & (-0.83, 0.92) & 14 & 0 & 8699 & 500\\
5& no & (2.22, 3.16) & (-1.82, 1.96) & 12 & 0 & 4195 & 500\\
50& no & (4.59, 8.95) & (-4.90, 5.65) & 18 & 0 & 1629 & 500\\
\hline\hline
\multicolumn{8}{|c|}{\textbf{Probit BVS model}}\\
\multicolumn{8}{|l|}{Add/delete sampler}\\
\hline
0.05& no & (0.36, 0.60) & (-0.39, 0.42) & 11 & 0 & 185 & 478\\
0.5& yes & N/A & N/A &N/A&N/A&N/A&N/A\\
5& yes & N/A & N/A &N/A&N/A&N/A&N/A\\
50&yes & N/A & N/A &N/A&N/A&N/A&N/A\\
\hline
\multicolumn{8}{|l|}{Neighbourhood sampler ($Pcor90$, univariate Gibbs within neighbourhoods)}\\
\hline
0.05& no & (0.35, 0.60) & (-0.38, 0.43) & 14 & 0 & 10020 & 500\\
0.5& no & (103.38, 6.00$e4$) & (-5.64$e5$, 1.09$e9$) & 49 & 0 & 9116 & 500\\
5& no & (2815.56, 1.59$e8$) & (-1.42$e9$, 2.02$e12$) & 25 & 0 & 8873 & 500\\
50& no & (236.82, 4.13$e5$) & (-6.60$e10$, 2.17$e12$) & 13 & 1 & 8246 & 500\\
\hline
\multicolumn{8}{l}{
\footnotesize{$^\sharp$ $\bs{I} =\{i: (\sum_{m=1}^M{\gamma_{i,m}}) > 0\}$, i.e. number of variables for which $\gamma_i
= 1$ in at least one MCMC iteration}} \\
\multicolumn{8}{l}{
\footnotesize{$^\dag$false positives and false negatives if cut-off at ratio of posterior to prior $> 5$ (i.e. $\hat p(\gamma_i = 1| \bs{x},\bs{y}) > 0.05$)}}\\
\multicolumn{8}{l}{
\footnotesize{$^\ddag$only for variables, which were visited at least once by the Markov chain}}\\
\multicolumn{8}{l}{
\footnotesize{$^\flat$does not include $\beta_{\gamma 3}$, which was never visited by the Markov chain}}\\
\multicolumn{8}{l}{
\footnotesize{$^\natural$N/A = not applicable}}
\end{tabular}
\end{center}

\end{table}

\clearpage
\section{Application to ovarian cancer gene expression data\label{realdata}}

We apply the Bayesian variable selection logistic model to an
ovarian cancer gene expression data set \citep{schwartz02} in order
to classify between intrinsically chemotherapy-resistant tumours and
more responsive histologies. The gene expression data were generated
by Affymetrix HuGeneFL gene chips which contain 7129 probe sets,
each corresponding to a specific gene. Here, $p=4000$ of these gene
variables are used after univariate unspecific filtering. Data are available
for $n=104$ ovarian cancer tissue samples, including 18 mucinous and
clear-cell samples, which are inherently resistant against the standard
platinum-based chemotherapeutic drug, while the 86 serous and
endometrioid tumours are usually more responsive to treatment. The
microarray data are background-correction by the RMA procedure \citep{irizarry03} and loess-normalised within each array\citep{cleveland79}. In addition, all gene variables are centred and
scaled to have zero mean and unit variance. In the Bayesian logistic
variable selection model the sparsity-inducing hyper-parameters are
set to the values $\pi_i = 5/p = 0.00125$ for all $i=1...,p$, so that $5$
variables are expected to be selected \emph{a priori}. The value $c^2 = 10$ is larger than in the simulation examples in Section \ref{sim} to account for the fact that now the true $\beta_i$ values are unknown and not set within the range $[-2,2]$, as it was the case with the simulation data.

We compare the performances of four MCMC algorithms for
sampling from the logistic BVS model: the vanilla add/delete
Metropolis-Hastings sampler, a neighbourhood MCMC sampler (\emph{Pcor}, C=99\%) and in addition both samplers in
combination with a parallel tempering algorithm. Parallel tempering is implemented such that only
neighbouring Markov chains in the temperature ladder are
proposed for state swaps in a Metropolis-Hastings algorithm.
We use five parallel Markov chains and a geometric temperature ladder $\{1, \tau, \tau^2, \tau^3,
\tau^4\}$ with $\tau = 1.2$. All parallel Markov chains are run un-coupled, i.e. without
state swaps, for $B^{PT} = 50,000$ iterations before starting
the parallel tempering algorithm proper to allow the Markov chains to move towards their target distribution before starting exchange moves between chains. An alternative approach could be the all-exchange parallel tempering
scheme by \citet{calvo05}, where all possible pairwise swap
acceptance probabilities are computed for all parallel chains
in each iteration and the pair of chains, that is to be
swapped, is sampled according to this probability distribution.
Both algorithms have been implemented in the \texttt{MATLAB} toolbox \texttt{BVS} available from \texttt{http://www.bgx.org.uk/software.html}. Results are compared with our previous analysis based on lasso logistic
regression \citep{tibshirani96}, where five genes were found to be
especially strongly linked to the response \citep{zucknick08}. Between one (untempered $AD$) and four (both MCMC runs with parallel tempering) of these genes are recovered here (see Table \ref{datamcmc}).

\begin{table}[!ht]
\caption[Diagnostic measures for Markov chain mixing with
respect to $\bs{\gamma}$]{\small\label{datamcmc}Diagnostic measures
for Markov chain mixing with respect to $\bs{\gamma}$; results for
$M = 1,000,000$ post burn-in MCMC iterations (CPU time for
total iteration number $N = 1,100,000$).}
\begin{center}
\begin{tabular}{|l|r|r|r|r|r|}
\hline MCMC&CPU time&$\mbox{ESS}^*(\bs{\gamma})$&$\# \bs{I}^\sharp$&\# genes &\# genes not \\
sampler &$t$ (min)&&&in lasso$^\dag$&in lasso$^\ddag$\\
\hline $AD$&315&8&{198}&1&23\\
\hline $Pcor99$&1356&3,793&{2856}&3&6\\
\hline Parallel tempering&&&&&\\
with $AD$&1726&19,900&{1091}&4&15\\
with $Pcor99$&6601&41,985&{3752}&4&5\\
\hline
\multicolumn{6}{l}{\footnotesize{$^\sharp \bs{I} = \{i: (\sum_{m=1}^M{\gamma_{i,m}}) > 0\}$}}\\
\multicolumn{6}{l}{\footnotesize{$^\dag$How many of the five genes consistently selected by
lasso in \citep{zucknick08} are}}\\
\multicolumn{6}{l}{\footnotesize{recovered by BVS, if cut-off at posterior to prior prob.
$>10$, i.e. $\hat p(\gamma_i = 1|\bs{x},\bs{y}) > 0.0125$?}}\\
\multicolumn{6}{l}{\footnotesize{$^\ddag$How many genes are consistently selected besides these
five genes \citep{zucknick08}?}}
\end{tabular}
\end{center}
\end{table}

\begin{figure}[!ht]
\begin{center}
\includegraphics[width=0.47\textwidth,height=0.33\textwidth]{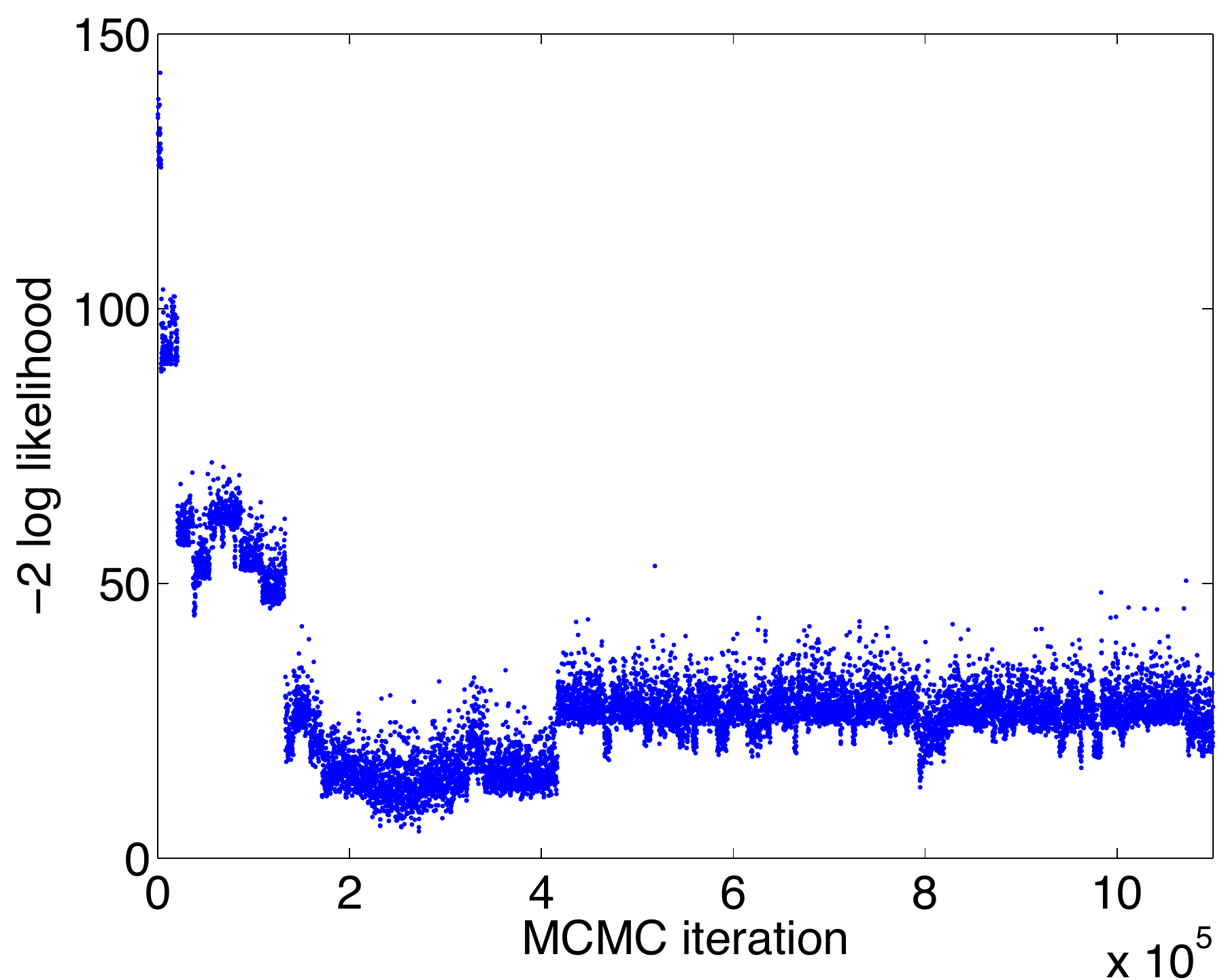}
\includegraphics[width=0.47\textwidth,height=0.33\textwidth]{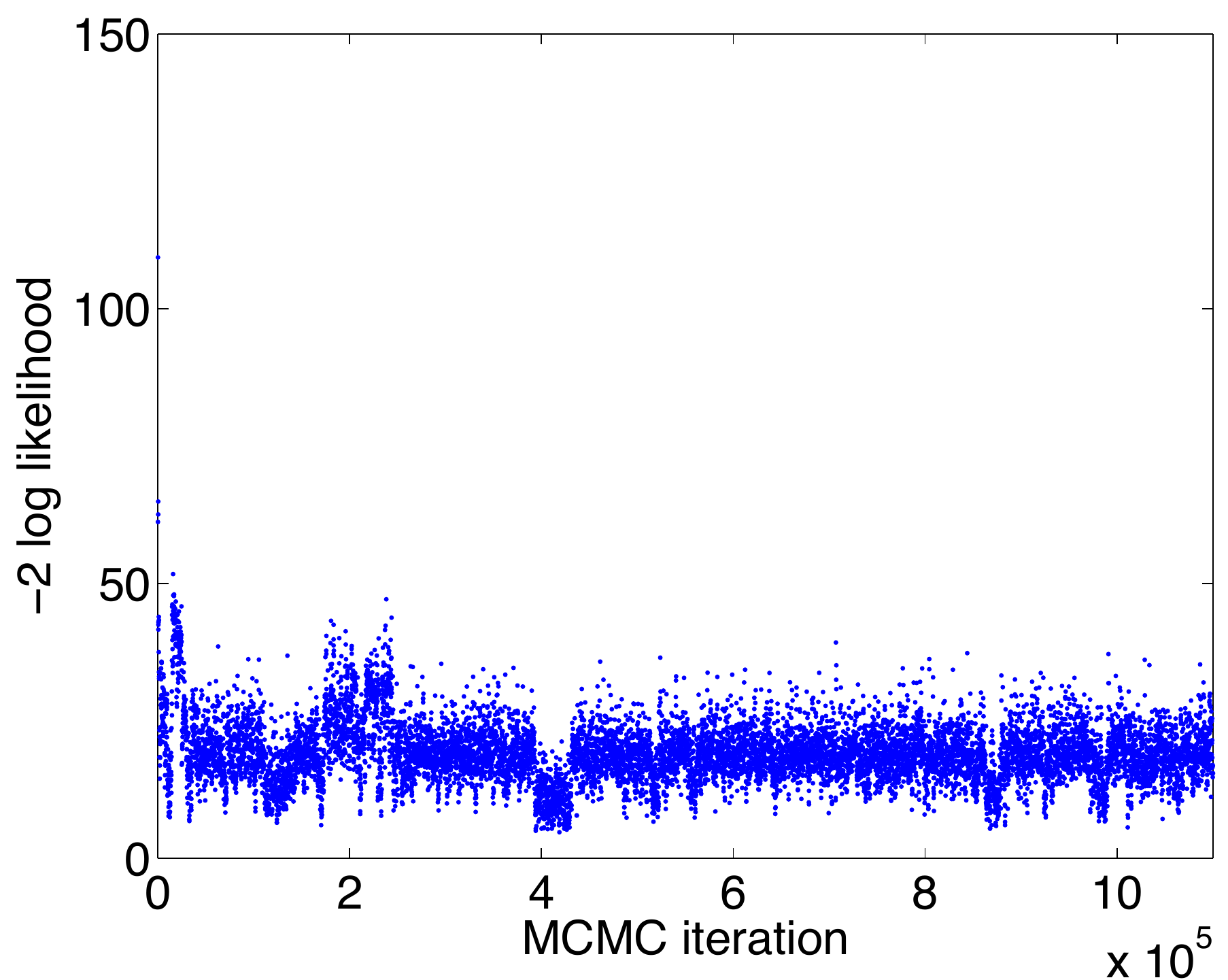}\\
\includegraphics[width=0.47\textwidth,height=0.33\textwidth]{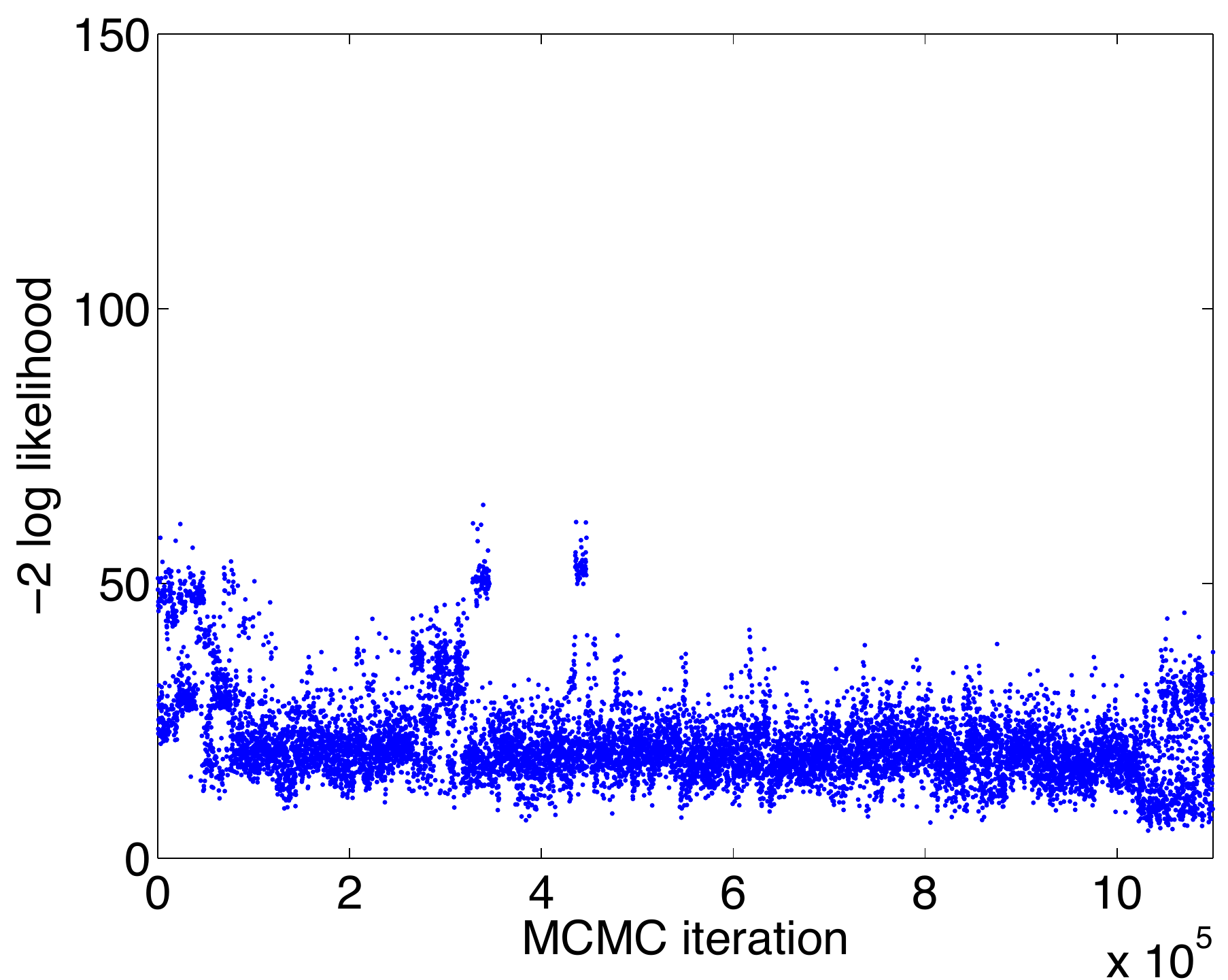}
\includegraphics[width=0.47\textwidth,height=0.33\textwidth]{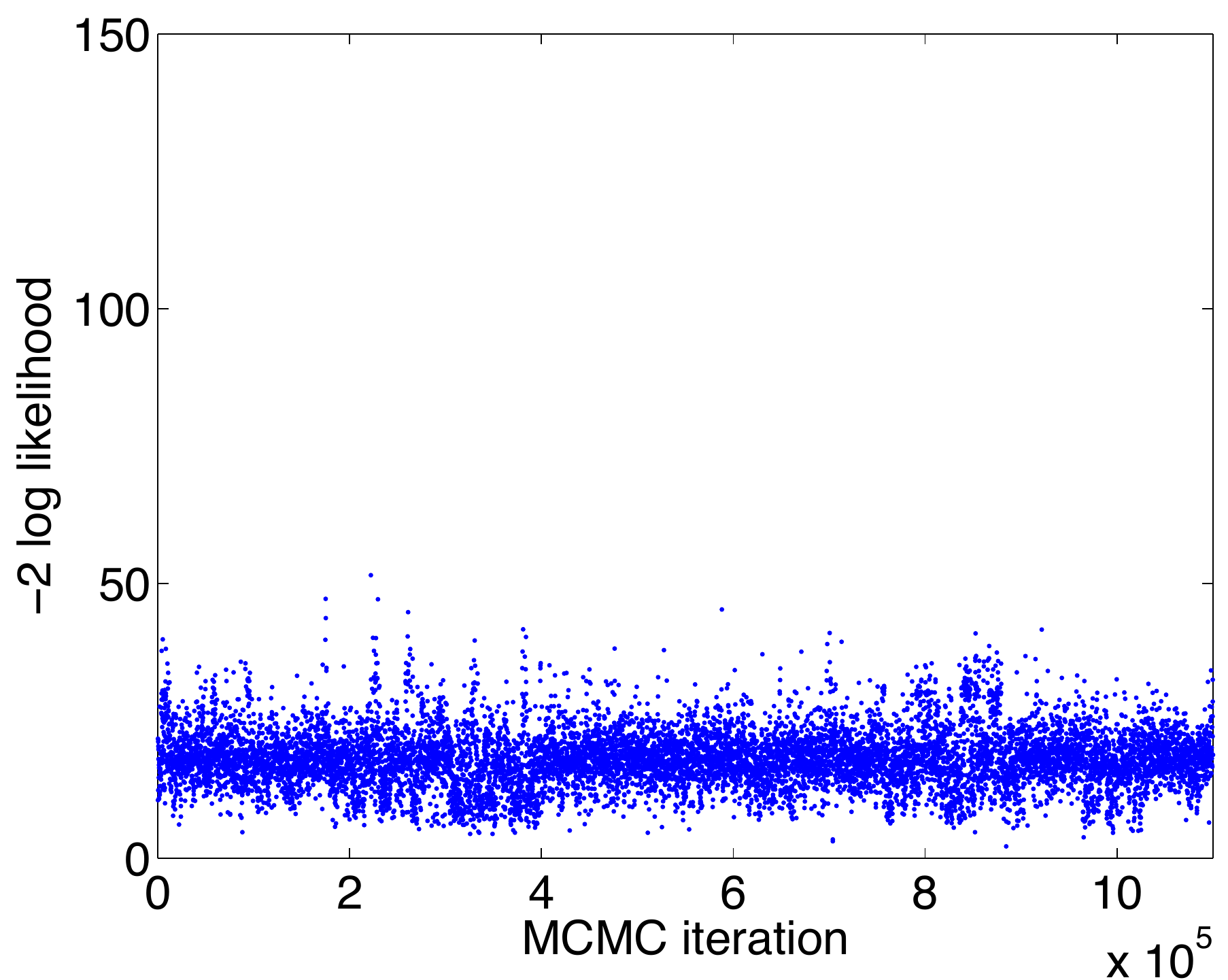}
\end{center}
\caption[Trace plots of model deviances for add/delete sampler and neighbourhood
sampler, with and without parallel tempering applied to \citet{schwartz02} data]
{\small\label{datadev}Trace plots of model fit in terms of
model deviances for add/delete sampler (left) and neighbourhood
sampler (\emph{Pcor99}) (right), with (bottom) and without (top) parallel tempering
in application to gene expression data \citep{schwartz02}.}
\end{figure}

\begin{figure}[!ht]
\begin{center}
\includegraphics[height=0.34\textwidth]{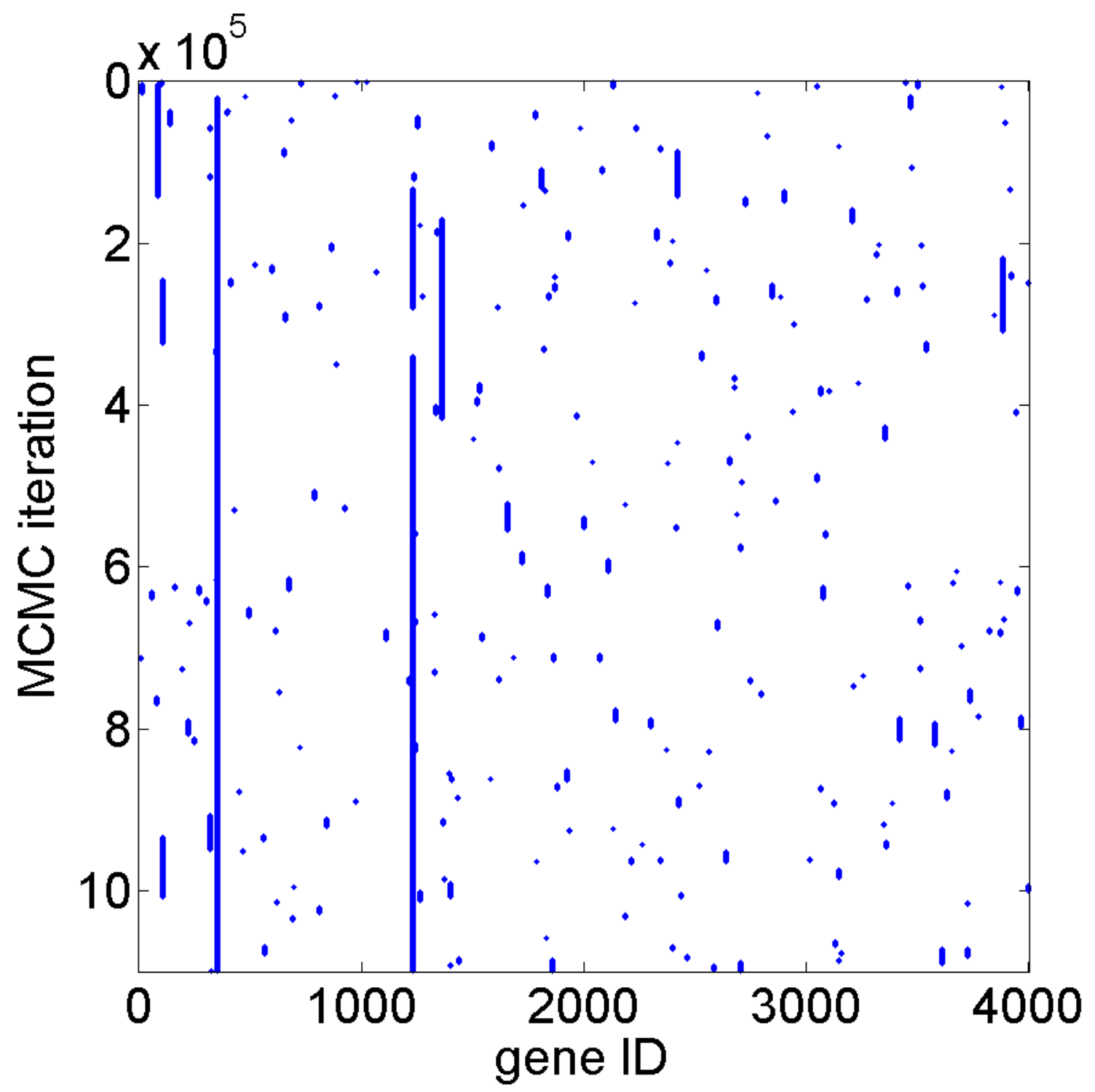}
\includegraphics[height=0.34\textwidth]{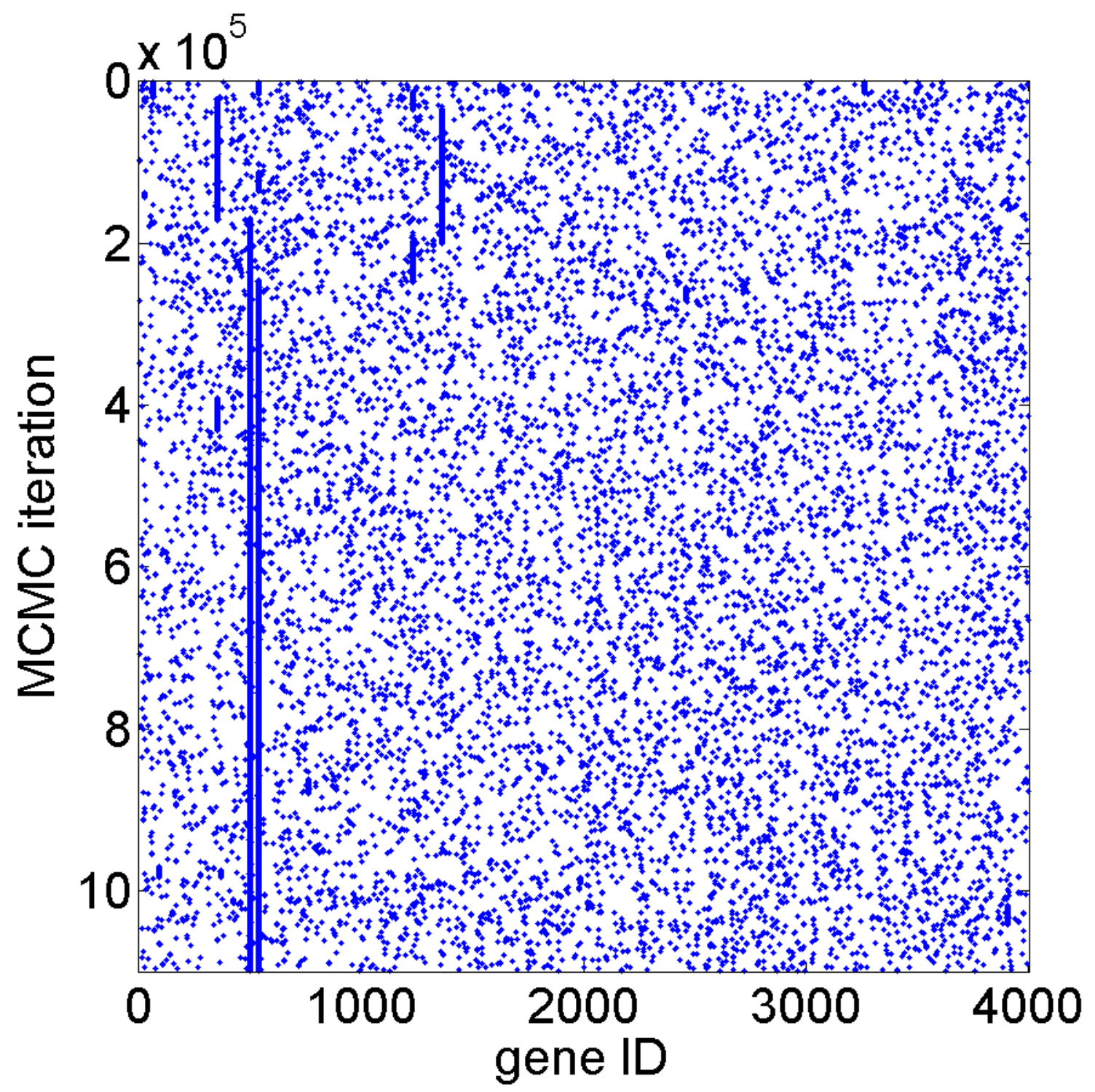}\\
\includegraphics[height=0.34\textwidth]{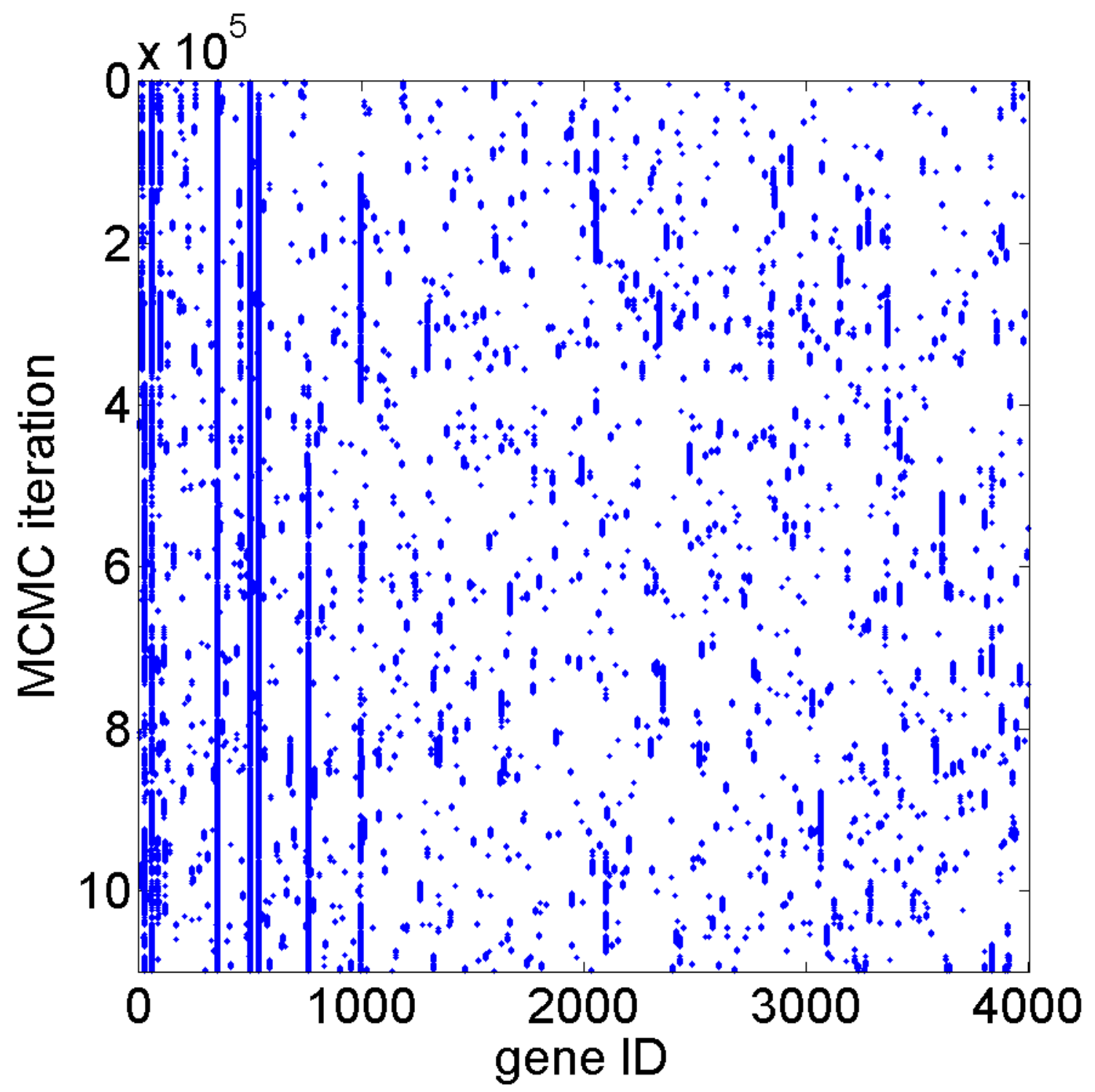}
\includegraphics[height=0.34\textwidth]{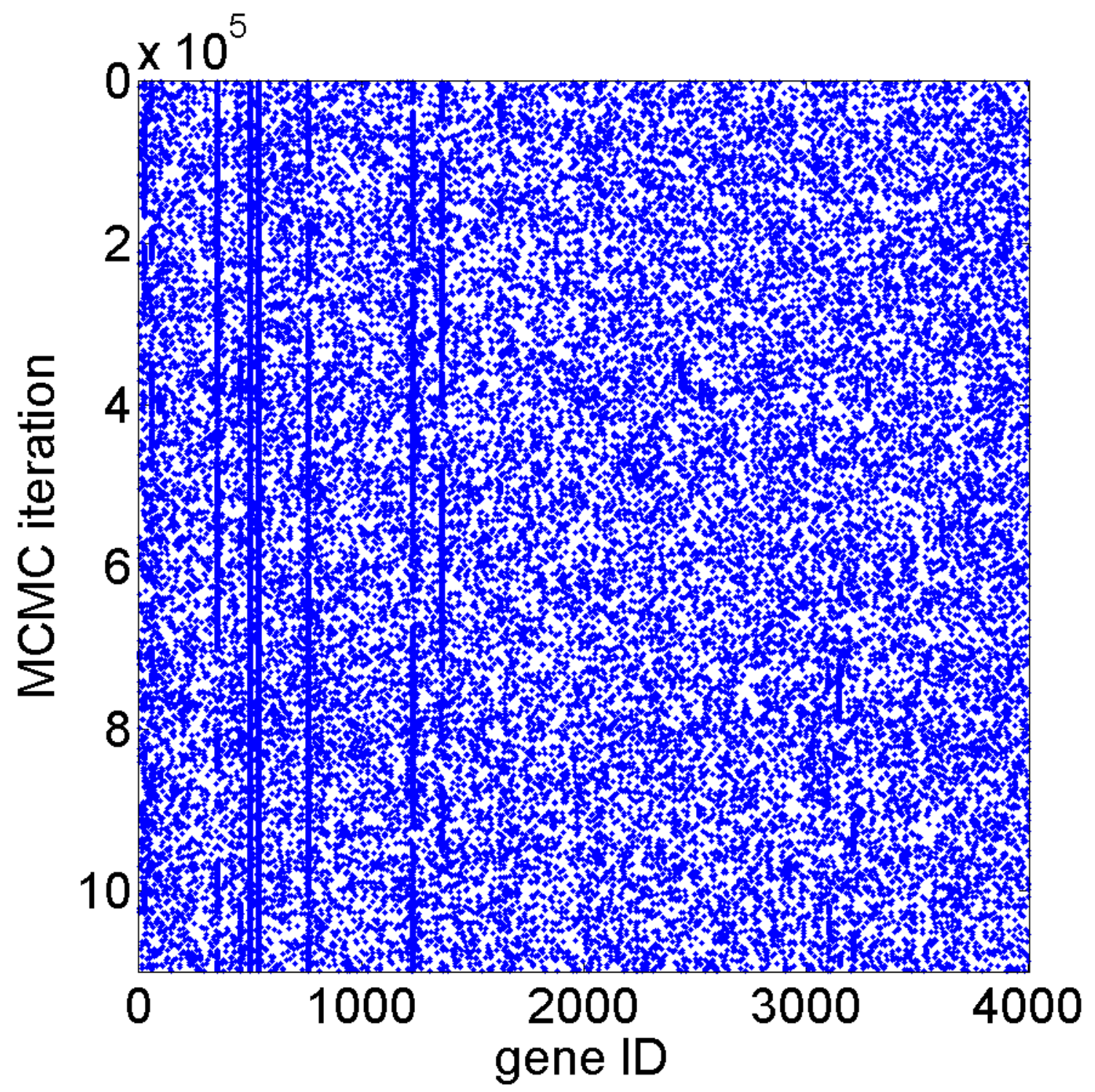}
\end{center}
\caption[Trace plots of $\bs{\gamma}$ vector for
add/delete sampler and neighbourhood sampler, with and without parallel tempering applied to \citet{schwartz02} data]
{\small\label{datatrace}Trace plots of $\bs{\gamma}$ vector for
add/delete sampler (left) and neighbourhood sampler (\emph{Pcor99})
(right), with (bottom) and without (top) parallel tempering
in application to gene expression data \citep{schwartz02}.}
\end{figure}

The add/delete sampler gets stuck with a model that has a worse fit in terms of model deviance than many other models (Figure
\ref{datadev}). This model consistently contains two variables with
IDs 354 (gene symbol \emph{ANX4}) and 1232 (\emph{TFF1}), so
that overall these are the only two variables with marginal posterior
probability estimates larger than $\hat{p}(\gamma_i = 1| \bs{x},\bs{y}) = 0.5$.
The other three MCMC algorithms all find that two other
variables are the only ones with marginal posterior probability
estimates larger than $0.5$, namely the genes with ID
501 (\emph{CYP2C18}) and 540 (\emph{SPINK1}). \emph{ANX4} is also selected by the other three MCMC samplers, while \emph{TFF1} gets quickly replaced by \emph{CYP2C18} and \emph{SPINK1}. \emph{ANX4}, \emph{CYP2C18} and
\emph{SPINK1} are all in the set of five genes found in our previous analysis of this data set \citep{zucknick08} by lasso logistic regression combined with a heuristic version of stability selection \citep{meinshausen10}. In addition, the parallel tempering algorithms also identify a fourth member of this set, namely gene \emph{ABP1} with ID 60.

The traces of the individual covariate indicator variables
$\gamma_i$ for all variables $i=1,...,p=4000$ are shown in
Figure \ref{datatrace}. The trace plots illustrate the
extremely slow mixing of the add/delete sampler at the level of
individual $\gamma_i$ variables. Mixing improves when adding
the parallel tempering algorithm, and also when replacing the
add/delete sampling algorithm by the neighbourhood sampler. Based on
the trace plots, mixing performance is best for the MCMC neighbourhood
sampler combined with parallel tempering. Diagnostic measures for Markov chain mixing listed in Table
\ref{datamcmc} confirm this impression. The effective sample size is largest for the parallel
tempering algorithm when combined with the neighbourhood sampler
($\mbox{ESS}^*(\gamma) = 41,985$), and about half that with when combined with the
add/delete sampler ($\mbox{ESS}^*(\gamma) = 19,900$). Compared to
this acceptable result, the effective sample size is only
$\mbox{ESS}^*(\gamma) = 8$ for the add/delete sampler
without parallel tempering, which is clearly not sufficient
for valid posterior inference about the $\gamma$ vector. Thus,
the improvement in effective sample size from the introduction
of parallel tempering is huge for the add/delete
Metropolis-Hastings algorithm. It is not as large
for the neighbourhood sampler, but the effective sample size still
increases about eleven-fold from $\mbox{ESS}^*(\gamma) = 3,793$
for neighbourhood sampling without parallel tempering, which means that
it is still advantageous to perform the parallel tempering
algorithm, since the computation time only increases about
five-fold due to having to run five Markov chains rather than
just one. Note that our parallel tempering implementation is
serial, but could of course be done in parallel. In a parallel implementation, the
computation time would only increase slightly, so that adding
parallel tempering to the MCMC algorithm can increase chain
mixing dramatically with near to no cost in terms of increased
computation time.

The improvement in mixing by introducing neighbourhood sampling and
parallel tempering is also seen in the number of $\gamma_i$
variables, which are visited at least once by the MCMC
samplers. The parallel tempering with neighbourhood sampling approach
visits $\# I = 3752$ variables out of all $4000$. The neighbourhood sampler without the added
parallel tempering scheme already results in good mixing and
visits $\# I = 2856$ variables, whereas the mixing of
the add/delete sampler is very poor and visits only
$\# I = 198$ variables. 

\section{Discussion}

Variable selection is a common task for large-scale genomic
applications where many thousands of biological entities such as
gene expression values or genetic markers are screened in order to
identify a very small number of variables which might be linked to the
disease or phenotype of interest. In this context Bayesian variable
selection methods have the advantage that sparsity can be enforced by the
choice of hyper-priors. Also, it has the advantage over non-Bayesian
methods that posterior distributions are estimated for all variables. In addition to marginal inference to identify individual variables with large posterior inclusion probabilities, inference based on the joint posterior probabilities of these models allows us to identify combinations of variables that appear frequently together, providing a start for more detailed exploration of the model space.

However, MCMC sampling from the posterior distribution of a Bayesian
variable selection model is computationally very demanding for
large-scale $p>>n$ applications. In previous publications
\citep[e.g.][]{brown98,brown98_2,lee03,sha04} the Gibbs sampler and
the add/delete(/swap) Metropolis-Hastings sampler have been used for
sampling the indicator variable $\bs{\gamma}$ that determines the model
space. However as we have seen, full Gibbs sampling is
computationally very demanding, 
and while the add/delete sampler is much faster, 
very slow mixing is a problem, not just in terms of how many iterations it takes to
convergence, but also because the sampler can get seriously stuck,
as seen in the data application in Section \ref{realdata}. 


We proposed and explored a simple way to account for most of
the dependence structure among covariates to create a neighbourhood
sampler which improves mixing and reduces the probability of
the sampler getting stuck in a local optimum, but which is not
as computationally demanding as a full Gibbs sampler. In two
simulation studies we compared the neighbourhood samplers derived from correlation or partial correlation matrices. We compared the mixing performances as assessed by the
effective sample size measure $\mbox{ESS}^*(\bs{\gamma})$ and its
relation to the required computation time per 10,000 iterations. In our simulation studies the add/delete
sampler always performed worst. The performance of our neighbourhood
samplers improved with increased threshold $C$ until it levelled off at a point when the neighbourhood size became so large that the
additional gain in mixing was not big enough anymore to offset the
increased computation time per iteration. In simulation scenario 1, both correlation- and
partial-correlation-based neighbourhood samplers outperformed full
Gibbs sampling when the average neighbourhood size was large enough,
while in scenario 2 only the samplers with neighbourhood construction
based on partial correlations outperformed the full Gibbs
sampler. Note that none of the MCMC algorithms have been optimised with respect to
computation time and that results might change with optimised
samplers.

In summary, our neighbourhood sampling method is easy to implement, and it
is successful in speeding up mixing relative to computation time per
iteration compared to standard full Gibbs sampling and the add/delete
Metropolis-Hastings sampler. A further advantage is that is does not
impose any structure on the data because the neighbourhoods are only used as
a guide for the MCMC sampler and are not part of the model. A
potential disadvantage is that it is quite heuristic, meaning that
it is not known in advance how big the threshold for correlation or
partial correlation values should be should be to achieve optimal
mixing improvements. However, some improvement is easily achieved if
the threshold $C$ is not too small. If available, prior knowledge
about the average expected number of neighbours for the variables
can be used. Such knowledge can be available for example from known
biological networks for gene expression data, or from knowledge
about the average extent of linkage disequilibrium for genomic
markers such as SNP data. The threshold $C$ can
be set to a value so that the average neighbourhood size equals that
expected neighbourhood size to ensure that enough of the dependence
structure is captured in order to improve mixing sufficiently.

Note that the performance of Bayesian variable selection methods is
not just influenced by the choice of the MCMC algorithm, which has
been the focus of this paper. Other factors are the choice of prior
for the regression coefficients $\bs{\beta}$ and also the prior for the
indicator variable $\bs{\gamma}$. 
For example, instead of using the independence prior $p(\bs{\beta}) = N(\bs{b} = \bs{0}_p, v = c^2\bs{I}_p)$, we could have used the g-prior $c^2(\bs{x}'\bs{x})^{-1}$ \citep[e.g.][]{lee03,bottolo10}. Arguments for both priors can
be found in \citet{brown02} and \citet{bottolo10}. 
And instead of fixing the prior probabilities $\pi_i$ ($i=1,...,p$) in the binomial prior $p(\bs{\gamma}) = \prod_{i=1}^p \pi_i^{\gamma_i} (1 - \pi_i)^{1-\gamma_i}$, we could have assigned a hyper-prior distribution to $\bs{\pi}$, which would result in a 
beta-binomial prior for $\bs{\gamma}$. \Citet{bottolo10} have performed extensive
experiments in the linear regression context to investigate and
compare the performances of different choices of priors for $\bs{\beta}$.

A sensitivity analysis for the choice of $c^2$ has shown that
while the estimates of $\bs{\beta}_\gamma$ are influenced by the
choice of $c^2$, the estimates of $\bs{\gamma}$ - which is what we
are most interested in here - are not influenced much in the
logistic BVS model. So, with respect to the main interest of
finding variables and models with high posterior probability
for being linked to the response, the logistic BVS model can be
applied without the need for extensive fine-tuning of the prior
covariance parameter $c^2$. Contrary to that, in the probit
regression model posterior inference of $\bs{\gamma}$ was highly
sensitive to the choice of $c^2$, and if $c^2$ was large then
the add/delete MCMC sampler broke down completely due to the previously reported problem \citep[][and others]{hans07}, that the acceptance probability for deleting variables will tend to zero for small prior variable inclusion probabilities.

Finally, the neighbourhood updates proposed in this manuscript
can be readily combined with other methods for improving Markov
chain mixing, e.g. parallel tempering or evolutionary Monte
Carlo, as we have demonstrated in the application to a gene expression data set where we combined the samplers with a
parallel tempering algorithm involving five Markov chains tempered with a geometric temperature ladder. Further improvements might for example be possible by using the evolutionary stochastic search algorithm (ESS) by \citet{bottolo10}. ESS improves ordinary parallel tempering by combining the running of multiple tempered Markov chains in parallel with sophisticated local and global exchange moves (ideas adopted from genetic algorithms) and an automatic adaptation of the temperature ladder during the burn-in phase.

\bibliographystyle{plainnat}
\bibliography{literature}

\clearpage
\renewcommand\thesection{\Alph{section}}
\section{Appendix: Sampling from the Bayesian logistic variable selection model}
\subsection{\label{appGibbs}Gibbs sampling algorithm}
In this section the Gibbs algorithm to sample from the logistic
BVS model as proposed by \citet{holmes06} is presented in
detail. The joint
distribution is given as (see equation \ref{jointpost})
\begin{equation}
p(\bs{\beta}_\gamma,\bs{\gamma},\bs{z},\bs{\lambda}|\bs{x},\bs{y}) \propto
p(\bs{y}|\bs{z})p(\bs{z}|\bs{\lambda},\bs{\beta},\bs{\gamma},\bs{x})p(\bs{\beta}_\gamma|\bs{\gamma})p(\bs{\gamma})p(\bs{\lambda}),
\end{equation}
where
\begin{eqnarray*}
p(\lambda_{jj}) &\sim& \frac{1}{4\sqrt{\lambda_{jj}}}\mbox{KS}(0.5\sqrt{\lambda_{jj}})\quad\mbox{and}\\
p(\bs{z}|\bs{\lambda},\bs{\beta},\bs{\gamma},\bs{x}) &=& N(\bs{x}_\gamma\bs{\beta}_\gamma,\bs{\lambda}).
\end{eqnarray*}
$KS()$ denotes the Kolmogorov-Smirnov distribution. It is
proposed to sample from this distribution (equation \ref{jointpost}) via the full
conditionals $p(\bs{z},\bs{\lambda}|\bs{\beta},\bs{\gamma},\bs{x},\bs{y})$ and
$p(\bs{\beta}_\gamma,\bs{\gamma}|\bs{z},\bs{\lambda},\bs{x})$. These distributions are
as follows
\begin{eqnarray}
p(\bs{z},\bs{\lambda}|\bs{\beta},\bs{\gamma},\bs{x},\bs{y}) &=&
p(\bs{z}|\bs{\beta},\bs{\gamma},\bs{x},\bs{y})p(\bs{\lambda}|\bs{z},\bs{\beta},\bs{\gamma},\bs{x})\nonumber\\
p(z_j|\bs{\beta},\bs{\gamma},\bs{x},\bs{y}) &=&
\left\{\begin{array}{ll}\mbox{Logistic}(\bs{x}_{\gamma j}\bs{\beta}_\gamma, 1) I(z_j > 0),    &y_j=1\\
                        \mbox{Logistic}(\bs{x}_{\gamma j}\bs{\beta}_\gamma, 1) I(z_j \leq 0), &y_j=0
\end{array}\right.\label{pz}\\
p(\lambda_{jj}|\bs{z},\bs{\beta},\bs{\gamma},\bs{x}) &\propto& p(z_j|\bs{\lambda}, \bs{\beta},\bs{\gamma},\bs{x})
p(\lambda_{jj}) = N(\bs{x}_{\gamma j} \bs{\beta}_\gamma, \lambda_{jj})
\frac{1}{4\sqrt{\lambda_{jj}}}KS(0.5 \sqrt{\lambda_{jj}})\label{plam}\\
\nonumber
\end{eqnarray}
and
\begin{eqnarray}
p(\bs{\beta}_\gamma,\bs{\gamma}|\bs{z},\bs{\lambda},\bs{x}) &=&
p(\bs{\gamma}|\bs{z},\bs{\lambda},\bs{x})p(\bs{\beta}_\gamma|\bs{\gamma},\bs{z},\bs{\lambda},\bs{x})\nonumber\\
p(\bs{\gamma}|\bs{z},\bs{\lambda},\bs{x}) &\propto& p(\bs{z}|\bs{\lambda},\bs{\gamma},\bs{x})p(\bs{\gamma}) =
N(\bs{0}_{n}, \bs{\lambda} + \bs{x}_\gamma \bs{v}_\gamma \bs{x}_\gamma')
\prod_{i=1}^p{\pi_i^{\gamma_i}(1-\pi_i)^{1-\gamma_i}}\label{pgam}\\
p(\bs{\beta}_\gamma|\bs{\gamma},\bs{z},\bs{\lambda},\bs{x}) &=& N(\bs{B}_\gamma,\bs{V}_\gamma)\label{pbet}\\
\bs{B}_\gamma &=& \bs{V}_\gamma \bs{x}_\gamma'\bs{\lambda}^{-1}\bs{z}\nonumber\\
\bs{V}_\gamma &=& (\bs{x}_\gamma'\bs{\lambda}^{-1}\bs{x}_\gamma +
\bs{v}_\gamma^{-1})^{-1}.\nonumber\\
\nonumber
\end{eqnarray}
From the conditional distributions (\ref{pz}), (\ref{pgam}) and
(\ref{pbet}) we can sample directly, with various algorithms
available for updating $\bs{\gamma}$, described in Sections
\ref{mcmcalgo} and \ref{blocks}. Distribution (\ref{plam}) can be sampled
from efficiently in the following way, using a rejection
algorithm introduced by \citet{holmes06}.

The acceptance probability is given as $\alpha(\lambda_{jj}) =
\frac{\ell(r_j^2,\lambda_{jj})p(\lambda_{jj})}{M g(\lambda_{jj})}$ with
$r_j^2 = (z_j - \bs{x}_{\gamma j}\bs{\beta}_\gamma)^2$ and
$\ell(r_j^2,\lambda_{jj}) = p(z_j|\bs{x}_{\gamma
j},\bs{\beta}_{\gamma},\lambda_{jj}) = N_{z_j}(\bs{x}_{\gamma
j}\bs{\beta}_{\gamma},\lambda_{jj})$. Here, $g(\lambda_{jj})$ is the
rejection sampling density $$g(\lambda_{jj}) =
c(|r_j|)\lambda_{jj}^{-1/2}\exp(-0.5(\frac{r_j^2}{\lambda_{jj}} +
\lambda_{jj}))$$ with $c(|r_j|)$ being a normalising constant not
dependent on $\lambda_{jj}$. This corresponds to a Generalised
Inverse Gaussian distribution $GIG(0.5,1,r_j^2) =
|r_j|/IG(1,|r_j|)$, where $IG$ denotes an Inverse Gaussian
distribution with probability density function
\citep[][p.148]{devroye86}
$$
p(X) = \sqrt{\frac{|r_j|}{2\pi X^3}}\exp{-\frac{|r_j|(X-1)^2}{2X}}\quad (X\geq 0).
$$
This choice of rejection distribution leads to
\begin{equation} \alpha(\lambda_{jj}) = \exp(0.5
\lambda_{jj})p(\lambda_{jj}) = \exp(0.5 \lambda_{jj}) c
h(0.5\sqrt{\lambda_{jj}})(1 - a_1(0.5\sqrt{\lambda_{jj}}) +
a_2(0.5\sqrt{\lambda_{jj}}) - ...),
\end{equation}
which is an alternate series expansion representation of
$KS(0.5\sqrt{\lambda_{jj}})$, i.e. the Kolmogorov-Smirnov density,
with terms $ch(.)$ and $a_1(.),a_2(.),...$ as in
\citet{devroye86} (pp.161-167) and \citet{holmes06}, which
allows for efficient sampling.

\subsection{\label{appalpha}Metropolis-Hastings acceptance probability for sampling from $p(\bs{\beta}_\gamma,\bs{\gamma}|\bs{z},\bs{\lambda},\bs{x})$}
In the following, we derive the acceptance probability
$\alpha(\bs{\gamma},\bs{\beta}_\gamma)$ for sampling from the conditional
distribution $p(\bs{\beta}_\gamma,\bs{\gamma}|\bs{z},\bs{\lambda},\bs{x}) =
p(\bs{\gamma}|\bs{z},\bs{\lambda},\bs{x})p(\bs{\beta}_\gamma|\bs{\gamma},\bs{z},\bs{\lambda},\bs{x})$ in the
logistic BVS model by \citet{holmes06}, using the
\emph{add/delete Metropolis-Hastings sampler}. In the
add/delete sampler, one indicator variable $k$ is selected at
random from $\{1,...,p\}$ and it is proposed to change its state: the proposal
distribution $q(\bs{\gamma}^*)$ is given by
\begin{equation}
q(\gamma^*_i) = \left\{\begin{array}{ll}
\gamma_i & \mbox{if }i\ne k\\
1 & \mbox{if }i=k \mbox{ and } \gamma_k = 0\\
0 & \mbox{if }i=k \mbox{ and } \gamma_k = 1\end{array}\right. \mbox{ for }i=1,...,p.
\end{equation}
Note that this implies
\begin{equation}
\frac{p(\bs{\gamma}^*)q(\bs{\gamma})}{p(\bs{\gamma})q(\bs{\gamma}^*)} =
\left\{\begin{array}{ll} \frac{1 - \pi_k}{\pi_k} & \mbox{if }\gamma_k=1\\
                         \frac{\pi_k}{1 - \pi_k} & \mbox{if }\gamma_k=0\end{array}\right.,
\end{equation}
if $p(\bs{\gamma}) =
\prod_{i=1}^p{\pi_i^{\gamma_i}(1-\pi_i)^{1-\gamma_i}}$ is the
prior distribution for $\bs{\gamma}$. This results in the following
acceptance probability for updating $(\bs{\gamma},\bs{\beta}_\gamma)$:
\begin{eqnarray}\label{alphaderive}
\alpha(\bs{\gamma}, \bs{\beta}_\gamma) &=& \min\left\{1,
\frac{p(\bs{\beta}^*,\bs{\gamma}^*|\bs{z},\bs{\lambda},\bs{x})}{p(\bs{\beta},\bs{\gamma}|\bs{z},\bs{\lambda},\bs{x})}\frac{q(\bs{\gamma},
\bs{\beta})}{q(\bs{\gamma}^*, \bs{\beta}^*)}\right\}\\\nonumber &=& \min\left\{1,
\frac{p(\bs{\gamma}^*|\bs{z},\bs{\lambda},\bs{x})p(\bs{\beta}^*|\bs{\gamma}^*,\bs{z},\bs{\lambda})}{p(\bs{\gamma}|\bs{z},\bs{\lambda},\bs{x})p(\bs{\beta}|\bs{\gamma},\bs{z},\bs{\lambda})}\frac{p(\bs{\beta}|\bs{\gamma},\bs{z},\bs{\lambda})
q(\bs{\gamma})}{p(\bs{\beta}^*|\bs{\gamma}^*,\bs{z},\bs{\lambda}) q(\bs{\gamma}^*)}\right\}\\\nonumber &=&
\min\left\{1,
\frac{p(\bs{z}|\bs{\lambda},\bs{\gamma}^*,\bs{x})}{p(\bs{z}|\bs{\lambda},\bs{\gamma},\bs{x})}\frac{p(\bs{\gamma}^*)q(\bs{\gamma})}{p(\bs{\gamma})q(\bs{\gamma}^*)}\right\}\\\nonumber
&=& \min\left\{1, \begin{array}{ll} \frac{\displaystyle
C(\bs{\gamma}^*)\exp(-0.5\bs{z}'(\bs{\lambda}+\bs{x}_{\gamma^*} \bs{v}_{\gamma^*}
\bs{x}'_{\bs{\gamma}^*})^{-1}\bs{z})}{\displaystyle C(\bs{\gamma})\exp(-0.5\bs{z}'(\bs{\lambda}+\bs{x}_\gamma
\bs{v}_\gamma \bs{x}'_\gamma)^{-1}\bs{z})}
\frac{\displaystyle 1 - \pi_k}{\displaystyle \pi_k} & \mbox{if }\gamma_k=1\\
\frac{\displaystyle C(\bs{\gamma}^*)\exp(-0.5\bs{z}'(\bs{\lambda}+\bs{x}_{\gamma^*} \bs{v}_{\gamma^*} \bs{x}'_{\gamma^*})^{-1}\bs{z})}{\displaystyle C(\bs{\gamma})\exp(-0.5\bs{z}'(\bs{\lambda}+\bs{x}_\gamma \bs{v}_\gamma \bs{x}'_\gamma)^{-1}\bs{z})}
\frac{\displaystyle \pi_k}{\displaystyle1 - \pi_k} & \mbox{if }\gamma_k=0\end{array}\right\},\nonumber
\end{eqnarray}
where $C(.)$ is a normalising constant, which will be defined
later. When we apply the Sherman-Morrison-Woodbury matrix
inversion formula \citep[e.g.][]{schott97} to the density
$p(\bs{z}|\bs{\lambda},\bs{x},\bs{\gamma})$ we get
\begin{eqnarray}
p(\bs{z}|\bs{\lambda},\bs{\gamma},\bs{x}) &=& C(\bs{\gamma})\exp(-0.5\bs{z}'(\bs{\lambda}+\bs{x}_\gamma \bs{v}_\gamma \bs{x}'_\gamma)^{-1}\bs{z})\\\nonumber
&=& C(\bs{\gamma})\exp(-0.5\bs{z}'(\bs{\lambda}^{-1} - \bs{\lambda}^{-1}\bs{x}_\gamma (\bs{v}_\gamma^{-1} + \bs{x}'_\gamma \bs{\lambda}^{-1}\bs{x}_\gamma)^{-1} \bs{x}'_\gamma\bs{\lambda}^{-1})\bs{z}).\nonumber
\end{eqnarray}
From here it follows, using the relation between prior covariance $\bs{v}_\gamma$
and posterior covariance $\bs{V}_\gamma$ of $\bs{\beta}_\gamma$ $\bs{V}_\gamma =
(\bs{v}^{-1}_\gamma + \bs{x}'_\gamma\bs{\lambda}^{-1}\bs{x}_\gamma)^{-1}$ (see (\ref{betapost})):
\begin{eqnarray}\label{distZ}
p(\bs{z}|\bs{\lambda},\bs{\gamma},\bs{x})
&=& C(\bs{\gamma})\exp(-0.5\bs{z}'(\bs{\lambda}^{-1} - \bs{\lambda}^{-1}\bs{x}_\gamma \bs{V}_\gamma \bs{x}'_\gamma\bs{\lambda}^{-1})\bs{z})\\\nonumber
&=& C(\bs{\gamma})\exp(-0.5\bs{z}'\bs{\lambda}^{-1}\bs{z})\exp(0.5\bs{z} '\bs{\lambda}^{-1}\bs{x}_\gamma \bs{V}_\gamma \bs{x}'_\gamma\bs{\lambda}^{-1} \bs{z})\\\nonumber
&=& C(\bs{\gamma})\exp(-0.5\bs{z}'\bs{\lambda}^{-1}\bs{z})\exp(0.5\bs{B}'_\gamma \bs{V}^{-1}_\gamma \bs{B}_\gamma),\nonumber
\end{eqnarray}
because $\bs{B}_\gamma = \bs{V}_\gamma \bs{x}'_\gamma \bs{\lambda}^{-1} \bs{z}$. The normalising constant $C(\bs{\gamma})$ in the
normal distribution $p(\bs{z}|\bs{\lambda},\bs{\gamma},\bs{x})$ is
\begin{eqnarray}\label{constZ}
C(\bs{\gamma}) &=& (2\pi)^{-n/2}|\bs{\lambda} + \bs{x}_\gamma \bs{v}_\gamma \bs{x}'_\gamma|^{-1/2}\\\nonumber
&=& (2\pi)^{-n/2}|\bs{\lambda}|^{1/2}|\bs{I} + \bs{\lambda}^{-1}\bs{x}_\gamma \bs{v}_\gamma \bs{x}'_\gamma|^{-1/2}\\\nonumber
&=& (2\pi)^{-n/2}|\bs{\lambda}|^{1/2}|\bs{I} + \bs{x}'_\gamma\bs{\lambda}^{-1}\bs{x}_\gamma \bs{v}_\gamma|^{-1/2}\\\nonumber
&=& (2\pi)^{-n/2}|\bs{\lambda}|^{1/2}|(\bs{v}^{-1}_\gamma + \bs{x}'_\gamma\bs{\lambda}^{-1}\bs{x}_\gamma) \bs{v}_\gamma|^{-1/2}\\\nonumber
&=& (2\pi)^{-n/2}|\bs{\lambda}|^{1/2}|\bs{V}^{-1}_\gamma \bs{v}_\gamma|^{-1/2} = (2\pi)^{-n/2}|\bs{\lambda}|^{1/2} \frac{|\bs{V}_\gamma|^{1/2}}{|\bs{v}_\gamma|^{1/2}}.\nonumber
\end{eqnarray}
Hence, following from (\ref{alphaderive}) by plugging in
(\ref{distZ}) and (\ref{constZ}), the acceptance probability is
given as
\begin{eqnarray}\label{alphafinal}
\alpha(\bs{\gamma},\bs{\beta}_\gamma) &=& \min\left\{1,
\begin{array}{ll}\frac{\displaystyle C(\bs{\gamma}^*)\exp(0.5\bs{B}'_{\gamma^*}\bs{V}^{-1}_{\gamma^*}\bs{B}_{\gamma^*})}
{\displaystyle C(\bs{\gamma})\exp(0.5\bs{B}'_{\gamma}\bs{V}^{-1}_{\gamma}\bs{B}_{\gamma})}
\frac{\displaystyle 1 - \pi_k}{\displaystyle \pi_k} & \mbox{if }\gamma_k = 1\\
\frac{\displaystyle C(\bs{\gamma}^*)\exp(0.5\bs{B}'_{\gamma^*}\bs{V}^{-1}_{\gamma^*}\bs{B}_{\gamma^*})}
{\displaystyle C(\bs{\gamma})\exp(0.5\bs{B}'_{\gamma}\bs{V}^{-1}_{\gamma}\bs{B}_{\gamma})}
\frac{\displaystyle \pi_k}{\displaystyle 1 - \pi_k} & \mbox{if }\gamma_k = 0\end{array}\right\}\\\nonumber
&=& \min\left\{1, \begin{array}{ll}
\frac{\displaystyle |\bs{V}_{\gamma^*}|^{1/2} |\bs{v}_{\gamma}|^{1/2}}{\displaystyle |\bs{V}_{\gamma}|^{1/2} |\bs{v}_{\gamma^*}|^{1/2}}
\frac{\displaystyle \exp(0.5\bs{B}'_{\gamma^*}\bs{V}^{-1}_{\gamma^*}\bs{B}_{\gamma^*})}{\displaystyle \exp(0.5\bs{B}'_{\gamma}\bs{V}^{-1}_{\gamma}\bs{B}_{\gamma})}
\frac{\displaystyle 1 - \pi_k}{\displaystyle \pi_k} & \mbox{if }\gamma_k = 1\\
\frac{\displaystyle |\bs{V}_{\gamma^*}|^{1/2} |\bs{v}_{\gamma}|^{1/2}}{\displaystyle|\bs{V}_{\gamma}|^{1/2} |\bs{v}_{\gamma^*}|^{1/2}}
\frac{\displaystyle\exp(0.5\bs{B}'_{\gamma^*}\bs{V}^{-1}_{\gamma^*}\bs{B}_{\gamma^*})}{\displaystyle\exp(0.5\bs{B}'_{\gamma}\bs{V}^{-1}_{\gamma}\bs{B}_{\gamma})}
\frac{\displaystyle\pi_k}{\displaystyle 1 - \pi_k} & \mbox{if }\gamma_k = 0\end{array}\right\}.
\end{eqnarray}
Note that the acceptance probability for a \emph{Gibbs
sampler}, updating either the complete $\bs{\gamma}$ vector or a
subset of components $\bs{\gamma}_I = (\gamma_i)_{i\in \bs{I}}$ by the
conditional distribution $p(\bs{\gamma}_I|\bs{\gamma}_{-I},\bs{z},\bs{\lambda},\bs{x})$,
is always equal to one:
\begin{eqnarray}
\alpha(\bs{\gamma},\bs{\beta}_\gamma) &=& \min\left\{1,
\frac{p(\bs{\beta}^*_\gamma,\bs{\gamma}^*| \bs{z},\bs{\lambda},\bs{x})}{p(\bs{\beta}_{\gamma},\bs{\gamma}| \bs{z},\bs{\lambda},\bs{x})}
\frac{q(\bs{\beta}_{\gamma},\bs{\gamma})}{q(\bs{\beta}^*_{\gamma},\bs{\gamma}^*)}
\right\}\nonumber\\
&=& \min\left\{1,
\frac{p(\bs{\beta}^*_\gamma|\bs{\gamma}^*,\bs{z},\bs{\lambda},\bs{x})p(\bs{\gamma}^*|\bs{z},\bs{\lambda},\bs{x})}
{p(\bs{\beta}_\gamma|\bs{\gamma},\bs{z},\bs{\lambda},\bs{x})p(\bs{\gamma}|\bs{z},\bs{\lambda},\bs{x})}
\frac{p(\bs{\beta}_{\gamma}|\bs{\gamma},\bs{z},\bs{\lambda},\bs{x})p(\bs{\gamma}_I|\bs{\gamma}_{-I},\bs{z},\bs{\lambda},\bs{x})}
{p(\bs{\beta}^*_{\gamma}|\bs{\gamma}^*,\bs{z},\bs{\lambda},\bs{x})p(\bs{\gamma}^*_I|\bs{\gamma}^*_{-I},\bs{z},\bs{\lambda},\bs{x})}
\right\}\nonumber\\
&=& \min\left\{1,
\frac{\prod_{i=1}^p \pi_i^{\gamma^*_i} (1 - \pi_i)^{1-\gamma^*_i}}{\prod_{i=1}^p \pi_i^{\gamma_i} (1 - \pi_i)^{1-\gamma_i}}
\frac{\prod_{i \in \bs{I}} \pi_i^{\gamma_i} (1 - \pi_i)^{1-\gamma_i}}{\prod_{i \in \bs{I}} \pi_i^{\gamma^*_i} (1 - \pi_i)^{1-\gamma^*_i}}
\right\} = 1.
\end{eqnarray}

\subsection{\label{appPT}Sampling from a tempered distribution}
When the parallel tempering algorithm is applied to the
logistic BVS model in Section \ref{realdata}, the hierarchical
model is as follows.
\begin{eqnarray*}
y_{j} & = & \left\{\begin{array}{ll}1 & \mbox{if }z_{\gamma j}>0\\0
& \mbox{otherwise}\end{array}\right.\\
z_{\gamma j} & = & x_{\gamma j}\bs{\beta}_\gamma + \epsilon_{j}\\
\epsilon_{j} & \sim & N(0,T\lambda_{jj})\\
\lambda_{jj} & = & (2\phi_{j})^2\\
\phi_{j} & \sim & \mbox{Kolmogorov-Smirnov (i.i.d.)}\\
\bs{\beta}_\gamma & \sim & N(\bs{b}_\gamma = \bs{0}_{p_\gamma}, \bs{v}_\gamma = c^2 \bs{I}_{p_\gamma})\\
\bs{\gamma} & \sim & p(\bs{\gamma}) =
\prod_{i=1}^p{\pi_i^{\gamma_i}(1-\pi_i)^{1-\gamma_i}}.
\end{eqnarray*}
This corresponds to the joint posterior distribution
\begin{eqnarray*}
p_T(\bs{\beta}_\gamma, \bs{\gamma}, \bs{z}, \bs{\lambda} | \bs{x}, \bs{y}) & \propto &
p_T(\bs{\beta}_\gamma, \bs{\gamma}, \bs{z}, \bs{\lambda}, \bs{y} | \bs{x}) \\
& = & p(\bs{y}|\bs{z}) p_T(\bs{z}|\bs{\lambda}, \bs{\beta}, \bs{\gamma}, \bs{x}) p(\bs{\beta}_\gamma | \bs{\gamma})
p(\bs{\gamma}) p(\bs{\lambda}),
\end{eqnarray*}
where the prior distributions $p(\bs{\gamma})$, $p(\bs{\beta}_\gamma)$,
and $p(\lambda_{jj}) \sim \frac{1}{4\sqrt{\lambda_{jj}}}KS(0.5
\sqrt{\lambda_{jj}})$ are as previously, but the likelihood
$p_T(\bs{z}|\bs{\lambda}, \bs{\beta}, \bs{\gamma}, \bs{x}) = N_z(\bs{x}_\gamma \bs{\beta}_\gamma,
T\bs{\lambda})$ is tempered with a temperature parameter $T > 1$,
which is a scalar factor multiplied to the diagonal covariance
matrix $\bs{\lambda}$. Sampling is done via a Gibbs algorithm
corresponding to the algorithm described in Appendix
\ref{appGibbs}, i.e. we sample from the full conditionals
$p_T(\bs{z},\bs{\lambda}|\bs{\beta},\bs{\gamma},\bs{x},\bs{y})$ and
$p_T(\bs{\beta}_\gamma,\bs{\gamma}|\bs{z},\bs{\lambda},\bs{x})$ with
\begin{itemize}
\item $p_T(\bs{z},\bs{\lambda}|\bs{\beta},\bs{\gamma},\bs{x},\bs{y}) =
    p_T(\bs{z}|\bs{\beta},\bs{\gamma},\bs{x},\bs{y})p_T(\bs{\lambda}|\bs{z},\bs{\beta},\bs{\gamma},\bs{x})$
    with
\begin{itemize}
\item $p_T(z_j|\bs{\beta},\bs{\gamma},\bs{x},\bs{y}) =
\left\{\begin{array}{ll}\mbox{Logistic}(\bs{x}_{\gamma j}\bs{\beta}_\gamma, \sqrt{T}) I(z_j > 0),    &y_j=1\\
                        \mbox{Logistic}(\bs{x}_{\gamma j}\bs{\beta}_\gamma,\sqrt{T})I(z_j\leq 0),&y_j=0
\end{array}\right.$

\item $p_T(\lambda_{jj}|\bs{z},\bs{\beta},\bs{\gamma},\bs{x}) \propto
    p_T(z_j|\bs{\lambda}, \bs{\beta},\bs{\gamma}, \bs{x})p(\lambda_{jj}) =
    N(\bs{x}_{\gamma j} \bs{\beta}_\gamma, T\lambda_{jj})
    \frac{1}{4\sqrt{\lambda_{jj}}}KS(0.5
    \sqrt{\lambda_{jj}})$
\end{itemize}

\item $p_T(\bs{\beta}_\gamma,\bs{\gamma}|\bs{z},\bs{\lambda},\bs{x}) =
    p_T(\bs{\gamma}|\bs{z},\bs{\lambda},\bs{x})p_T(\bs{\beta}_\gamma|\bs{\gamma},\bs{z},\bs{\lambda},\bs{x})$
    with
\begin{itemize}
\item $p_T(\bs{\gamma}|\bs{z},\bs{\lambda},\bs{x}) \propto
    p_T(\bs{z}|\bs{\lambda},\bs{\gamma},\bs{x})p(\bs{\gamma}) = N(\bs{0}_{n}, T\bs{\lambda} +
    \bs{x}_\gamma \bs{v}_\gamma \bs{x}_\gamma')
    \prod_{i=1}^p{\pi_i^{\gamma_i}(1-\pi_i)^{1-\gamma_i}}$

\item $p_T(\bs{\beta}_\gamma|\bs{\gamma},\bs{z},\bs{\lambda},\bs{x}) =
    N(\bs{B}_\gamma, \bs{V}_\gamma)$

where $\bs{B}_\gamma = \bs{V}_\gamma \bs{x}_\gamma' (T\bs{\lambda})^{-1}\bs{z}$ and $\bs{V}_\gamma
=(\bs{x}_\gamma'(T\bs{\lambda})^{-1}\bs{x}_\gamma + \bs{v}_\gamma^{-1})^{-1}$. Note that $N(\bs{0}_{n},
T\bs{\lambda} + \bs{x}_\gamma \bs{v}_\gamma \bs{x}_\gamma') = N(\bs{0}_{n}, T(\bs{\lambda}^{-1} - \bs{\lambda}^{-1}
\bs{x}_\gamma T^{-1}\bs{V}_\gamma \bs{x}_\gamma'\bs{\lambda}^{-1})^{-1})$ according to the
Sherman-Morrison-Woodbury matrix inversion formula \citep{schott97}.
\end{itemize}
\end{itemize}
For a derivation of these formulae refer to the
paragraph below. When proposing to exchange the values of
$\theta_1 = (\bs{\beta}_{\gamma 1},\bs{\gamma}_1,\bs{\lambda}_1,\bs{z}_1)$ sampled
from the distribution of temperature $T_1$ and $\theta_2 =
(\bs{\beta}_{\gamma 2},\bs{\gamma}_2,\bs{\lambda}_2,\bs{z}_2)$ from the
distribution of temperature $T_2$, then the acceptance
probability $\alpha_{12}$ is given as
\begin{eqnarray*}
\alpha_{12} &=& \min\left\{1, \frac{p_{T_1}(\bs{z}_2|\bs{\beta}_2,\bs{\gamma}_2,\bs{\lambda}_2,\bs{x}_2) p_{T_2}(\bs{z}_1|\bs{\beta}_1,\bs{\gamma}_1,\bs{\lambda}_1,\bs{x}_1)}
{p_{T_2}(\bs{z}_2|\bs{\beta}_2,\bs{\gamma}_2,\bs{\lambda}_2,\bs{x}_2)p_{T_1}(\bs{z}_1|\bs{\beta}_1,\bs{\gamma}_1,\bs{\lambda}_1,\bs{x}_1)}\right\} \\
&=& \min\left\{1, \frac{N_{\bs{z}_2}(\bs{x}_{\gamma 2}\bs{\beta}_{\gamma 2}, T_1\bs{\lambda}_2)
N_{z_1}(\bs{x}_{\gamma 1}\bs{\beta}_{\gamma 1}, T_2\bs{\lambda}_1)} {N_{z_2}(x_{\gamma
2}\bs{\beta}_{\gamma 2}, T_2\bs{\lambda}_2)
N_{z_1}(\bs{x}_{\gamma 1}\bs{\beta}_{\gamma 1}, T_1\bs{\lambda}_1)}\right\} \\
&=& \min\left\{1,\frac{((2\pi)^n|T_1\bs{\lambda}_2|)^{-1/2} \exp(-\frac{1}{2T_1}(\bs{z}_2
- \bs{x}_{\gamma 2}\bs{\beta}_{\gamma 2})'\bs{\lambda}_2^{-1}(\bs{z}_2 - \bs{x}_{\gamma 2}\bs{\beta}_{\gamma
2}))}{((2\pi)^n|T_2\bs{\lambda}_2|)^{-1/2} \exp(-\frac{1}{2T_2}(\bs{z}_2 - \bs{x}_{\gamma
2}\bs{\beta}_{\gamma 2})'\bs{\lambda}_2^{-1}(\bs{z}_2 - \bs{x}_{\gamma
2}\bs{\beta}_{\gamma 2}))} \times \right.\\
&& \hspace{3.5cm}\left.\frac{((2\pi)^n|T_2\bs{\lambda}_1|)^{-1/2}
\exp(-\frac{1}{2T_2}(\bs{z}_1 - \bs{x}_{\gamma 1}\bs{\beta}_{\gamma 1})'\bs{\lambda}_1^{-1}(\bs{z}_1 -
\bs{x}_{\gamma 1}\bs{\beta}_{\gamma 1}))}{ ((2\pi)^n|T_1\bs{\lambda}_1|)^{-1/2}
\exp(-\frac{1}{2T_1}(\bs{z}_1 - \bs{x}_{\gamma 1}\bs{\beta}_{\gamma
1})'\bs{\lambda}_1^{-1}(\bs{z}_1 - \bs{x}_{\gamma 1}\bs{\beta}_{\gamma 1}))}\right\} \\
&=& \min\left\{1, \exp\left((\frac{1}{T_1} - \frac{1}{T_2})\right.\right.\\
&\times& \quad\quad\left.\left.(-\frac{1}{2}(\bs{z}_2 - \bs{x}_{\gamma 2}\bs{\beta}_{\gamma
2})'\bs{\lambda}_2^{-1}(\bs{z}_2 - \bs{x}_{\gamma 2}\bs{\beta}_{\gamma 2}) + \frac{1}{2}(\bs{z}_1 -
\bs{x}_{\gamma 1}\bs{\beta}_{\gamma 1})'\bs{\lambda}_1^{-1}(\bs{z}_1 - \bs{x}_{\gamma 1}\bs{\beta}_{\gamma
1}))\right)\right\}.
\end{eqnarray*}

In this paragraph we derive how the conditional distributions
in the Gibbs sampler change from the untempered to the tempered
distribution (as outlined above).
\begin{itemize}
\item $\mathbf{p_T(z_j|\bs{\beta},\bs{\gamma},\bs{x},\bs{y})}$: We want to show that (for any vector $\bs{m}$ of length $n$)
\begin{eqnarray}
p_T(\bs{z}|\bs{m})&=&\int{p_T(\bs{z}|\bs{m},\bs{\lambda})p(\bs{\lambda})d\bs{\lambda}}
=\int{N_z(\bs{m},T\bs{\lambda})\frac{1}{4\sqrt{\bs{\lambda}}}KS(0.5\sqrt{\bs{\lambda}})d\bs{\lambda}}\label{pt1}\\
&=&\mbox{Logistic}(\bs{m}, \sqrt{T})
=\frac{1}{\sqrt{T}}\exp(-\frac{\bs{z}-m}{\sqrt{T}})(1 +
\exp(-\frac{\bs{z}-m}{\sqrt{T}}))^{-2}:\label{pt2}
\end{eqnarray}
It is known that
\begin{eqnarray}
p(\bs{z})&=&\int{p(\bs{z}|\bs{\lambda})p(\bs{\lambda})d\bs{\lambda}}
=\int{N_z(\bs{0}_{n},\bs{\lambda})\frac{1}{4\sqrt{\bs{\lambda}}}KS(0.5\sqrt{\bs{\lambda}})d\bs{\lambda}}\label{pt3}\\
&=&\mbox{Logistic}(0, 1) =\exp(-\bs{z})(1 + \exp(-\bs{z}))^{-2}.\label{pt4}
\end{eqnarray}
It is easy to see that the variable $\bs{z}^* = \sqrt{T} \bs{z} + \bs{m}$ has the density from
equation (\ref{pt2}) if $\bs{z}$ has the standard logistic density (equation
\ref{pt4}). Hence, the same variable transformation will change the integral in
equation (\ref{pt3}) to the form in equation (\ref{pt1}).

\item $\mathbf{p_T(\lambda_{jj}|\bs{z},\bs{\beta},\bs{\gamma},\bs{x})}$: Rejection
    sampling with acceptance probability

$\alpha(\lambda_{jj}) =
\frac{\ell(r_j^2,\lambda_{jj})p(\lambda_{jj})}{M g(\lambda_{jj})}$
with $r_j^2 = (z_j - \bs{x}_{\gamma j}\bs{\beta}_\gamma)^2$ and
$\ell(r_j^2,\lambda_{jj}) = p_T(z_j|\bs{x}_{\gamma
j},\bs{\beta}_{\gamma},\lambda_{jj}) = N_{z_j}(\bs{x}_{\gamma
j}\bs{\beta}_{\gamma},T\lambda_{jj})$.

Here, $g(\lambda_{jj})$ is the rejection sampling density
$g(\lambda_{jj}) = T
c(|r_j|)(T\lambda_{jj})^{-1/2}\exp(-0.5(\frac{r_j^2}{T\lambda_{jj}}
+ T\lambda_{jj}))$ with $c(|r_j|)$ being a normalising
constant not dependent on $T$ and $\lambda_{jj}$ with linear
transformation $\lambda_{jj} = T^{-1}X$ where $X \sim
GIG(0.5,1,r_j^2) = |r_j|/IG(1,|r_j|)$, $GIG$ denoting the
Generalised Inverse Gaussian and $IG$ the Inverse Gaussian
densities.

This leads to
\begin{eqnarray*}
\alpha(\lambda_{jj}) &=& \exp(0.5 T\lambda_{jj})T^{-1}p(\lambda_{jj})\\
&=& \exp(0.5 T\lambda_{jj})T^{-1} c h(0.5\sqrt{\lambda_{jj}})(1 -
a_1(0.5\sqrt{\lambda_{jj}}) + a_2(0.5\sqrt{\lambda_{jj}}) - ...),
\end{eqnarray*}
where $c h$ and $a_1, a_2,...$ are from an alternate series
expansion for the Kolmogorov-Smirnov density
$KS(0.5\sqrt{\lambda_{jj}})$, equivalently to the untempered
situation.

\item $\mathbf{p_T(\bs{\gamma}|\bs{z},\bs{\lambda},\bs{x})}$: With
    $p_T(\bs{z}|\bs{\gamma},\bs{\beta},\bs{\lambda},\bs{x}) =
    N_z(\bs{x}_\gamma\bs{\beta}_\gamma, T\bs{\lambda})$ and
    $p(\bs{\beta}_\gamma|\bs{\gamma}) = N_{\bs{\beta}_\gamma}(\bs{b}_\gamma=\bs{0}_{p_\gamma},
    \bs{v}_\gamma=c^2 \bs{I}_{p_\gamma})$ it follows for the marginal
    distribution \citep[e.g.][]{lindley72}
\begin{eqnarray*}
p_T(\bs{z}|\bs{\gamma},\bs{\lambda},\bs{x}) =
\int{p_T(\bs{z}|\bs{\beta},\bs{\gamma},\bs{\lambda},\bs{x})p(\bs{\beta}_\gamma|\bs{\gamma})d\bs{\beta}_\gamma} =
N_z(\bs{0}_{n}, T\bs{\lambda} + \bs{x}_\gamma \bs{v}_\gamma \bs{x}_\gamma').
\end{eqnarray*}

\item $\mathbf{p_T(\bs{\beta}_\gamma|\bs{\gamma},\bs{z},\bs{\lambda},\bs{x})}$: With
    $p_T(\bs{z}|\bs{\beta},\bs{\gamma},\bs{\lambda},\bs{x}) =
    N_z(\bs{x}_\gamma\bs{\beta}_\gamma, T\bs{\lambda})$ and
    $p(\bs{\beta}_\gamma|\bs{\gamma}) = N_{\bs{\beta}_\gamma}(\bs{b}_\gamma=\bs{0}_{p_\gamma},
    \bs{v}_\gamma=c^2 \bs{I}_{p_\gamma})$ it follows for the
    posterior distribution \citep{lindley72}
\begin{eqnarray*}
p_T(\bs{\beta}_\gamma|\bs{\gamma},\bs{z},\bs{\lambda},\bs{x}) &=&
N(\bs{B}_\gamma,\bs{V}_\gamma) \\
\bs{V}_\gamma &=& (\bs{x}_\gamma'(T\bs{\lambda})^{-1}\bs{x}_\gamma +
\bs{v}_\gamma^{-1})^{-1} \\
\bs{B}_\gamma &=& \bs{V}_\gamma \bs{x}_\gamma' (T\bs{\lambda})^{-1}\bs{z}.
\end{eqnarray*}
\end{itemize}

\end{document}